\documentclass[aps,prb,twocolumn,floats,nofootinbib]{revtex4-2}
\usepackage{epsfig}
\usepackage{tikz}
\usepackage{amsmath, amssymb}
\usepackage{graphicx}
\usepackage{soul}
\usepackage{braket}
\usepackage[colorlinks=true,linkcolor=blue]{hyperref}
\usepackage{subfigure}
\usepackage{color}
\usepackage{orcidlink}

\newcommand{\blue}{\textcolor{blue}}

\begin{document}
\title{ Quantum Tunneling-induced Hybridization and Coherent Dynamics of Jackiw-Rebbi Zero Modes in a Modified Su-Schrieffer-Heeger Chain}
\author{Surajit Mandal$^{1,2}$\orcidlink{0000-0002-9091-0539}}
\email{surajitmandalju@gmail.com}
\homepage{https://orcid.org/0000-0002-9091-0539}
\affiliation{$^1$Department of Physics, Jadavpur University, Kolkata - 700032, West Bengal, India\\
$^2$Department of Physics, AKPC Mahavidyalaya, Bengai - 712611, West Bengal, India}

\begin{abstract}
We investigate analytically and numerically the tunneling-induced hybridization and coherent dynamics of Jackiw-Rebbi zero modes in a modified Su-Schrieffer-Heeger (SSH) model. Unlike the conventional SSH model, this modified system possess two bulk gap-closing points, namely the quadratic-type gap-closing point at momenta $k_0=0$ and the Dirac-type gap-closing points at $k_0=\pm\pi/4a$. Although the quadratic point does not support a topological domain wall due to the absence of mass inversion, the low-energy Dirac theory around $k_0=\pm\pi/4a$ predicts an effective mass that changes sign at two spatially separated domain walls under a kink profile, generating a pair of JR bound states with each states localized at one of those domain walls. We show that the finite spatial overlap between the JR zero modes lifts the zero-energy degeneracy through quantum tunneling, producing symmetric-antisymmetric hybridized states. Using an effective two-level description, we demonstrate that hybridized JR modes undergo coherent tunneling oscillations between the two domain walls, accompanied by periodic transfer of sublattice polarization between the (A,~C) and (B,~D) sectors. Our results establish a unified framework connecting JR zero modes, quantum tunneling, and coherent dynamics in modified SSH systems, offering a promising platform for controllable topological quantum-state transfer in engineered lattice structures.
\end{abstract}
\maketitle
\section{Introduction}\label{sec1}
The discovery of localized zero-energy states (ZESs) bound to topological defects represents one of the most profound developments at the interface of quantum field theory and condensed-matter physics\cite{rebbi}. In 1976, a seminal work by Jackiw and Rebbi demonstrated that a one-dimensional (1D) massive Dirac fermion coupled to a spatially varying soliton mass field can support a localized topological state at the mass domain wall, familiar as the Jackiw–Rebbi modes (JRMs)\cite{rebbi}, which can show fractional charge excitations\cite{rebbi,jackiw,jackiw1,goldstone,jackiw2}. In recent years, the study of JRMs has attracted considerable interest in modern condensed matter physics because of their exotic nature and fractional charge excitations. In particular, they can be a feasible candidate for performing topological quantum computation\cite{nayak}, and their non-Abelian braiding properties have been investigated in several theoretical studies\cite{rev1,rev2,rev3,rev4}. Moreover, the JRMs subsequently emerged as a paradigmatic example of a topological defect state and have found manifestations in a wide range of physical systems, including conducting polymers\cite{heeger,heeger1}, topological insulators\cite{hasan,qi}, photonic lattices\cite{photon1,photon2}, as well as photonic and acoustic metamaterials\cite{m1,m2,m3,m4}. Recently, the realization of JRMs has been extended beyond lattice systems to non-uniform topological insulator nanowires with external magnetic flux\cite{arijit}.

The most popular condensed-matter realization of the JRMs is provided by the Su-Schrieffer-Heeger (SSH) model, which was originally introduced in the context of soliton excitations in polyacetylene. Owing to its chiral symmetry and dimerized structure, the model results in two degenerate ground states\cite{heeger1,su1,su2}. A domain wall connecting regions with opposite dimerized phases acts as a lattice analogue of the mass kink in the Jackiw–Rebbi (JR) problem and hosts exponentially localized midgap ZESs. The SSH model has therefore become a cornerstone in the study of 1D topological phases, offering a simple yet powerful platform to explore the interplay between symmetry, topology, and localized quantum states\cite{asboth,ryu,kiteav,scopa}. The outcome of having static domain walls (DWs) at an interface of the SSH chain\cite{d0} and some other models\cite{d1,d2,d3} has been probed to some extent in the literature.

In this respect, it is important to add here that the domain-wall bound state for the original SSH model can be understood from the low-energy Dirac theory, around the Dirac-type critical momentum $k_0=\pi/a$, where the bulk gap closes at the topological phase transition\cite{kane}. The resulting sign reversal of the Dirac mass or changing hopping pattern across the DW\cite{kane,mandal,scollon} gives rise to a localized JR zero mode. In contrast, the present model possess both a quadratic-type band-touching point at $k_0=0$ and Dirac-type critical points at $k_0=\pm\pi/4a$. In the pertinent study documented in Ref.~\cite{mandal}, the authors numerically examined the emergence of ZESs at the DW position for different commensurate frequencies. They did not found any zero mode at the DW location for $\theta=\pi/2$ (corresponding to the modified SSH model considered in the present work, to be discussed later). Instead, zero modes remain localized at the two ends of the finite chain. In the present work, we provide a clear theoretical understanding of the same. Specifically, we analytically show that the effective mass of the continuum Dirac Hamiltonian around the quadratic point does not undergo a topological sign inversion under the kink profile and therefore cannot support a JRM. Thus, in Ref.\cite{mandal}, the authors numerically studied the system in the parameter regime corresponding to the absence of mass inversion at the quadratic band-touching point. In Particular, we analytically show that the continuum Hamiltonian around the Dirac-type critical points yields an effective Dirac mass that undergoes two successive sign inversions near the chain center, giving rise to the formation of two DWs capable of supporting two zero-energy JRMs. Additionally, we study the same by numerical diagonalization of the real-space Hamiltonian.

However, the hybridization of spatially separated DW states, the associated bonding–antibonding energy splitting, and the emergence of effective low-energy models have been investigated in earlier studies. For example, Ref.~\cite{d1} developed an effective bipartite lattice description of a periodic array of DW states and explored the topological properties of that bipartite lattice. The relation between chiral solitons, bonding-antibonding combinations, and fractional boundary charges in polyacetylene and graphene nanoribbons was discussed in Ref.~\cite{soliton-poly}. These studies establish important aspects of domain-wall and soliton physics. Furthermore, the hybridization and energy splitting of topological bound states have emerged as central themes in studies of Majorana modes and related topological excitations, highlighting the broader relevance of mode hybridization in topological matter and quantum information platforms\cite{kiteav,alicea,majorana,majorana1,majorana2,majorana3}. In particular, the energy splitting of near-zero-energy Majorana modes has been observed experimentally, exhibiting an exponential suppression of the splitting with increasing wire length\cite{mzm1}. Moreover, the finite overlap of the two Majorana modes results in hybridized finite-energy Bogoliubov quasiparticle states\cite{mzm2}.

In contrast, the present work considers zero-energy JRMs generated by mass inversions at two spatially separated DWs in a modified SSH chain, rather than the periodic set of DWs in a bipartite lattice\cite{d1} or self-conjugate Majorana quasiparticles of a topological superconducting system\cite{mzm2}. In particular, low-energy sectors of the model around $k_0=\pm\pi/4a$ exhibit two spatially separated JRMs under the kink profile. Their finite spatial overlap gives rise to the hybridized states and an associated tunnel splitting. We further establish a quantitative connection between the DWs separation, overlap, tunnel splitting, the resulting coherent oscillations and sublattice polarization between the two JRMs. Although the study of spatial overlap-induced hybridization of JRMs is analogous to that encountered for spatially separated Majorana zero modes, the hybridization due to coherent quantum tunneling in the present modified system has not been explored in this context. Specifically, the  detailed dependence of the tunnel splitting on the DW separation is determined by the spatial structure and spatial overlap between the JRMs and exhibits Gaussian and asymptotic exponential regimes in the present system. Therefore, while the hybridization of spatially separated JRMs is conceptually related to Majorana-mode hybridization, the microscopic origin, symmetry structure, spatial dependence of the tunnel splitting, and resulting coherent dynamics studied here are specific to the modified SSH system. More recently, transport-based identification of braiding fidelity has been investigated, highlighting the relevance of hybridization between JRMs~\cite{ge-wu}; this further motivates us to investigate the hybridization and coherent dynamics of the JRMS in the modified SSH chain. Additionally, most previous studies have addressed the domain-wall physics in SSH-like systems\cite{mandal}, extended SSH chains\cite{hemant}, and Majorana-mode hybridization\cite{hodge,bena}, to some extent. Nevertheless, to the best of our knowledge, the tunneling dynamics of JRMs in the context of the modified SSH model with four-site periodicity has not yet been reported.

Thus, in contrast to these studies, the present work focuses on the hybridization due to quantum tunneling and tunneling dynamics of JRMs in a modified SSH chain by analyzing both the effective two-state Hamiltonian and numerical diagonalization of real-space Hamiltonian. Specifically, we demonstrate that the symmetric-antisymmetric JRMs in the spatial overlap regime form a topological analogue of bonding-antibonding states in a quantum-mechanical double-well system, where finite tunneling transforms two degenerate zero modes into bonding-antibonding states with small energy splitting\cite{q1,q2,griffiths,shankar}. Moreover, the coherent oscillations between the JRMs, analogous to Rabi oscillations in a two-level system\cite{qm,dynamic}, provide a direct dynamical signature of tunneling between topological JRMs. Remarkably, quantum tunneling in double-well potentials provides one of the most fundamental paradigms for understanding coherent dynamics in two-level quantum systems\cite{t1,t2,t3}. The phenomenon underlies a broad range of physical processes spanning vastly different length scales, from microscopic examples such as the inversion dynamics of the ammonia molecule\cite{t4} and electron transfer between coupled quantum dots\cite{t5} to macroscopic manifestations including Josephson oscillations in Bose–Einstein condensates\cite{t6,t7,t8,t9} and quantum tunneling in superconducting quantum interference devices\cite{t10,t11,t12,t13}. Crucially, the two JRMs bound together by quantum tunneling can be viewed as a “{\it Jackiw-Rebbi molecule}”, analogous to the bonding and antibonding molecular orbitals in the molecular ion $H_{2}^{+}$\cite{q1,q2,griffiths,grosso}.

The paper is organized as follows. In Sec. \ref{sec2}, we present a brief introduction to the modified SSH chain. This section also includes the phase transition of this modified SSH system. Section \ref{secJR} discusses zero-energy JRMs both analytically and numerically. This section shows the existence of isolated zero-energy JRMs for sufficiently large DW separation and also illustrates how JR zero modes become near-zero-energy hybridized JRMs, via quantum tunneling, for finite DW separation. In Sec. \ref{tunneling oscillations}, we study the coherent tunneling dynamics of hybridized JRMs via real-time evolution. Finally, in Sec. \ref{summary}, we summarize and conclude our findings and discuss possible future directions and applications of our work.

\section{Modified SSH Chain}\label{sec2}
The Hamiltonian for the 1D SSH chain of $L=M*N$ (in which $M$ and $N$ denote the number of sublattices and unit cells, respectively) sites with periodically modulated hopping strength can be defined, under open boundary conditions (OBC), as\cite{mandal,topology}:
\begin{equation}\label{1}
  \mathcal{H}_{\operatorname{SSH}}=\sum_{i}^{L-1}(t+\delta_{i})[\ket{i}\bra{i+1}+{\rm H.c.}],
\end{equation}
where $\delta_{i}=\Delta\cos[(i-1)\theta]$ with $i=1,2,3,......, n$ specifies the periodic modulation in the nearest-neighbor hopping strength $t$ and $\rm H.c.$ denotes the Hermitian conjugate. Generally, one obtains $\delta_{i+1}=\Delta\cos(2\pi i/M)$ for a commensurate frequency $\theta=\frac{2\pi}{M}$ with $\Delta$ representing the hopping modulation parameter. Therefore, the system is represented by a $M\times M$ Hamiltonian matrix having $M$ number of eigenmodes. Crucially, the commensurate frequencies such as $\theta=\pi,~\pi/2~\pi/3$ and $\pi/4$ produce, respectively, the two, four, six, and eight sublattices in a unit cell, giving the increasing periodicity of the system for decreasing frequency. For the sake of simplicity, we consider $t=1$ as the unit of energy, which further makes the unit of physical parameters dimensionless throughout this work. However, all the parameters considered here are real.

\begin{figure}
     \vskip -.1 in
   \begin{picture}(100,100)
     \put(-90,0){
  \includegraphics[width=.52\linewidth, height=1.28 in]{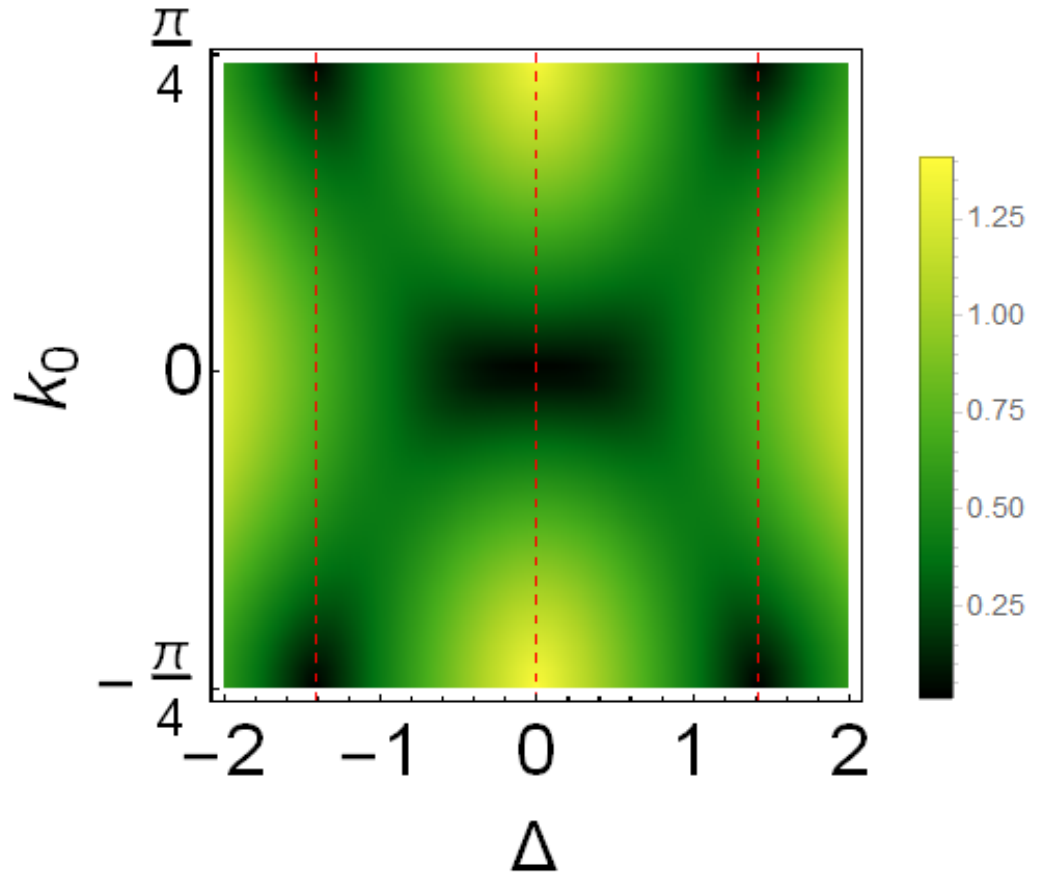}
  \includegraphics[width=.52\linewidth, height=1.25 in]{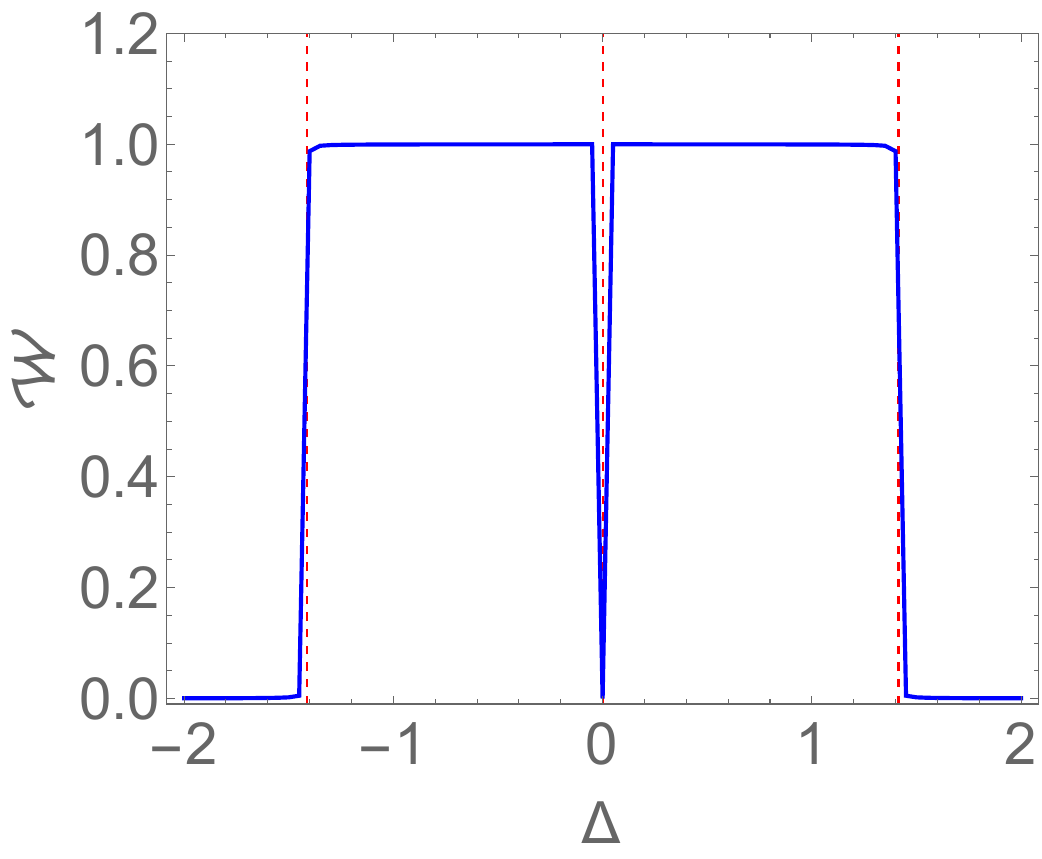}}
     \put(-40,65){(a)}
     \put(100,60){(b)}
    \end{picture}
\caption{(a) {\bf Bulk gap closing} in ($\Delta-k_0$)- parameter space. (b) The variation {\bf bulk winding number} with $\Delta$. Here, the vertical red dashed line indicates the location of the Dirac-type gap-closing point at $|\Delta/t|=\sqrt{2}$ and the quadratic-type gap-closing point at $\Delta=0$. The black region in the color bar in (a) corresponds to a vanishing bulk gap, indicating the gap-closing points.} 
\label{gap}
\end{figure}
We should mention here that the above model (Eq. (\ref{1})) for $\theta=\pi$ corroborates with the original SSH model having two atom basis unit cells\cite{su1,mandal,topology,kar}. However, the lattice periodicity increases twice for $\theta=\pi/2$, with the number of sublattices now becoming four. We shall discuss the rest of the physics by considering this four-sublattice basis SSH chain, called as the modified SSH model. Before studying the quantum tunneling, appearance of hybridized JRMs and coherent tunneling dynamics of these JRMs, we shall briefly discuss about the phase transition of this modified SSH model. To do so, the Bloch Hamiltonian in the $k$-space for the model under periodic boundary conditions (PBC) yields\cite{mandal}
\begin{equation}\label{11}
\mathcal{H}(k) = 
\begin{pmatrix}
0 & (t+\Delta) & 0 & te^{-i4ak} \\
(t+\Delta) & 0 & t & 0 \\
0 & t &0 & (t-\Delta) \\
te^{i4ak} & 0 & (t-\Delta) & 0
\end{pmatrix},
\end{equation}
Now, the dispersion relation turns out to be
\begin{align}\label{12}
E(k)=\pm\sqrt{2t^2+\Delta^2\pm t\sqrt{2t^2+6\Delta^2+2(t^2-\Delta^2) \cos4ak}},
\end{align}
thus, the bulk band touches at $E(k)=0$ for the momenta given by $k_0=\pm\frac{1}{4a} \operatorname{arc}\cos\Big[\frac{2t^2-2t^2\Delta^2}{2t^2(t^2-\Delta^2)}\Big]$ with $a$ denoting the lattice spacing. Therefore, the bulk energy gap-closing point can be observed not only at the boundary of the reduced first Brillouin zone (BZ) ($k_0=\pm\pi/4a$) at $|\Delta/t|=\sqrt{2}$, but also at the BZ center ($k_0=0$) for $\Delta=0$ (see Fig. \ref{gap}(a)). The energy band gap is $E_{gap}\sim|2t^2-\Delta^2|$ at $k_0=\pm\pi/4a$ for $|\Delta/t|\ne\sqrt{2}$ and $E_{gap}\sim2\sqrt{2}ta|k_0\mp\pi/4a|$ at $|\Delta/t|=\sqrt{2}$ for $k_0\ne\pm\pi/4a$, whereas it becomes $E_{gap}\sim |\Delta|^2$ at $k_0=0$ when $\Delta\ne0$ and $E_{gap}\sim 4ta|k_0|$ for $\Delta=0$ when $k_0\ne0$. Thus, under PBC, the gap closes quadratically exactly at $\Delta=0$ in $E-\Delta$ diagram, although Dirac-type linear band touching in noticed at $k_0=0$ in $E-k$ diagram. In contrast, a linear band touching is noticed both in the $E-\Delta$ (at $|\Delta/t|=\sqrt{2}$) and the $E-k$ diagram (at $k_0=\pm\pi/4a$). While under OBC, a Dirac-type linear gap-closing can be observed at $|\Delta/t|=\sqrt{2}$, and a quadratic-type gap-closing is discernible exactly at $\Delta=0$ in the $E-\Delta$ diagram\cite{crg}. Importantly, for the system under OBC, one can notice a small gap at these two points due to finite-size effects, which, however, vanishes in the thermodynamic limit\cite{soliton-poly,mandal}. It is exemplary that the Hamiltonian in Eq. (\ref{11}) respects the chiral symmetry $\{\Gamma_{4},\mathcal{H}(k)\}=0$ with the chiral operator $\Gamma_{4}=\operatorname{diag}\{1,-1,1,-1\}$. Additionally,  the Hamiltonian also respects the particle-hole symmetry along with time-reversal symmetry and falls in the BDI class in a tenfold way classification scheme\cite{mandal,bdi,bdi1}. Thus, the topology of the model can be characterized by the global bulk topological invariant, the winding number\cite{w1,w2,w3,w4}.

The bulk winding number changes on either side of $|\Delta/t|=\sqrt{2}$, leading to the visibility of a ``first-order'' (as the modulation parameter $\Delta$ enters linearly in $E_{gap}$ or effective mass) non-trivial topological phase transition point only at $k_0=\pm\pi/4a$. On the other hand, the winding number does not change in the vicinity of the $\Delta=0$ point, reflecting it as a trivial metallic point due to a ``second-order'' (because $\Delta$ enters quadratically in $E_{gap}$) trivial topological phase transition\cite{crg}. One can notice the non-trivial topological phase with ${\mathcal W}=1$ within the range $0<|\Delta/t|<\sqrt{2}$ while the trivial topological phase with ${\mathcal W}=0$ occurs when $\Delta=0$ or $|\Delta/t|>\sqrt{2}$ (notice Fig. \ref{gap}(b))\cite{mandal}. Similar to the usual SSH chain ($\theta=\pi$ case for Hamiltonian Eq. (\ref{1})), two ZES can be found here; however, unlike that scenario, this configuration also admits two mid-gap states having non-zero energy\cite{mandal}. The features of the ZES have been previously studied in Ref.\cite{mandal}, which exhibit their edge localization. 

We now consider the DW configuration and study the tunneling-induced hybridization of the JRMs, if any, in the following sections.

\section{Study of Zero Energy Jackiw-Rebbi Modes}\label{secJR}
In this section, we seek to investigate the zero-energy JRMs, which, in the context of the SSH chain is one of the most beautiful and fundamental concepts in topological condensed matter physics. It represents a lattice-bound realization of a famous particle physics phenomenon: fractional quantum numbers and zero-energy bound states\cite{jackiw}. We should mention here that the topologically protected zero-energy JRMs can be discernible at the topological defect \cite{dw}, which can be created by changing the hopping arrangement at a certain lattice site or by flipping the fermion mass at a certain point\cite{rebbi,jackiw,jackiw1,scollon,mandal}. The celebrated JR mode can be achieved both analytically and numerically, as discussed below. Specifically, the analytical study is based on the continuum limit of the momentum-space Hamiltonian within the JR framework, while the numerical investigation is carried out by exact diagonalization of the corresponding real-space Hamiltonian.

\begin{figure}
   \vskip -.1 in
   \begin{picture}(100,100)
     \put(-70,0){
  \includegraphics[width=.45\linewidth, height=1.25 in]{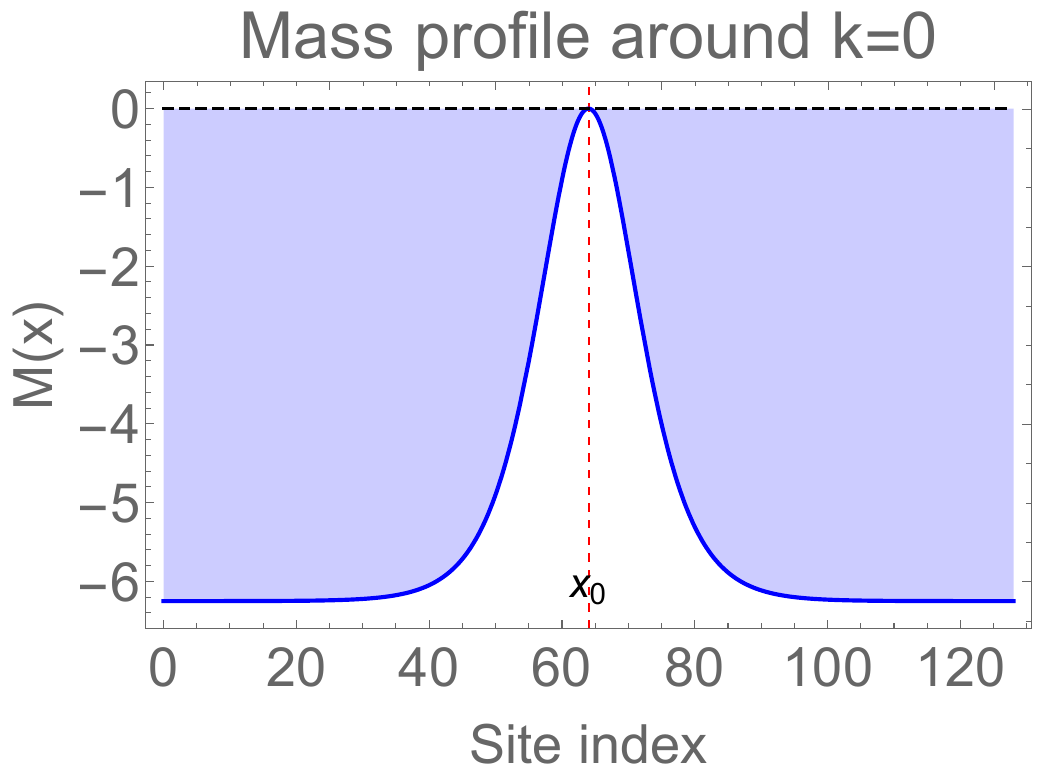}
  \includegraphics[width=.45\linewidth, height=1.25 in]{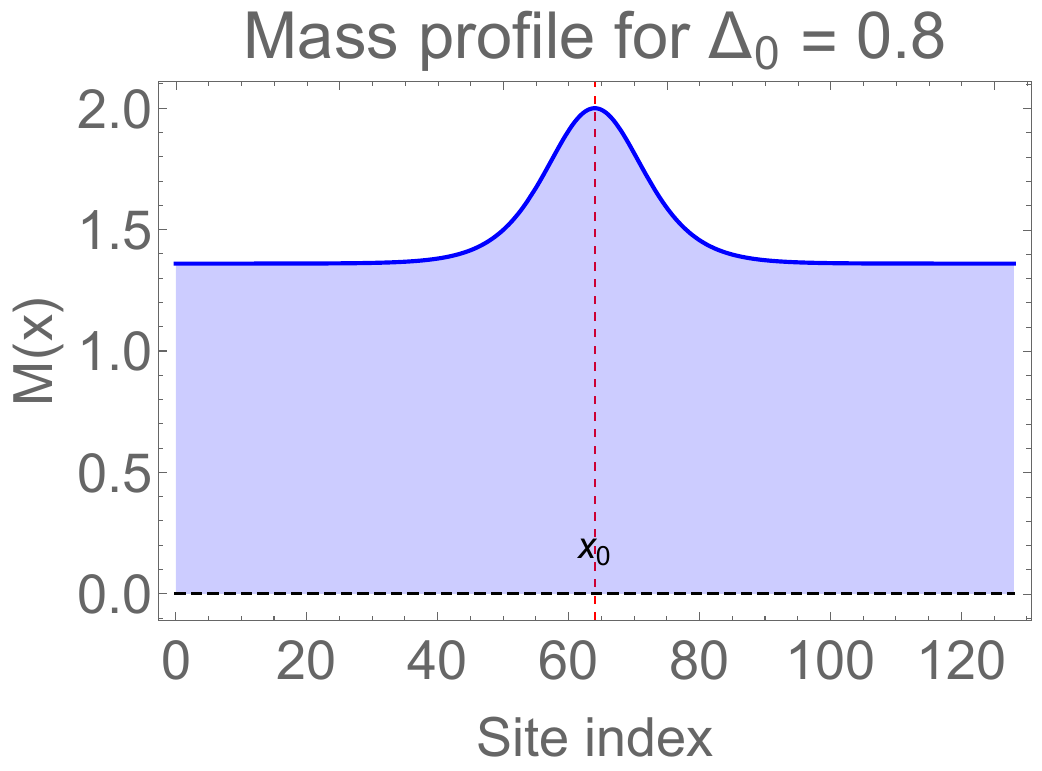}}
     \put(-40,66){(a)}
     \put(125,66){(b)}
    \end{picture}\\
     \vskip -.00004 in
   \begin{picture}(100,100)
     \put(-70,0){
\includegraphics[width=.45\linewidth,height=1.35 in]{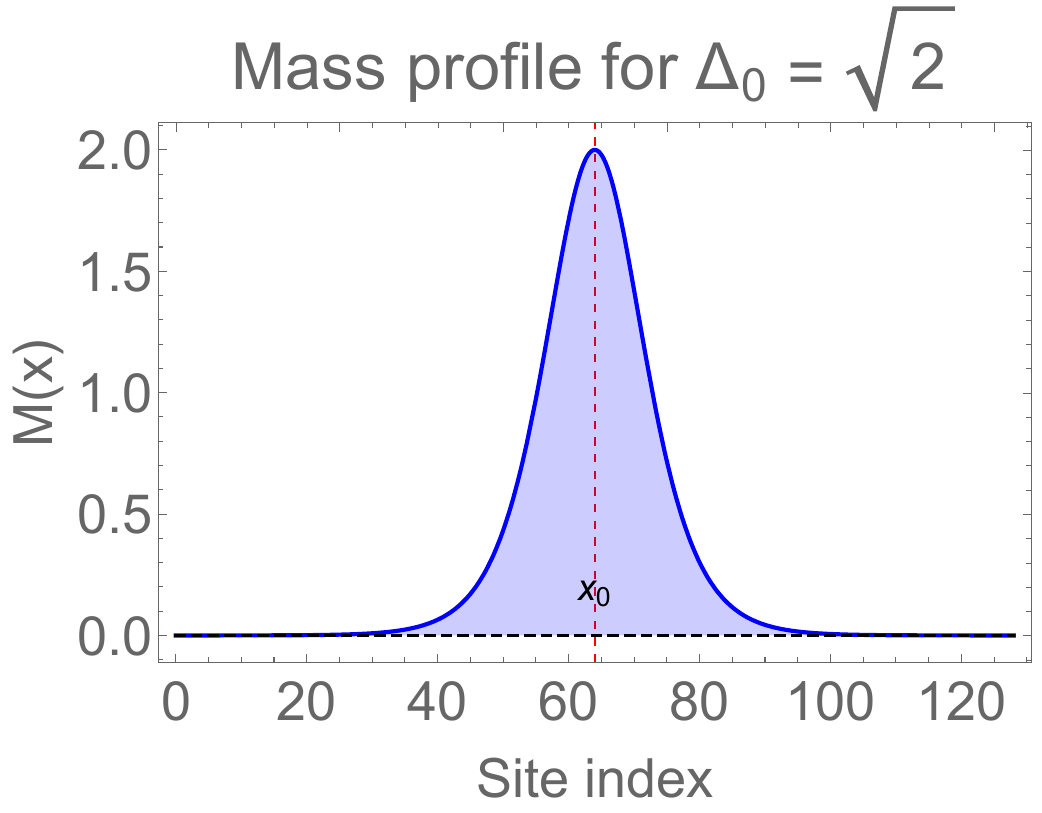}
    \includegraphics[width=.45\linewidth, height=1.28 in]{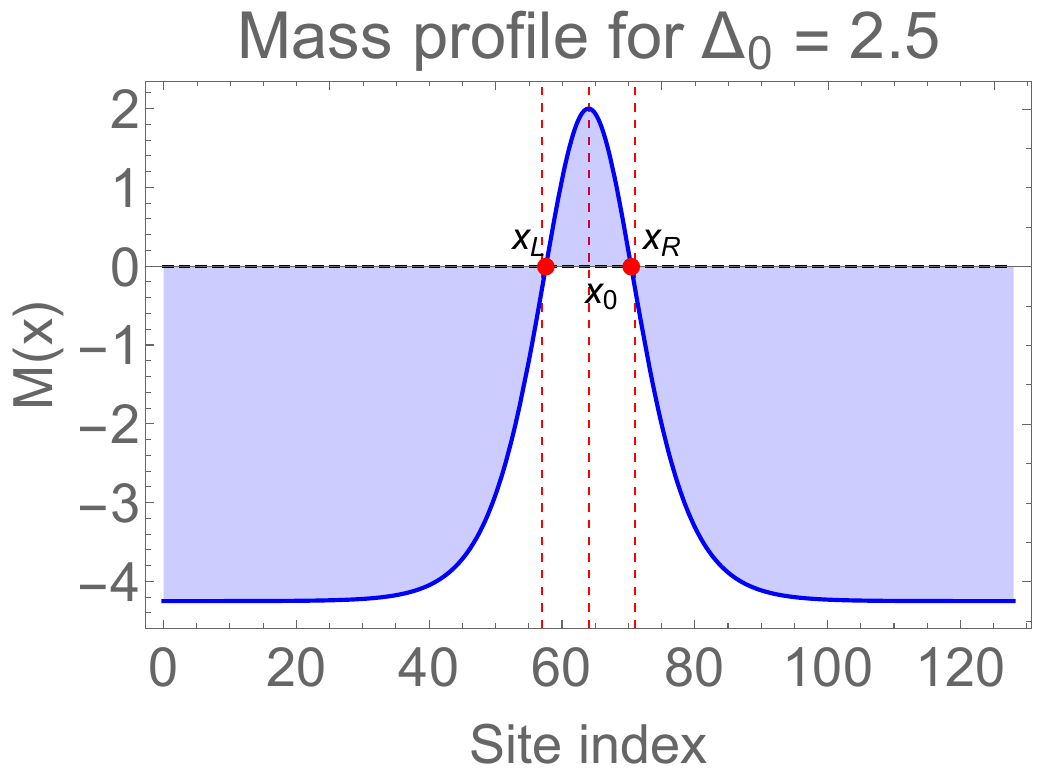}}
   \put(15,66){(c)}
    \put(125,68){(d)}
   \end{picture}
  \vskip -0.1 in
\caption{The profile for effective Dirac mass, $M(x)=-\Delta^2(x)$ around $k_0=0$ (a) and for effective Dirac mass $M(x)=2t^2-\Delta_{0}^2\tanh^2(\frac{x-x_{0}}{\xi/a})$ near $k_0=\pm\pi/4a$ with (b) $\Delta_{0}<\sqrt{2}t$, (c) $\Delta_{0}=\sqrt{2}t$, and (d) $\Delta_{0}>\sqrt{2}t$. In (d), $M(x)$ is not like the standard JR mass kink (Eq. (\ref{kink}) of Appendix \ref{app:analytical}); instead, it is symmetric about $x_{0}$, producing two mass zeros at $x_{L}$ and $x_{R}$, and it is like a double DW (kink-antikink pair) mass profile. Other parameters are $L=128$, $\xi/a=10$.} 
\label{dw2}
\end{figure}
\subsection{Analytical Study}\label{ana}
For the analytical study, we utilize the JR method, which considers the continuum limit of the momentum-space Hamiltonian as discussed in Appendix~\ref{app:analytical}. Before proceeding further, we need to mention that the present model includes both a quadratic band-touching point at $k_0=0$ and a Dirac-type critical point at $k_0=\pm\pi/4a$. Now, expanding the Bloch Hamiltonian (Eq. (\ref{11})) around $k_0=0$ yields a position-dependent effective Dirac mass $M(x)=-\Delta^2(x)$, which does not change sign for a kink-type profile (Eq. (\ref{kink}) of Appendix \ref{app:analytical}) as shown in Fig. \ref{dw2}(a). Therefore, a topologically protected JRM is not expected at the DW position. The JRMs, instead, arise from the Dirac theories around $k_0=\pm\pi/4a$ as discussed below.

We should mention here that although the modified SSH chain has a four-sublattice basis unit cell, the JR method (discussed in Appendix~\ref{app:analytical}) can still be applied to the present model. This is because the Bloch Hamiltonian in Eq. (\ref{11}) can be brought into block off-diagonal form due to the preservation of chiral symmetry\cite{mandal}. The determinant of the off-diagonal block defines the complex topological function ${\bf Det}[V(k)]=h_{x}(k)-ih_{y}(k)$, from which an effective two-band Hamiltonian can be constructed as (see Appendix~\ref{hamila} for the detailed discussion)
\begin{equation}\label{block}
\mathcal{H_{\mathrm{eff}}}(k)=h_{x}(k)\sigma_{x}+h_{y}(k)\sigma_{y}
\end{equation}

In the continuum limit, expanding Eq.~(\ref{block}) about each of the low-energy points $k_0=\pm\pi/4a$ and replacing $(k-k_0)\rightarrow-i\partial_x$ yields the effective Dirac Hamiltonian as (see Appendix~\ref{hamila} for the derivation)
\begin{equation}\label{low}
\mathcal{H}_{\mathrm{eff}}^D(x)=M(x)\sigma_{x}-i\hbar v\sigma_{y}\partial_{x}+\mathcal{O}(\partial_x^2)
 \end{equation}
where the velocity is $v=4at^2/\hbar$ and the spatially dependent effective Dirac mass $M(x)=2t^2-\Delta^2(x)$. Equivalently, the low-energy Dirac Hamiltonian can also be obtained by projecting the Bloch Hamiltonian of Eq.~\eqref{11} onto the two-dimensional low-energy subspace as discussed in Appendix~\ref{hamila}. Remarkably, the quadratic dependence of the effective mass on hopping modulation follows directly from the determinant of the chiral off-diagonal block rather than from an arbitrary choice of basis, making it unique and basis-independent (see Appendix~\ref{hamila}). This contrasts with the conventional SSH model, in which the effective Dirac mass is typically linear in the dimerization parameter near the topological transition\cite{kane}.

Here, we impose the kink-shaped mass profile on $\Delta(x)$, rather than directly on the Dirac mass $M(x)$. Following Eq. (\ref{kink}), we take $\Delta(x)=\Delta_{0}\tanh\Big(\frac{x-x_{0}}{\xi/a}\Big)$ and thus the Dirac mass to be $M(x)=2t^2-\Delta_{0}^2\tanh^2\Big(\frac{x-x_{0}}{\xi/a}\Big)$. For DW amplitude $\Delta_{0}<\sqrt{2}t$, $M(x)$ remains positive and does not undergo a sign change about the center of the chain (i.e., at $x=x_0$ as we fix it at the chain center), thereby creating no DWs, while $M(x)$ for $\Delta_{0}=\sqrt{2}t$ approaches zero asymptotically as $|x|\rightarrow\infty$, indicating the bulk gap-closing point and the DW positions move to infinity. Interestingly, the effective mass changes sign twice in the vicinity of the chain center for $\Delta_{0}>\sqrt{2}t$. Thus, two DWs are discernible at $x_{L}=x_{0}-x_{\star}$ and $x_{R}=x_{0}+x_{\star}$ and their separation becoming $D_{s}~=(x_{R}-x_{L})=2x_{\star}$ with $x_{\star}=\frac{\xi}{a} \operatorname{arctanh}
\left(
\frac{\sqrt{2}\,t}{\Delta_0}
\right)$. A distinctive feature of the present modified SSH chain is that the inhomogeneous effective Dirac mass has a quadratic dependence on $\Delta$, which generates a pair of DWs rather than the explicit construction of a single DW as in the conventional SSH chain\cite{kane}.

We now present the mass profile for different values of $\Delta_{0}$ in Fig. \ref{dw2}. Fig. \ref{dw2}(b) depicts a positive mass but without sign inversion, while Fig. \ref{dw2}(c) illustrates that the mass curve touches zero asymptotically, thereby making $\Delta_0=\sqrt{2}t$ a transition point. In contrast, Fig. \ref{dw2}(d) shows the appearance of two DWs exactly at $x_{L}$ and $x_{R}$. Interestingly, the existence of two DWs is analogous to the double-well potential in quantum mechanics\cite{qm}. Moreover, the finite positive mass within the region $x_{L}<x<x_{R}$ mimics the finite potential barrier in the double-well system\cite{qm}.

Therefore, the search for ZESs here is essential because the inversion of the sign is noticed at the positions $x_{L}$ and $x_{R}$ for $\Delta_{0}>\sqrt{2}t$. The ZESs, with domain wall configuration, come in pairs $\pm E$ due to the preservation of chiral symmetry\cite{kane}. To investigate the local structure of the ZESs, we take into account the linearized mass near each DW as $M(x)=M^{\prime}(x_{j})(x-x_{j})$ with $j=L,R$. Eq. (\ref{low}) then reads
\begin{equation}\label{linear}
 \mathcal{H}_{j}=M^{\prime}(x_{j})(x-x_{j})\sigma_{x}-iv\sigma_{y}\partial_{x}
 \end{equation}
 
Writing the envelope spinor as $\phi_j=\begin{pmatrix}
\phi_{j,1}(x) & \phi_{j,2}(x)\end{pmatrix}^T$,  the zero-energy equation $\mathcal{H}_{j}\phi_{j}=0$ admits normalizable zero-mode solutions with a definite chirality, corresponding to eigenstates of the effective chiral operator $\Gamma_{\rm eff}=\sigma_z$ with eigenvalues $\pm 1$. The zero-energy equation yields two independent solutions as
\begin{equation}
\begin{aligned}
\phi_{j,1}(x)&=A_j\exp\Bigg[-\frac{M'(x_j)}{2\hbar v_D}(x-x_j)^2\Bigg],\\
\phi_{j,2}(x)&=B_j\exp\Bigg[+\frac{M'(x_j)}{2\hbar v_D}(x-x_j)^2\Bigg].
\end{aligned}
\end{equation}
where $A_j$ and $B_j$ are normalization constants when the corresponding solutions are normalizable. Normalizability selects the first component at the left DW where $M'(x_L)>0$, and the second component at the right DW where $M'(x_R)<0$ and thus only one solution is normalizable at each DW. Accordingly, writing $\phi_{j}=f_{j}(x)\chi_{s_j}$, where $\sigma_{z}\chi_{s_j}=s\chi_{s_j}$ with $s_j=\pm 1$, one obtains the normalized envelope at either DW as $f_{j}(x)=\Big(\frac{\alpha}{\pi}\Big)^{1/4} \exp\Big[-\frac{\alpha}{2}(x-x_{j})^2\Big]$. Therefore, the normalized solutions for the two-component envelope spinor can be estimated as
\begin{equation}\label{envelope}
 \phi_{j}(x)=\Big(\frac{\alpha}{\pi}\Big)^{1/4} \exp\Big[-\frac{\alpha}{2}(x-x_{j})^2\Big]\chi_{s_j}
 \end{equation}
with $\alpha=\frac{|M^{\prime}(x_{j})|}{v}$, and the localization length being $l_G=1/\sqrt{\alpha}=\sqrt{\frac{\sqrt{2}t\xi\Delta_0}{(\Delta_0^2-2t^2)}}$. However, the chirality of Eq. (\ref{envelope}) can be determined by the sign of the mass gradient through the normalizability condition for the Gaussian envelope $f_j(x)$. Specifically, for the Dirac Hamiltonian in Eq.~(\ref{linear}), the normalizable solution satisfies $s_j=\operatorname{sgn}[M'(x_j)]$. Since $M'(x_L)>0$ and $M'(x_R)<0$, the corresponding chiral spinors are $\chi_{L}=\begin{pmatrix}
1 & 0\end{pmatrix}^T$ and $\chi_{R}=\begin{pmatrix}
0 & 1\end{pmatrix}^T$, respectively.

\begin{figure}
   \vskip -.1 in
   \begin{picture}(100,100)
     \put(-80,0){
  \includegraphics[width=.46\linewidth, height=1.25 in]{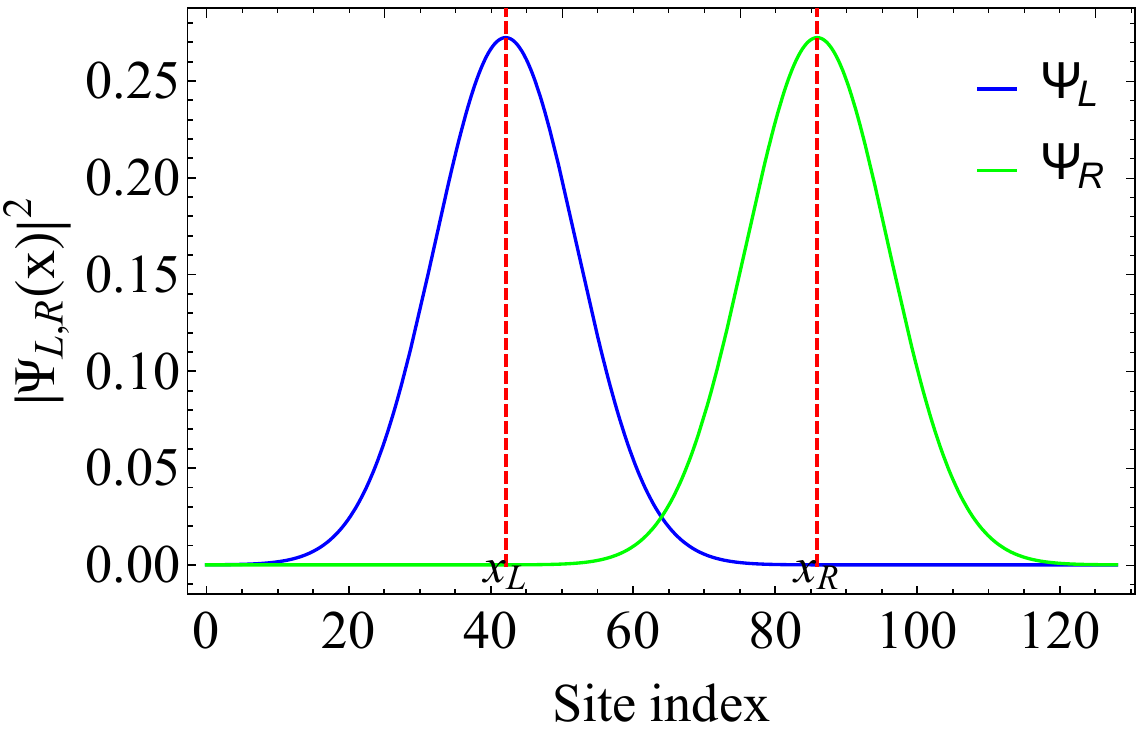}
  \includegraphics[width=.45\linewidth, height=1.25 in]{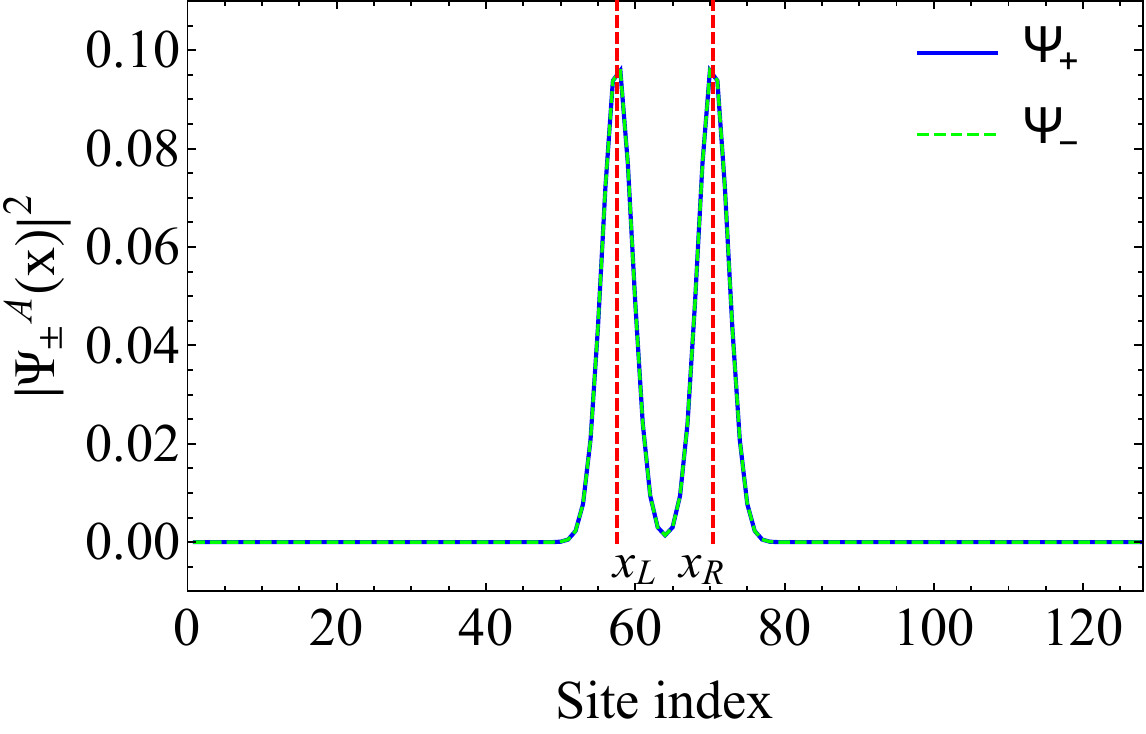}}
     \put(-50,68){(a)}
     \put(80,68){(b)}
    \end{picture}\\
     \vskip -.00004 in
   \begin{picture}(100,100)
     \put(-80,0){
\includegraphics[width=.46\linewidth,height=1.25 in]{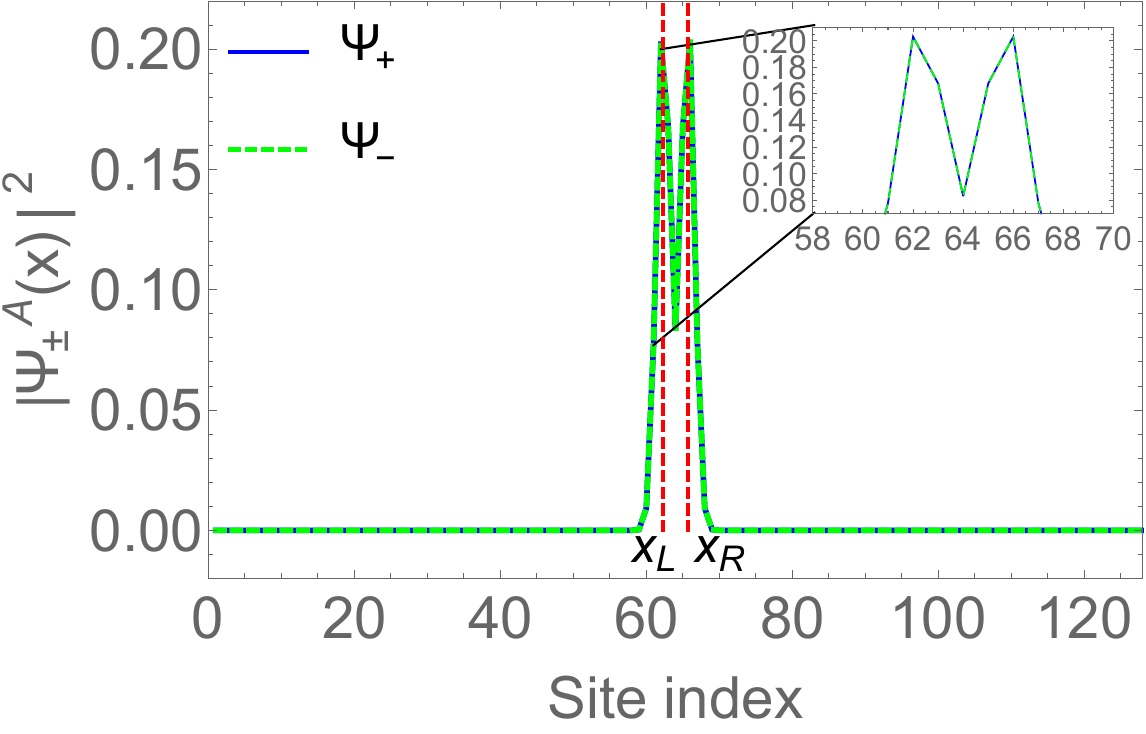}
\includegraphics[width=.46\linewidth,height=1.25 in]{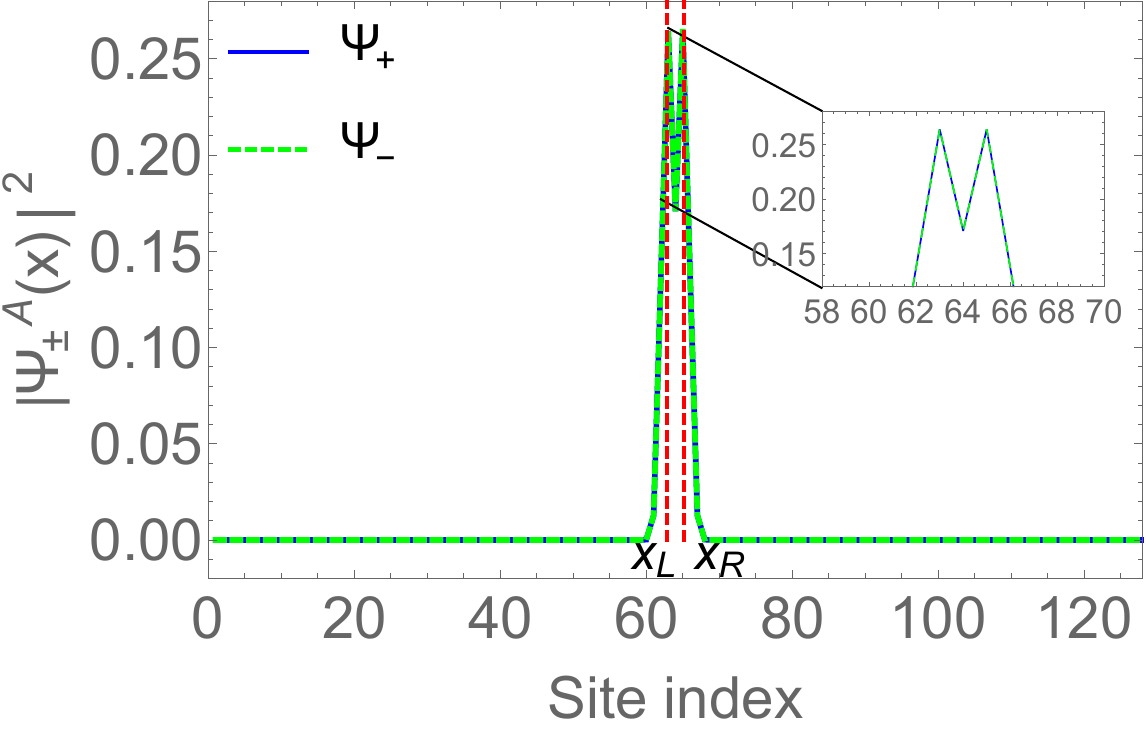}}
   \put(-40,50){(c)}
   \put(80,50){(d)}
   \end{picture}
  \vskip -0.1 in
\caption{{\bf The analytical probability density of zero energy JRMs $\Psi_{L,R}(x)$} (Eq.~\eqref{JRsolution}) with $\Delta_{0}=1.45$, $D_{s}=43.8248$ (a) and  {\bf two hybridized JRMs} $\Psi_{\pm}(x)$ (Eq. \ref{hybridized}) (b)--(d). In (b), $\Delta_{0}=2.5$, $D_{s}=12.8231$, (c)  $\Delta_{0}=8.5$, $D_{s}=3.35879$ while in (d) $\Delta_{0}=12.5$, $D_{s}=2.27247$ . The figure shows that each peak is positioned at $x_{L}$ and $x_{R}$. However, insets in panels (c) and (d) provide a magnified view in the positive mass region, illustrating the enhanced overlap of the JRMs as the DW separation decreases. For each plot, we set $L=128$ and $\xi/a=10$.} 
\label{figzesaa}
\end{figure}

Now, in the microscopic four-sublattice basis, the interface states become (see Appendix~\ref{hamila})
\begin{equation}
\left.
\begin{aligned}
\Psi_{L}(x)&=f_{L}(x)u_{\operatorname{AC}}^{(-)}\\
\Psi_{R}(x)&=f_{R}(x)u_{\operatorname{BD}}^{(+)}
\end{aligned}
\right \}
\label{JRsolution}
\end{equation}
It is evident from Eq.~\eqref{JRsolution} that the mode $\Psi_{L}(x)$ has a Gaussian envelope $f_{L}(x)$ that peaks at $x_{L}$ and carries weight on the A and C sublattices. While $\Psi_{R}(x)$ has an envelope $f_{R}(x)$ that peaks at $x_{R}$ and is entirely confined to the B and D sublattices. The states $\Psi_{L}(x)$ and $\Psi_{R}(x)$ are the zero modes obtained by solving the Dirac equation near each DW independently. Importantly, these two modes are four-component independent (or isolated) JRMs having exact zero-energy when the DWs are infinitely far apart, i.e., $D_{s}\rightarrow\infty$, and these localized modes are eigenstates of the chiral operator $\Gamma_{4}$. To observe the localization behavior of the zero modes, $\Psi_{L,R}(x)$, we plot Fig. \ref{figzesaa}. Specifically, Fig. \ref{figzesaa}(a) reveals the individual localization of $\Psi_{L}(x)$ and $\Psi_{R}(x)$ at the left DW $x_{L}$ and right DW $x_{R}$, respectively, but a very negligible spatial overlap between the wavefunction tails is also visible in the intermediate positive mass region. This is solely due to the much larger separation $\operatorname{D_{s}}$ than the Gaussian localization length $l_G$. As soon as we move towards $\Delta_{0}\rightarrow\sqrt{2}t^{+}$, two peaks move infinitely far apart as $x_{\star}\rightarrow\infty$ and $D_{s}\rightarrow\infty$. Then the two modes would have strictly zero spatial overlap, exactly like bound states in two infinitely separated quantum wells\cite{q1,q2,shankar}.

We need to mention here that although the effective Dirac Hamiltonian $\mathcal H_{\mathrm{eff}}^{D}$ is represented by a $2\times2$ matrix acting on the two-component low-energy Dirac spinor
$\phi(x)=\begin{pmatrix}
\phi_{1}(x) & \phi_{2}(x)\end{pmatrix}^T$,
it defines a spatially extended continuum theory with an infinite-dimensional Hilbert space. In the following, we project $\mathcal H_{\rm eff}^D$ onto the subspace spanned by the two orthonormal localized JR states, $\{|\phi_L\rangle,|\phi_R\rangle\}$. This reduces the infinite-dimensional Hilbert space of the Dirac theory to a finite-dimensional $2\times2$ effective Hamiltonian in the localized JR state basis, with its off-diagonal matrix element describing their tunneling strength, as implemented below.
\\

{\bf Quantum Tunneling, Energy splitting, and Hybridized States: Effective two-level description:}
Once the separation between the DWs is finite, the Gaussian envelopes of two approximately independent JRMs, namely $f_{L}(x)$ and $f_{R}(x)$, generally exhibit a finite spatial overlap $S_{\rm env}=\int dx f_{L}^*(x)f_{R}(x)=e^{-\alpha x_*^2}$. The finite spatial overlap facilitates tunneling between the two DWs and contributes to the hybridization of the JRMs. The corresponding tunneling strength obtained from the matrix element $T=\langle\phi_L|\mathcal{H}_{\rm eff}^D|\phi_R\rangle$, where $\mathcal{H}_{\rm eff}^D$ is the continuum two-component Dirac Hamiltonian (see Eq. \eqref{low}) and $|\phi_{L}\rangle$, and $|\phi_{R}\rangle$ denote the two-component localized JR states (see Eq. \eqref{envelope})\cite{devices1}. Although the envelope overlap is nonzero, the full inner product of these two-component JR states vanishes because $\langle\chi_L|\chi_R\rangle=\chi_L^\dagger\chi_R=0$. Therefore, the localized zero-energy JR states $\{|\phi_{L}\rangle,|\phi_{R}\rangle\}$ form an orthonormal basis, as they are individually  normalized $\langle\phi_L|\phi_L\rangle=\langle\phi_R|\phi_R\rangle=1$, and their mutual overlap disappears $S=\langle\phi_L|\phi_R\rangle=0$. Now, projecting $\mathcal{H}_{\rm eff}^D$ onto the subspace spanned by the two localized JR states, $\{|\phi_{L}\rangle,|\phi_{R}\rangle\}$, gives the matrix elements $(\mathcal{H}_{\rm eff}^{P})_{ij}
=\langle\phi_i|\mathcal{H}_{\rm eff}^D|\phi_j\rangle$.

Noticeably, the continuum Hamiltonian $\mathcal{H}_{\rm eff}^D$ preserves chiral symmetry and the localized JR states
$|\phi_L\rangle$ and $|\phi_R\rangle$ are eigenstates of the effective chiral operator $\Gamma_{\rm eff}$ having opposite eigenvalues. Hence, their diagonal matrix elements vanish, while the off-diagonal matrix elements describe tunneling between the two JRMs. Consequently, the projected Hamiltonian, similar to the two-level Hamiltonian of a particle in a double well\cite{shankar}, leads to an effective two-state Hamiltonian as: $\mathcal{H}_{\mathrm{eff}}^P=\begin{pmatrix}
0 & T \\
T^{\star} & 0 
\end{pmatrix}$ 
with energies $E_\pm=\pm T$ for real and positive $T$ and an energy splitting $\Delta E=2T$, where $T$ denotes the tunneling strength between the two JRMs (see Appendix \ref{projection}). Notice that symmetric energy splitting arises from the chiral symmetry of $\mathcal{H}_{\rm eff}^D$\cite{devices1}. The energy splitting indicates the hybridization of the two isolated JRMs, $\phi_L(x)$ and $\phi_R(x)$, giving rise to symmetric and antisymmetric combinations of these JR states, which in the microscopic four-sublattice basis can be expressed as\cite{pollak,devices1}
\begin{equation}\label{hybridized}
|\Psi_{\pm}(x)\rangle=\frac{|\Psi_{L}(x)\rangle\pm|\Psi_{R}(x)\rangle}{\sqrt{2}},
\end{equation}
where $`+'~(`-')$ stands for symmetric (antisymmetric) state with normalization constant $1/\sqrt{2}$. We mention here that the microscopic JRMs are also individually normalized and mutually orthogonal. Importantly, changing the representation of the states does not, by itself, change the tunneling amplitude. It is worth noting that the effective Hamiltonian $\mathcal{H}_{\rm eff}^P$ also respects chiral symmetry with the effective chiral operator $\Gamma_{\rm eff}=\sigma_z$ in the $\{|\phi_{L}\rangle,|\phi_{R}\rangle\}$ subspace and the effect of chirality on the hybridized states is discussed later.
Although $\Psi_{\pm}(x)$ possess finite weight at both DWs, their sublattice polarization remains spatially separated. Near $x_{L}$, where $f_{L}(x_{L})\gg f_{R}(x_{L})$, the JRMs are dominated by the (A,~C) sector. In contrast, near $x_{R}$, where $f_{R}(x_{R})\gg f_{L}(x_{R})$, the JRMs reside predominantly in the (B,~D) sublattices. To illustrate the behavior of $\Psi_{\pm}(x)$ graphically, we plot Figs. \ref{figzesaa}(b)--(d), which exhibit weight at both DWs along with finite tunneling in the intermediate region $x_{L}<x<x_{R}$.

To estimate the tunneling strength between the two JRMs, we now focus on the spatial overlap region, where their envelopes can be approximated by Gaussian profiles. The tunneling strength can be estimated either by evaluating the integral, $T=\langle\phi_L|\mathcal{H}_{\rm eff}^D|\phi_R\rangle=\int_{-\infty}^{\infty}\phi_L^\dagger(x)\mathcal{H}_{\rm eff}^D(x)\phi_R(x)dx$\cite{d1,devices1}, explicitly or by using dimensional analysis of the characteristic energy scale of the continuum Dirac Hamiltonian\cite{scaling-a} (see Appendix \ref{projection} for details). Thus, the tunneling strength can be scaled as $T\sim \frac{v}{l_G}e^{-\alpha x_{\star}^2}$ with $\alpha=\frac{|M^{\prime}(x)|}{v}$ and $l_G=\alpha^{-1/2}$. Plugging $v$ and $M^{\prime}$ yields the following approximate form for the tunneling strength:
\begin{equation}\label{Tmatrix}
T\sim\sqrt{\frac{8\sqrt{2}a^2\Delta_{0}t^3}{\xi}\Big(1-\frac{2t^2}{\Delta_{0}^2}\Big)}\exp\Big[-\frac{\Delta_{0}}{\sqrt{2}t\xi}\Big(1-\frac{2t^2}{\Delta_{0}^2}\Big)x_{\star}^2\Big],
\end{equation} 
Equation~\eqref{Tmatrix} describes the tunneling coupling in the Gaussian-overlap regime. Its validity requires the existence of two hybridized JRMs, which occurs for $\Delta_0>\sqrt{2}t$, along with the continuum description, requiring the Gaussian localization length to be larger than the lattice spacing, $l_G\gg a$. Additionally, the two-level description remains valid as long as the hybridized low-energy doublet remains well separated from the bulk spectrum, i.e., $|T|\ll \Delta E_{\rm bulk}$, with $\Delta E_{\rm bulk}$ denoting the energy gap from near-zero-energy modes to the bulk modes, rather than by the magnitude of the spatial overlap alone.

\begin{figure}[htbp]
    \centering
    \subfigure[]{
\includegraphics[width=0.46\columnwidth,height=1.25in]{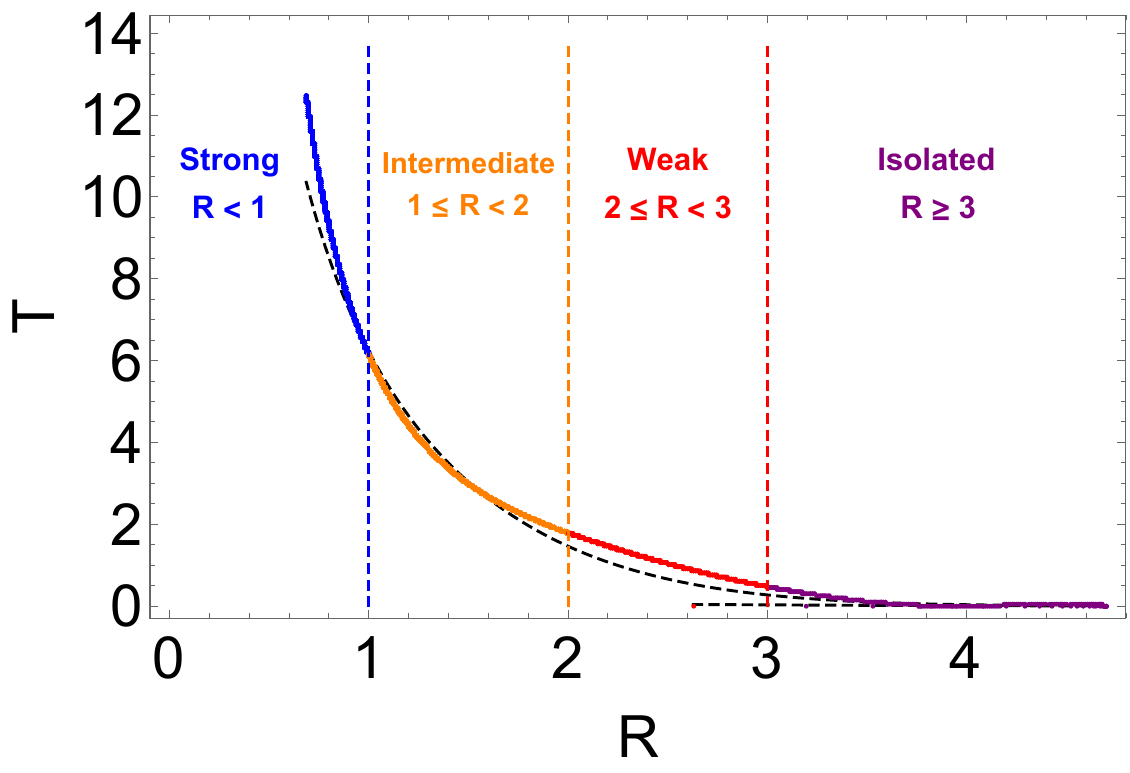}
        \label{fig:e1}}
    \hfill
    \subfigure[]{
\includegraphics[width=0.46\columnwidth,height=1.25in]{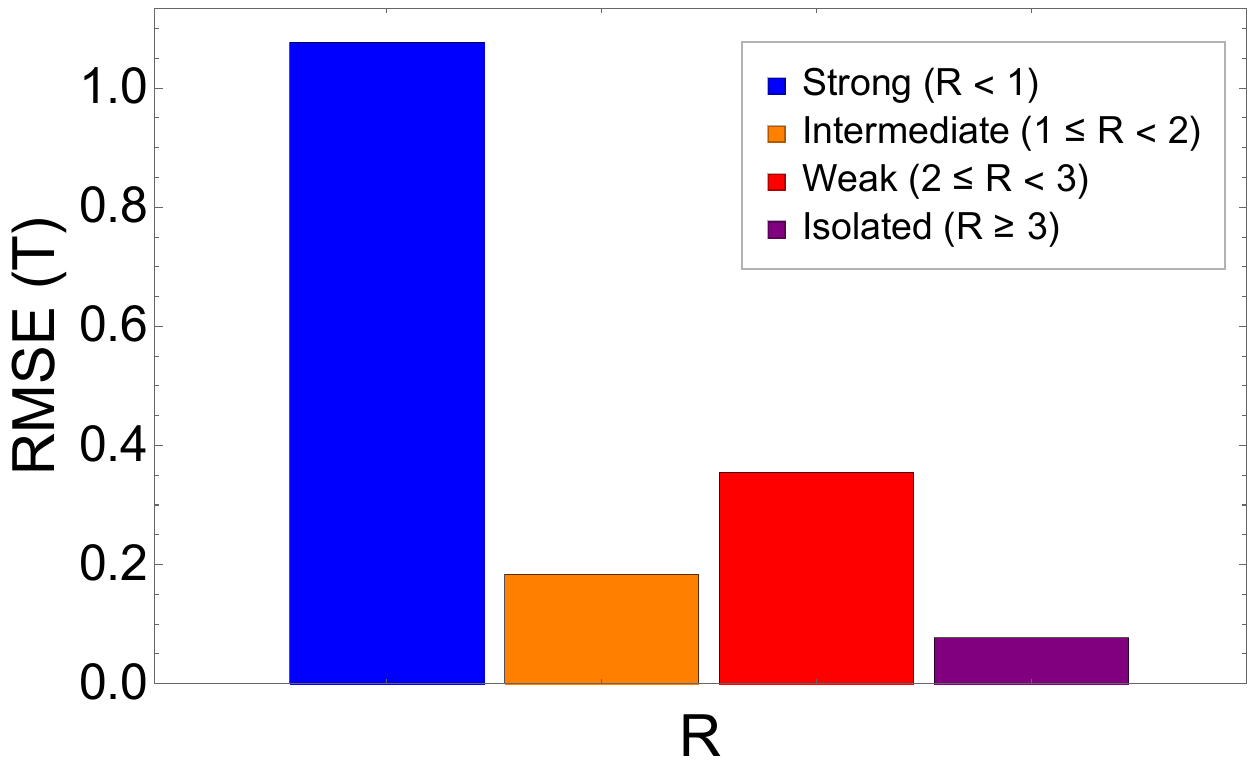}
        \label{fig:e1a}}
\caption{(a) {\bf Tunneling strength $T$} as a function of the dimensionless separation $R=D_s/l_G$ for $\xi=10$ and $L=128$. The color curve (represented by symbols) is obtained by exact diagonalization of Eq. \eqref{1} for $\theta=\pi/2$, and the black dashed curve is based on the analytical result predicted in Eq. \eqref{Tmatrix}. The calculations are performed by varying the DW amplitude over $\Delta_0\in[\sqrt{2}t+0.02t,120]$, thereby excluding the critical point $\Delta_0=\sqrt{2}t$, at which the DW separation $D_s$ diverges. The vertical dashed lines at $R= 1, 2, 3$ indicate phenomenological crossover criteria distinguishing different degrees of spatial overlap between the two JRMs; they do not represent phase transitions or sharp physical boundaries. (b) The {\bf RMSE-based goodness of agreement} between the analytical prediction and the exact diagonalization results across different overlap regimes. All other parameters are identical to (a).}
\label{splitting}
\end{figure}

The spatial overlap between the two JRMs decreases exponentially as $S_{\rm env}= e^{-R^2/4}$ with $R=D_s/l_G$ is a dimensionless ratio. However, Eq.~\eqref{Tmatrix} gives the tunneling strength as $T\sim T_0(\Delta_0)e^{-R^2/4}$, and hence $\Delta E\sim 2T_0(\Delta_0)e^{-R^2/4}$. Thus, $R$ provides a natural measure of the relative spatial separation between the two JRMs. In the limit $R\ll1$, the two JRMs exhibit strong spatial overlap and form a strongly hybridized low-energy doublet. In contrast, for $R\gg1$, their spatial overlap is exponentially suppressed, yielding an approximately isolated zero-energy doublet. For sufficiently large $R$, however, the spatial overlap is dominated by the asymptotic exponential tails of the JRMs rather than their Gaussian cores, resulting in $T\simeq T_0(\Delta_0) e^{- R l_G/l_D}$ and $\Delta E\simeq 2T_0(\Delta_0) e^{-R l_G/l_D}$, where $l_D=\kappa^{-1}$ characterizes the asymptotic localization length (see Appendix~\ref{asymptotic}). Moreover, the intermediate regime $R\sim1$ corresponds to a crossover from strong hybridization to weak overlap. For visualization, we further divide this crossover according to $R<1$, $1\le R<2$, $2\le R<3$, and $R\ge3$, referred to as strong, intermediate, weak, and effectively isolated spatial overlap regimes, respectively (see Fig. \ref{splitting}(a) for details).

Moreover, the hybridized low-energy doublet can remain well separated from the bulk spectrum even for appreciable spatial overlap, $R\lesssim1$. So $R\sim 1$ does not by itself signal a breakdown of the two-level description; instead, it marks the crossover between the weakly and strongly overlapping regimes. Particularly, in the limit $R\ll1$, corrections to the isolated JRMs approximations become increasingly important, and the continuum description may require additional corrections. Moreover, when the DW separation approaches the lattice scale, $D_s\lesssim a$, the continuum approximation ceases to be quantitatively reliable; therefore, the full lattice Hamiltonian is used to obtain the quantitative energy spectrum and hybridized states. Conversely, for sufficiently large $R$, the two-level description remains valid provided that the low-energy doublet remains well separated from the bulk states. In this weak-overlap regime, the Gaussian-core approximation to the tunneling is instead governed by the asymptotic exponential tails of the localized JRMs. Thus, the suppression of the tunneling strength in the Gaussian-overlap regime is directly controlled by the dimensionless ratio $R$, while in the asymptotic-tail regime the dependence is governed by $R$ together with the ratio $l_G/l_D$.

To validate the predicted exponential scaling laws for the analytical tunnel splitting, given in Eq. \eqref{Tmatrix}, we perform an exact diagonalization of the full lattice Hamiltonian mentioned in Eq. \eqref{1} for $\theta=\pi/2$. The numerical result is then compared directly with an analytical prediction, with all quantities, including the prefactor $T_0(\Delta_0)$ and the Gaussian exponent, evaluated from the model parameters without introducing any additional fitting parameters. Thus, to gain insight into the dependence of the tunneling strength (or energy splitting) on $R$, we plot Fig. \ref{splitting}(a). The figure clearly illustrates four phenomenological spatial overlap regimes, namely $R<1$, $1\le R<2$, $2\le R<3$, and $R\ge3$. These boundaries are phenomenological crossover criteria and are introduced only to quantify the degree of spatial overlap. The suppression of $T$ across these regimes, for increasing $R$, reflects the increasing relative separation between two JRMs. Nevertheless, the colored regions in Fig. \ref{splitting}(a) represent the degree of spatial overlap rather than specific intervals of $\Delta_0$, as each data point is determined solely by its actual value of $R$. 

Importantly, $D_s$ is a monotonic function of $\Delta_0$, whereas $R$ is not necessarily a monotonic function because $l_G$ also depends on $\Delta_0$. Consequently, the $T$ versus $R$ plot should be understood as a parametric representation obtained by varying $\Delta_0$, rather than as a one-to-one mapping between $R$ and $\Delta_0$. Thus, different values of $\Delta_0$ can therefore correspond to the same value of $R$ while yielding different tunnel splitting, which can result in a multivalued or branch-like representation. For the analytical result presented in Fig. \ref{splitting}(a), this branch-like behavior is further affected by the explicit $\Delta_0$-dependence prefactor $T_0(\Delta_0)$. In contrast, the numerical tunneling splitting is obtained directly from the lattice eigenvalue spectrum and does not involve such an explicit analytical decomposition into a Gaussian factor and a $\Delta_0$-dependent prefactor. Nevertheless, its representation as a function of $R$ can also exhibit a branch-like behavior because $R(\Delta_0)$ is non-monotonic. However, Fig. \ref{splitting}(a) presents excellent agreement between the analytical and numerical predictions.

Furthermore, to quantify the goodness of agreement between the analytical and the exact diagonalization results, we present a bar plot of the root mean square error (RMSE) of $T$ with $R$ in Fig.~\ref{splitting}(b). The RMSE varies with $R$ due to differences in the degree of hybridization and the resulting deviations from the analytical prediction. For $R\ll1$, as the continuum two-level description requires potential corrections, the analytical result may deviate more significantly from the exact diagonalization results, leading to a larger RMSE. In contrast, the smaller predicted error can be noticed in the isolated regime. However, the RMSE also depends on the magnitude and distribution of the tunneling amplitudes in each regime; therefore, a lower RMSE indicates a smaller absolute prediction error but does not necessarily imply better relative agreement.

\begin{figure}[htbp]
    \centering
    \subfigure[]{
\includegraphics[width=0.46\columnwidth,height=1.25in]{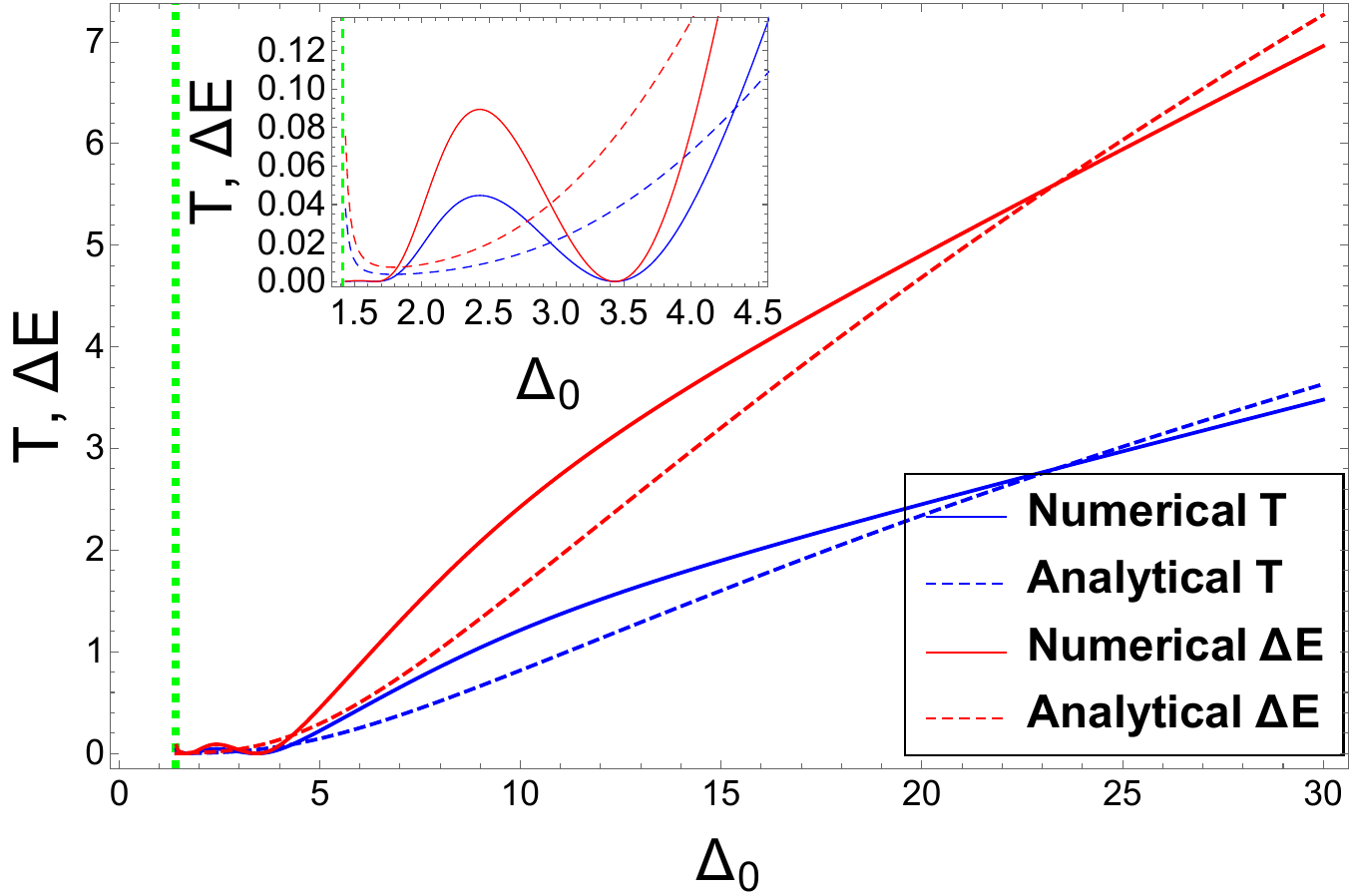}
        \label{fig:e1}}
    \hfill
    \subfigure[]{
\includegraphics[width=0.46\columnwidth,height=1.248in]{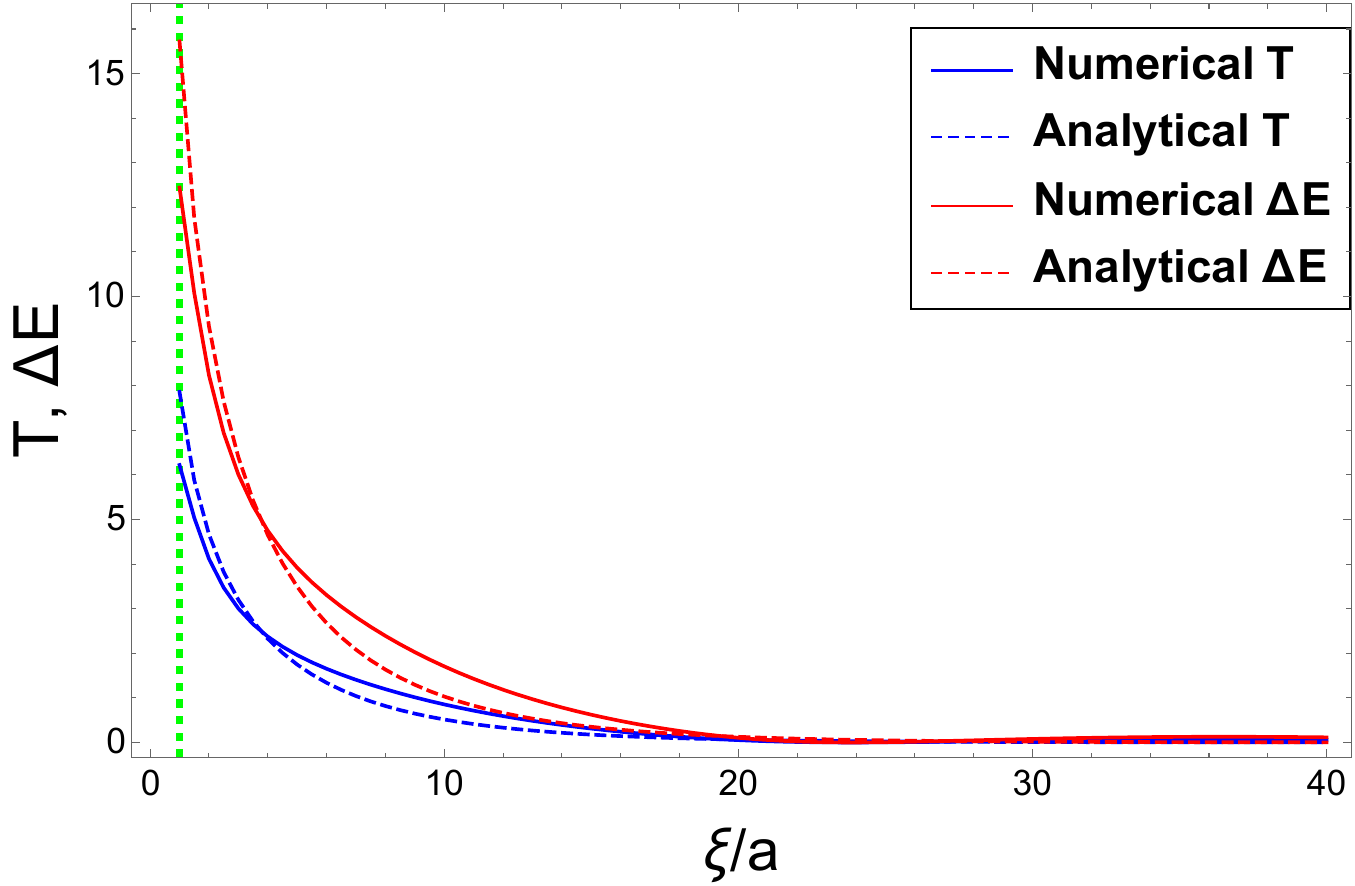}
        \label{fig:e1a}}
\caption{{\bf The tunneling strength and energy splitting} with respect to: (a) domain wall amplitude $\Delta_{0}$ ($\in[\sqrt{2}t+0.02t,30]$) for $\xi/a=10$, and (b) domain wall width $\xi/a$ ($\in[1,40]$) for $\Delta_0=8$. The solid curves are for numerical results, and the dashed curves are for analytical results for a chain with $L=256$. The inset of (a) is a magnified view of $T$ and $\Delta E$ for the smaller $\Delta_0$ region. The green dashed lines both in (a) and (b) are for $\Delta_0=\sqrt{2}$ and $\xi/a=1$, respectively.} 
\label{splitting1}
\end{figure}

Additionally, Fig. \ref{splitting1} presents the dependence of $T$ and $\Delta E$ on $\Delta_0$ and $\xi/a$, comparing the analytical predictions with the corresponding numerical results. Specifically, Fig. \ref{splitting1}(a) depicts the variation of energy splitting, or equivalently the tunneling matrix element, with respect to $\Delta_0$ for fixed $\xi/a$.  We notice that both $T$ and $\Delta E$ become negligible near the transition $\Delta_{0}\rightarrow\sqrt{2}t^{+}$. In this limit, two DWs are widely separated and diverge as $D_{s}\rightarrow\infty$\footnote{For $\Delta_0>\sqrt2t$, two DWs exist at finite positions. Their separation can be determined by $D_s=\frac{2\xi}{a}\,\operatorname{arctanh}\left(\frac{\sqrt2t}{\Delta_0}\right)$, which decreases monotonically with increasing $\Delta_0$, diverging as $\Delta_0\to\sqrt2t^+$ and approaching zero as $\Delta_0\to\infty$.}, causing the two JRMs to become completely isolated with $S_{\rm env}\to 0$ and recover their exact zero-energy character, as given by Eq.~\eqref{JRsolution}. Therefore, the isolated JRMs doublet lies phenomenologically in the $R\ge3$ window as presented in Fig. \ref{splitting}(a).

However, for $\Delta_{0}>\sqrt{2}t$, the two DWs exhibit a finite separation between them. Therefore, one can notice a finite tunneling matrix element, which permits the two independent JRMs to tunnel into each other, resulting in a finite spatial overlap. This results in a hybridization of two JRMs, producing a small energy splitting about zero-energy (notice Fig. \ref{splitting1}(a)). The isolated JRMs now become symmetric and antisymmetric combinations of $\Psi_{L}(x)$ and $\Psi_{R}(x)$, as given in Eq. \eqref{hybridized}. Importantly, Eq. (\ref{hybridized}) is valid when the two JRMs remain sufficiently separated, and their spatial overlap leads to appreciable hybridization. Therefore, based on the above discussion, Eq. (\ref{hybridized}) is expected to be valid for $1\le R<2$ (see Fig. \ref{splitting}(a)), where the low-energy sector is predominantly spanned by the two hybridized JRMs, thereby providing an appropriate two-level description of the system.

\begin{figure}[htbp]
\centering
\begin{minipage}[c]{0.485\columnwidth}
    \centering
    \includegraphics[
        width=\linewidth,
        height=1.75in
    ]{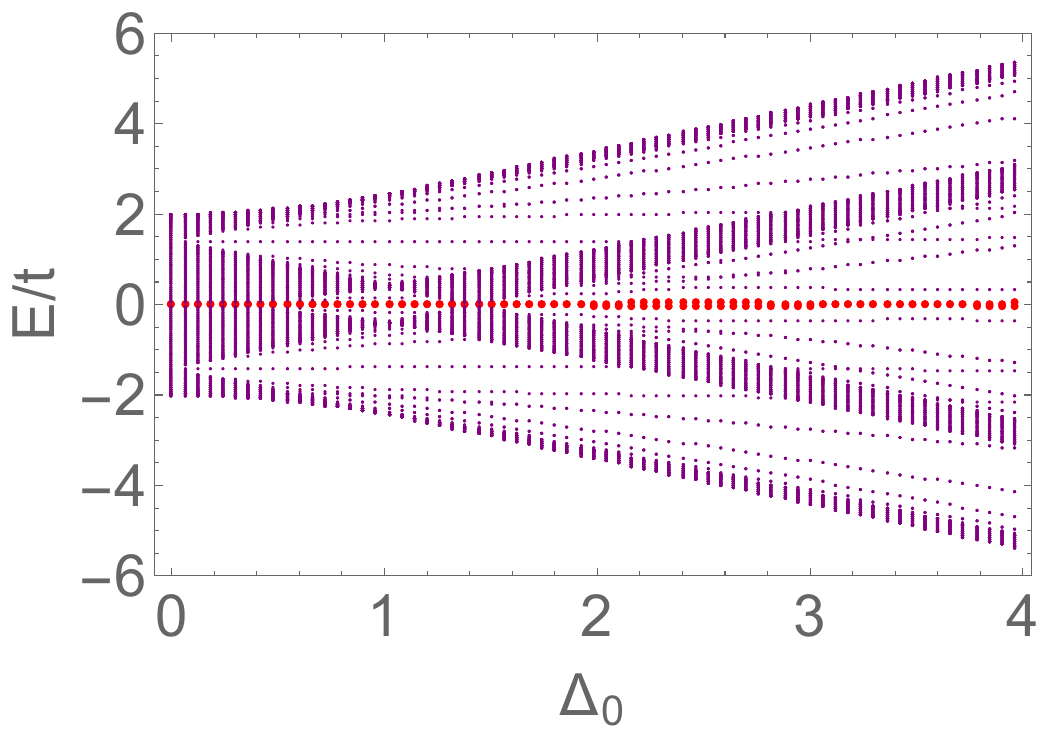}
    \vspace{2pt}
    \textbf{(a)}
\end{minipage}
\hfill
\begin{minipage}[c]{0.48\columnwidth}
    \centering
\includegraphics[width=\linewidth,height=1.15in]{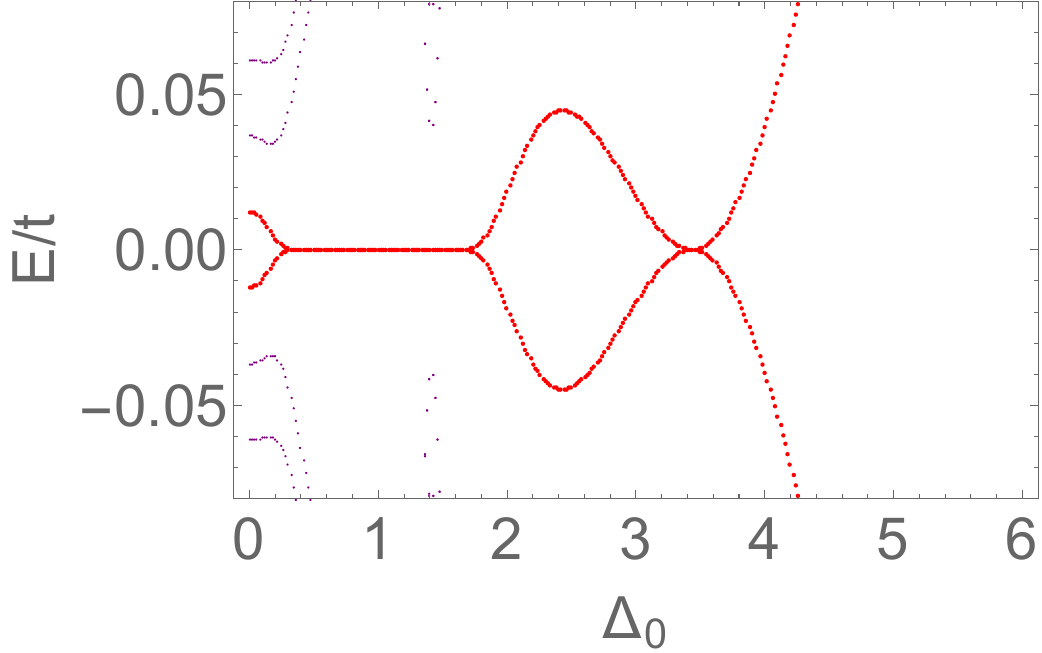}
    \vspace{0.08in}
    \textbf{(b)}
\includegraphics[width=\linewidth,height=1.15in]{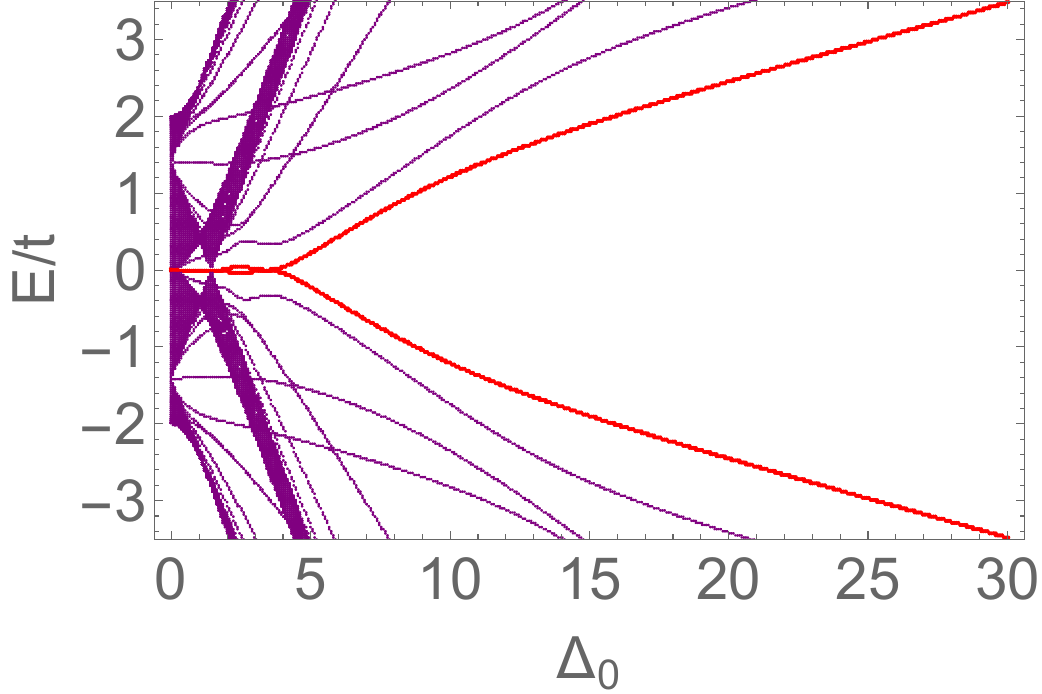}
\textbf{(c)}
\end{minipage}
\caption{(a) \textbf{Real-space energy spectra} as a function of domain wall amplitude $\Delta_0$ under OBC. (b)--(c) The corresponding {\bf low-energy spectra}. The low-energy spectra depict the energy splitting of zero-energy states with $\Delta_0$, which is marked in red color. Here, we set $L=256$, $\xi/a=10$, and $i_0=L/2$.}
\label{numericala}
\end{figure}

In the large amplitude limit, for $\Delta_0\gg\sqrt{2}t$, the separation between the two DWs decreas as $D_s\simeq \frac{2\sqrt{2}\,t\,\xi}{a\Delta_0}$ and hence $D_s\to 0$ as $\Delta_0\to\infty$. In this limit, however, $D_s$ eventually becomes comparable to the lattice spacing, beyond which the continuum approximation is no longer quantitatively reliable. It can be observed from Fig. \ref{splitting1}(a) that increasing $T$ leads to a greater energy splitting of JRMs. In this scenario, it can be clearly noticed from Figs. \ref{figzesaa}(c) and (d) that the intense overlap of exponentially decaying tails in the intermediate region gives rise to strong quantum tunneling for decreasing DW separation (see also $R<1$ domain in Fig. \ref{splitting}(a) for phenomenological demonstration). Consequently, one can witness the existence of strong hybridized JRMs. Moreover, the numerical energy splitting exhibits an oscillatory nature for smaller values of $\Delta_0$ (see inset of Fig. \ref{splitting1}(a), also see Fig. \ref{numericala}(b) for explicit energy splitting). This is due to spatially varying hopping modulation-dependent hybridization as discussed below while studying the numerical energy spectra.

We should note here that the nature of the quantum tunneling and hybridized states depends not only on $\Delta_{0}$ and $D_{s}$ but also explicitly on the width of the DWs $\xi$ (notice Eq. \eqref{Tmatrix} and Appendix~\ref{asymptotic} for explicit dependency on $\xi$). One may notice from Fig. \ref{splitting1}(b) that for fixed $\Delta_0$, decreasing $\xi/a$ reduces the separation (as $D_s\propto\xi$) between the two DWs and generally enhances the spatial overlap between the JRMs, thereby increasing the tunneling and energy splitting. Conversely, increasing $\xi$ increases $D_s$ and suppresses the hybridization and energy splitting. In the Gaussian-overlap regime, this behavior is determined by the spatial overlap of the localized JR cores, while in the large-separation regime the spatial overlap is governed by the exponentially decaying asymptotic tails of the JRMs. The exponential nature of the energy splitting as a function of $\Delta_0$ and $\xi/a$ is further verified by the numerical exact diagonalization method, as can be found in Fig. \ref{splitting1}.

\subsection{Numerical Study}
We begin by presenting the energy spectra in real space. We obtain real-space energy spectra by diagonalizing Eq. (\ref{1}) for $\theta=\pi/2$, with the periodic hopping modulation term now being replaced by $\Delta(x)$ (Eq. \ref{domain} of Appendix \ref{numerical}), and is plotted in Fig. \ref{numericala}. Fig. \ref{numericala}(a) clearly demonstrates the appearance of additional in-gap modes, including two nonzero energy in-gap modes. Notice that, due to the preservation of chiral symmetry, the energy spectrum is symmetric about zero-energy\cite{kane,bdi}. As we are interested in mid-gap zero-energy modes, we now focus on the low-energy spectra presented in Figs. \ref{numericala}(b) and (c), which show the evolution of ZESs with $\Delta_{0}$.

For $0<\Delta_{0}<\sqrt{2}t$, two mid-gap states are exactly at zero energy and remain localized at the two ends of the chain\cite{mandal}. This is because the hopping arrangement remains nearly unchanged around the chain center, preventing the formation of persistent DWs in this $\Delta_0$ regime. This is consistent with the analytical JR method, in which the absence of mass inversion precludes the formation of DWs, as illustrated in Fig. \ref{dw2}(b) (also see the winding number plot in Fig. \ref{winda}(a) of Appendix~\ref{wind}). For $\Delta_{0}=\sqrt{2}t$, the band gap closes, and the phase transition occurs. For $\Delta_{0}>\sqrt{2}t$, the hopping arrangements change within the lattice such that two DWs are created around the chain center (notice the corresponding Fig. \ref{dw2}(d)). In this situation, the exact zero-energy modes no longer exist; instead, we get two near-zero-energy states that exhibit a small energy splitting around $E=0$. In particular, no appreciable energy splitting is discernible when $\Delta_0$ is very close to $\sqrt{2}t$, as the spatial separation between the two DWs remains large (see Fig. \ref{numericala}(b)). Consequently, the spatial overlap as well as hybridization between the JRMs becomes exponentially small. For $\Delta_{0}\gg\sqrt{2}t$, more energy splitting is observed, making the result compatible with the above analytical findings for energy splitting.

Interestingly, the oscillatory behavior of the energy splitting is noticed for smaller $\Delta_0$ values (Fig. \ref{numericala}(b)) and no such behavior is observed for larger $\Delta_0$ values (see Fig. \ref{numericala}(c)). It originates from the spatially varying hopping modulation-dependent hybridization of the two localized JRMs. In our lattice model with $L=256$ and $\xi/a=10$, the spatially varying hopping amplitudes $t\pm\Delta_0\tanh\Big(\frac{i-L/2}{\xi/a}\Big)$ also influence the hybridization. For example, the first hopping $(t+\Delta(x))$ and third hopping $(t-\Delta(x))$ at sites $i=125$ and $i=131$ respectively (around the DW center $i=L/2=128$) involve the factor $(1-\Delta_0\tanh(0.3))$, taking the values 0.27, 0, and -1.33 for $\Delta_0=2.5$, $3.432$, and $8$, respectively. For $\Delta_0\simeq3.432$, the vanishing hopping at site $i=125$ ($i=131$) disconnect it from the site $i=126$ ($i=132$). The chain therefore separates into three segments: $\{1,.....125\}$, $\{126,.....131\}$ and $\{132,.....256\}$.

\begin{figure*}[t]
\centering
    \includegraphics[width=.4\linewidth, height=2.2 in]{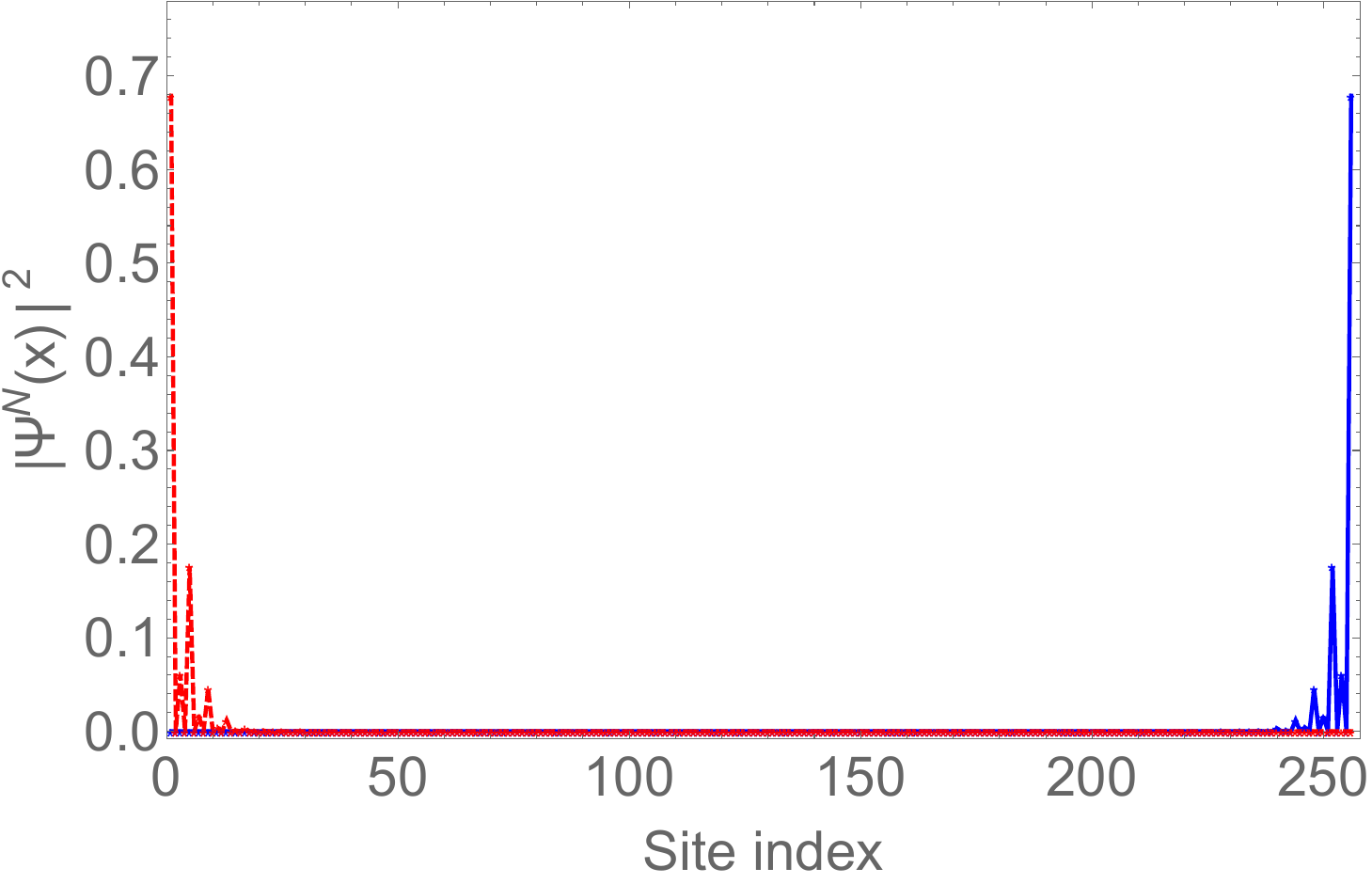}
  \includegraphics[width=.4\linewidth, height=2.2 in]{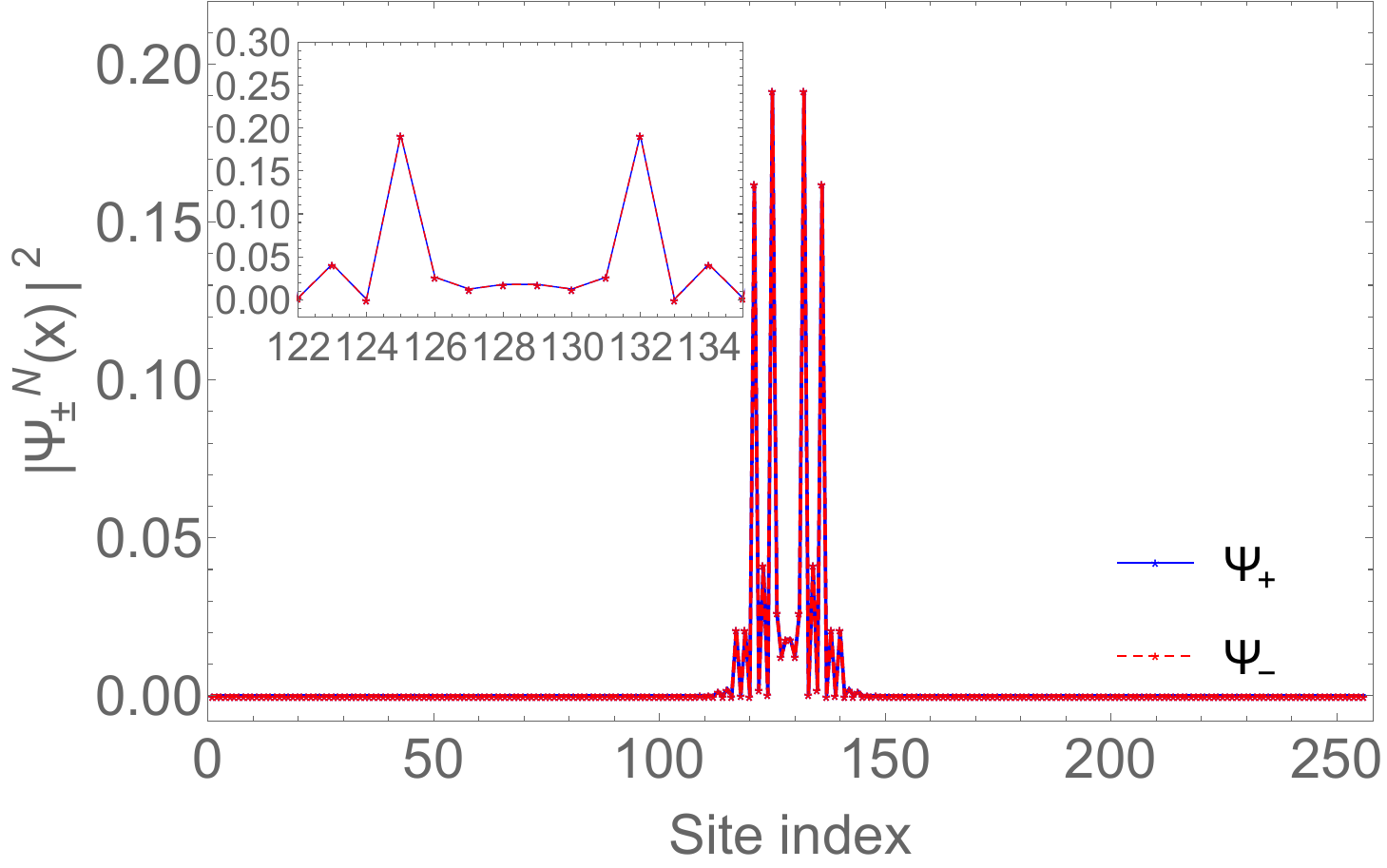}
     \put(-320,66){(a)}
     \put(-150,66){(b)}\\
    \includegraphics[width=.4\linewidth, height=2.2 in]{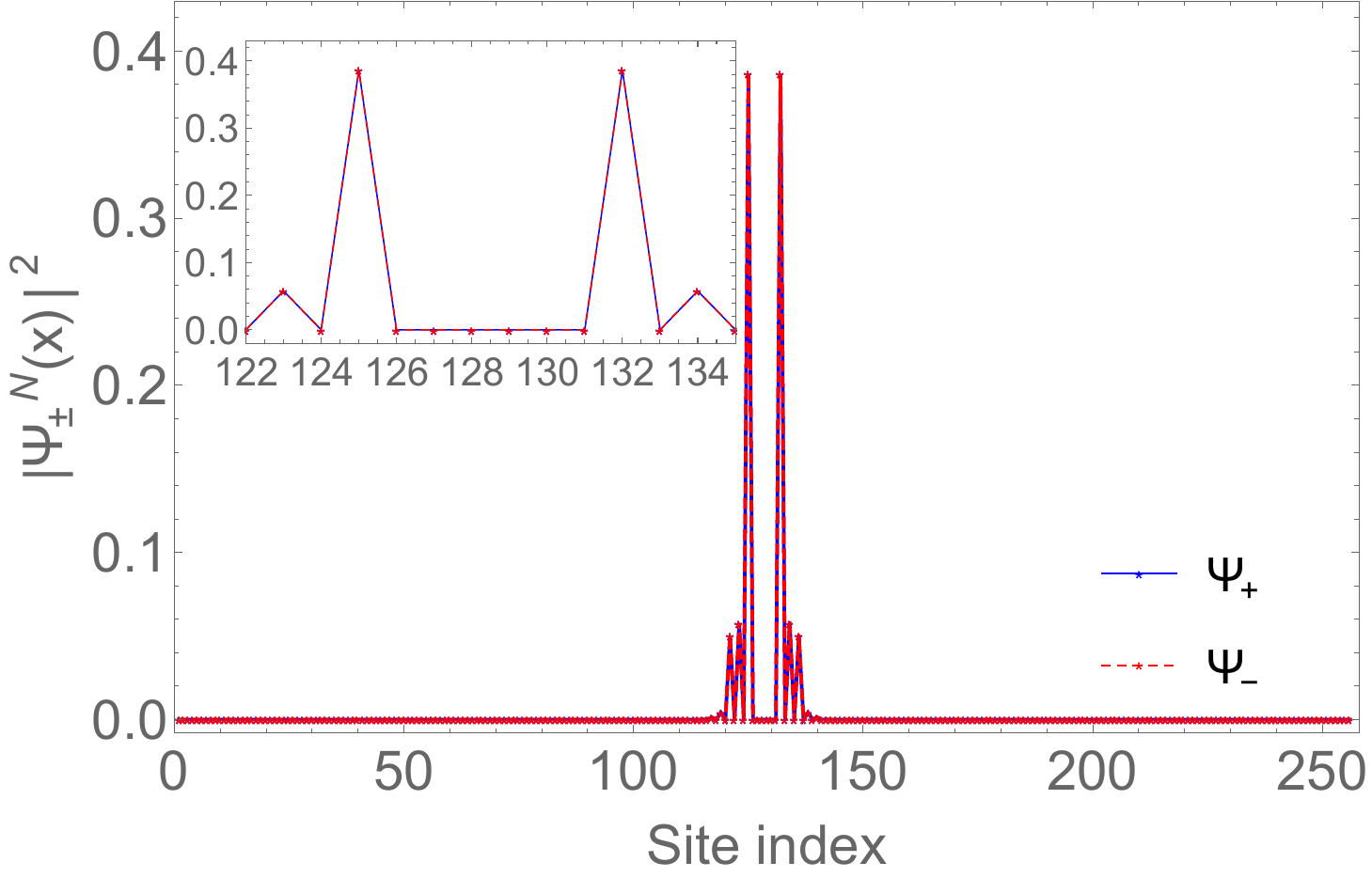}
    \includegraphics[width=.4\linewidth, height=2.2 in]{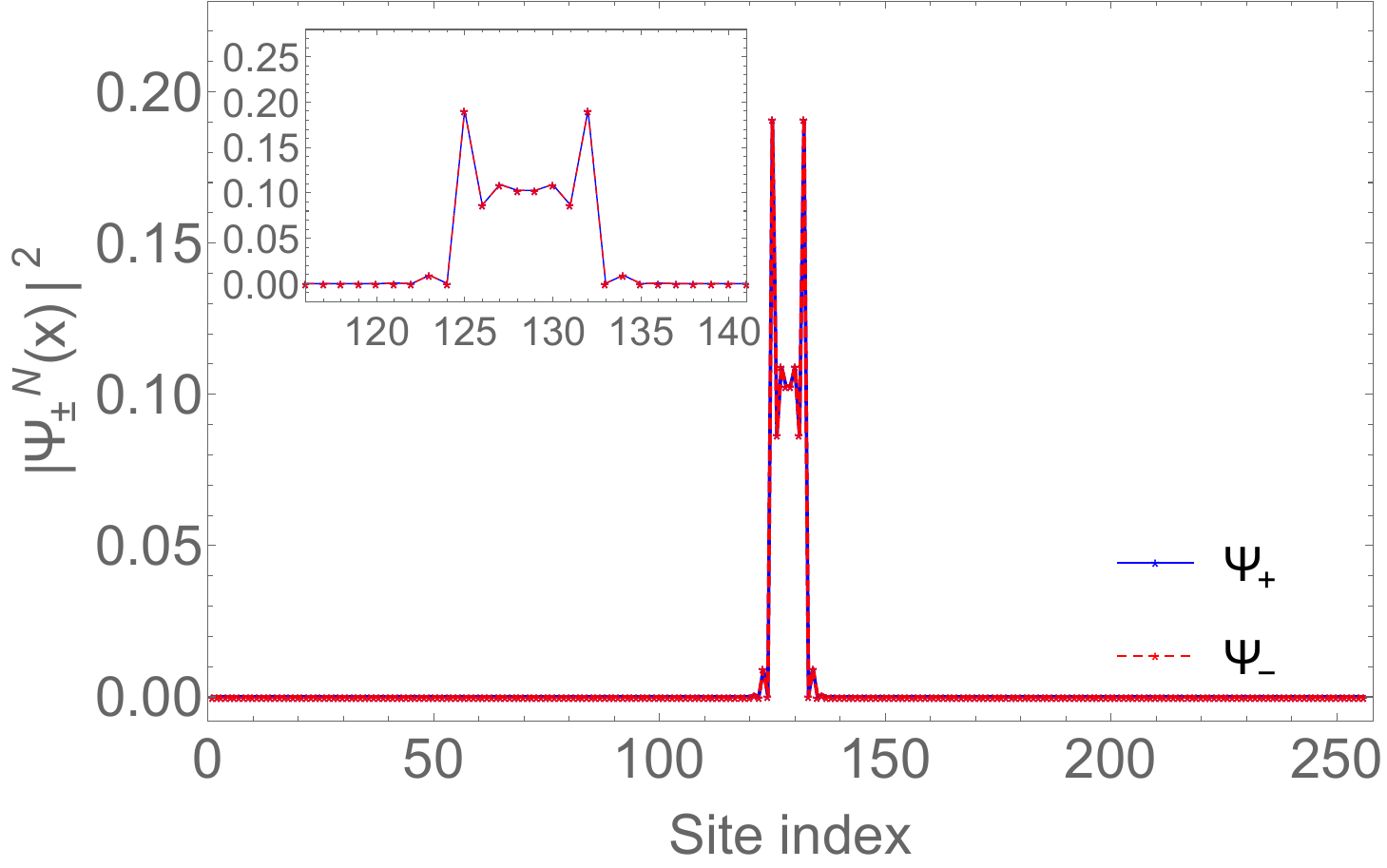}
   \put(-350,66){(c)}
    \put(-150,68){(d)}
\caption{{\bf The numerical probability density of mid-gap ZESs} with $\Delta_0=0.7$ (a) and {\bf hybridized JRMs} (b)--(d). Here, $\Delta_0=2.5,3.432$, and 8 for (b), (c) and (d), respectively. The inset shows a zoomed-in view of hybridized states around two DWs, $x_{L}$ and $x_{R}$, which depicts that the peak near the left DW appears only at odd site index positions, while near the right DW, it is visible only at even site index positions. We also fix $L=256$, and $\xi/a=10$.}
\label{figzesb}
\end{figure*}
The near-vanishing of the local hoppings around $\Delta_0\simeq3.432$ can strongly suppress the coupling between localized modes if that bond contributes significantly to their tunneling. For $\Delta_0>3.432$, the magnitude of the relevant hopping continues to increase, which can strengthen the coupling between the localized modes. Thus, the energy splitting grows monotonically over the parameter range shown in Fig.~\ref{numericala}(c). We elucidate that the local hopping analysis identifies a plausible contribution to the observed crossover near $\Delta_0\simeq3.432$. The persistence of this characteristic value, upon varying the chain length $L$, indicates that it is determined by the local hopping profile rather than by finite-size scaling. Nevertheless, it corresponds to the vanishing of the factor $(1-\Delta_0\tanh(0.3))$ at different sites around the DW center for different $L$ and does not, by itself, indicate a distinct topological phase transition. This is further confirmed numerically by the probability density plot of JRMs, as presented in Fig. \ref{figzesb}.

Furthermore, with the help of the numerical method studied in the Appendix \ref{numerical}, we plot the probability density of ZESs and hybridized JRMs in Fig. \ref{figzesb}. For $0<\Delta_0<2t$, each, among two, ZESs exhibit localization at either boundary of the chain (Fig.\ref{figzesb}(b)). However, for $\Delta_{0}>\sqrt{2}t$ we can witness of two JRMs at both DWs as studied analytically, confirming their domain-wall origin (Figs. \ref{figzesb}(b)--(d)). In particular, within the DW region, a small hybridization is discernible as a result of less energy splitting (see Fig. \ref{figzesb}(b)), while the hybridization is negligibly small due to the absence of energy splitting (see Fig. \ref{figzesb}(c)). However, the hybridization within the DWs region increases as a consequence of the increasing strength of energy splitting (see Fig. \ref{figzesb}(d)). This result is fully consistent with Fig. \ref{numericala}. We also notice from the zoomed-in view of Fig. \ref{figzesb} that the JRMs near the left DW have a nonzero amplitude only at odd site index locations, while they have a nonzero amplitude only at even site index locations near the right DW. It clearly reflects that the JRMs have weight on A,~C~(B,~D) sublattices near the left (right) DW. Nevertheless, the numerically calculated JRMs will also be strongly overlapped for $\Delta_{0}\gg\sqrt{2}t$. This numerical prediction is consistent with the analytical findings discussed earlier.

Based on the analytical and numerical study described above, we note here that, unlike the original SSH chain\cite{kane,mandal}, the JRMs are polarized in two sublattices in a way that no two consecutive sites have a nonzero amplitude of them. More explicitly, the JRMs are chiral modes having different chirality: at the left (right) DW they are noticeable only on (A,~C) ((B,~D)) sublattices\cite{soliton-poly}. Crucially, we might reasonably think of the hybridized JR states (Eq. \eqref{hybridized}) as a quantum-mechanical double-well system. In quantum mechanics\cite{q1,q2,griffiths,shankar}, considering a particle in a double-well potential with an infinite barrier width results in a states individually peaked around the left well and the right well. While the finite barrier width allows tunneling of the states, due to which the wave functions begin to overlap spatially and produce true eigenstates having symmetric and antisymmetric combinations\cite{qm}.

Importantly, the existence of zero-energy JRMs in the present model cannot be understood simply in terms of an isolated unpaired lattice site, as in the case of a conventional SSH chain\cite{kane}. Instead, the JR zero modes originate from a spatial inversion of the effective Dirac mass $M(x)=2t^2-\Delta^2(x)$. For $\Delta_{0}>\sqrt{2}t$, the mass profile generates two spatial locations where the sign reversal occurs, thereby creating two DWs between topologically distinct insulating phases. According to the JR mechanism, each mass inversion binds a localized zero-energy state\cite{jackiw}. Consequently, instead of a single, here a pair of exponentially localized JRMs emerges, one at each DW.

We should mention here that, when the two DWs are well separated, the spatial overlap between the corresponding JRMs is exponentially small. In this limit, the two isolated JRMs (see Eq. \eqref{JRsolution}) possess definite and opposite chiralities, $\Gamma_{4}|\Psi_{L}\rangle=+|\Psi_{L}\rangle$ and $\Gamma_{4}|\Psi_{R}\rangle=-|\Psi_{R}\rangle$, as a consequence of the opposite signs of the mass gradients at the two DWs, $M'(x_L)>0$ and $M'(x_R)<0$, and are pinned at zero energy ($E=0$). For finite separation, however, the spatial overlap between the two JRMs leads to their hybridization, producing bonding and antibonding states (see Eq. \eqref{hybridized}) with energies $E_{\pm}=\pm T$. The resulting hybridized states are no longer eigenstates of the chiral operator and therefore do not possess definite chirality. Instead, chiral symmetry maps one hybridized state onto the other, such that, $\Gamma_{4}| \Psi_{+} \rangle=|\Psi_{-}\rangle$ and $\Gamma_{4}| \Psi_{-} \rangle=|\Psi_{+}\rangle$, while the effective Hamiltonian continues to satisfy the exact chiral symmetry relation $\{\Gamma_{\rm eff},\mathcal{H}_{\rm eff}^P\}=0$. Consequently, the coupling between the oppositely chiral JRMs produces a finite-energy hybridized doublet with $E_\pm=\pm T$, while chiral symmetry ensures the spectral symmetry such that the two resulting energies remain symmetric about zero-energy (notice Fig. \ref{numericala}). Since JRMs are topologically protected and break the inversion symmetry\cite{dw}, their topological invariant can also be estimated by the change of local winding number across each interface\cite{teo} (see Appendix~\ref{wind} for more details). It is worth noting here that the topologically protected zero-energy modes trapped at the DWs are a local property around a real-space defect, while the topological insulator is a property associated with the whole band and the entire BZ, not only the low effective description near the gap\cite{dw}. 

\section{Coherent tunneling oscillations}\label{tunneling oscillations}
The study of coherent tunneling oscillations provides a direct dynamical probe of the hybridization between JR interface modes. Although the energy splitting of the hybridized JRMs reveals their spatial overlap spectrally, the oscillatory transfer of probability density between the interfaces offers a real-time manifestation of this hybridization. Following \cite{q1,q2,griffiths,shankar}, we start from an effective two-level description introduced above to investigate coherent tunneling between the JRMs. Although the effective two-level Hamiltonian is calculated in the two-component JR basis $\{|\phi_L\rangle,|\phi_R\rangle\}$, we study the tunneling dynamics in the microscopic four-sublattice basis to investigate the real-space probability density and sublattice-resolved dynamics. In the four-sublattice basis, the localized JR states and their relation to hybridized eigenstates have already been defined in Eq.~\eqref{hybridized}. We begin by considering the JRMs initially prepared at the left DW $| \Psi(x,0) \rangle=|\Psi_{L}(x)\rangle$ (see Eq.~\eqref{JRsolution}) with $|\Psi_{L}(x)\rangle=\frac{|\Psi_+(x)\rangle+|\Psi_-(x)\rangle}{\sqrt{2}}$. The corresponding right-localized state is defined as
$|\Psi_R(x)\rangle=\frac{|\Psi_+(x)\rangle-|\Psi_-(x)\rangle}{\sqrt{2}}$. The time evolution of the initially left-localized JRM is, therefore, given by
\begin{equation}\label{time}
|\Psi_L(x,t^{\prime})\rangle=\frac{e^{-iE_{+}t^{\prime}}|\Psi_{+}(x)\rangle+e^{-iE_{-}t^{\prime}}|\Psi_{-}(x)\rangle}{\sqrt{2}},
\end{equation}

After rewriting it on a localized basis, one readily finds the evolution of the left-localized JR state as
\begin{equation}\label{time1}
|\Psi_L(x,t^{\prime})\rangle=\cos(T t^{\prime})|\Psi_{L}(x)\rangle-i\sin(T t^{\prime})|\Psi_{R}(x)\rangle,
\end{equation}

From which the probabilities of finding the time-evolved state at the left and right DWs are obtained as
\begin{equation}
\left.
\begin{aligned}
P_{L}(t^{\prime})&=|\langle\Psi_{L}(x)|\Psi_L(x,t^{\prime})\rangle|^2=\cos^2(Tt^{\prime})\\
P_{R}(t^{\prime})&=|\langle\Psi_{R}(x)|\Psi_L(x,t^{\prime})\rangle|^2=\sin^2(Tt^{\prime})
\end{aligned}
\right\}
\label{prob}
\end{equation}
\begin{figure}
     \vskip -.1 in
   \begin{picture}(100,100)
     \put(-90,0){
  \includegraphics[width=.54\linewidth, height=1.25 in]{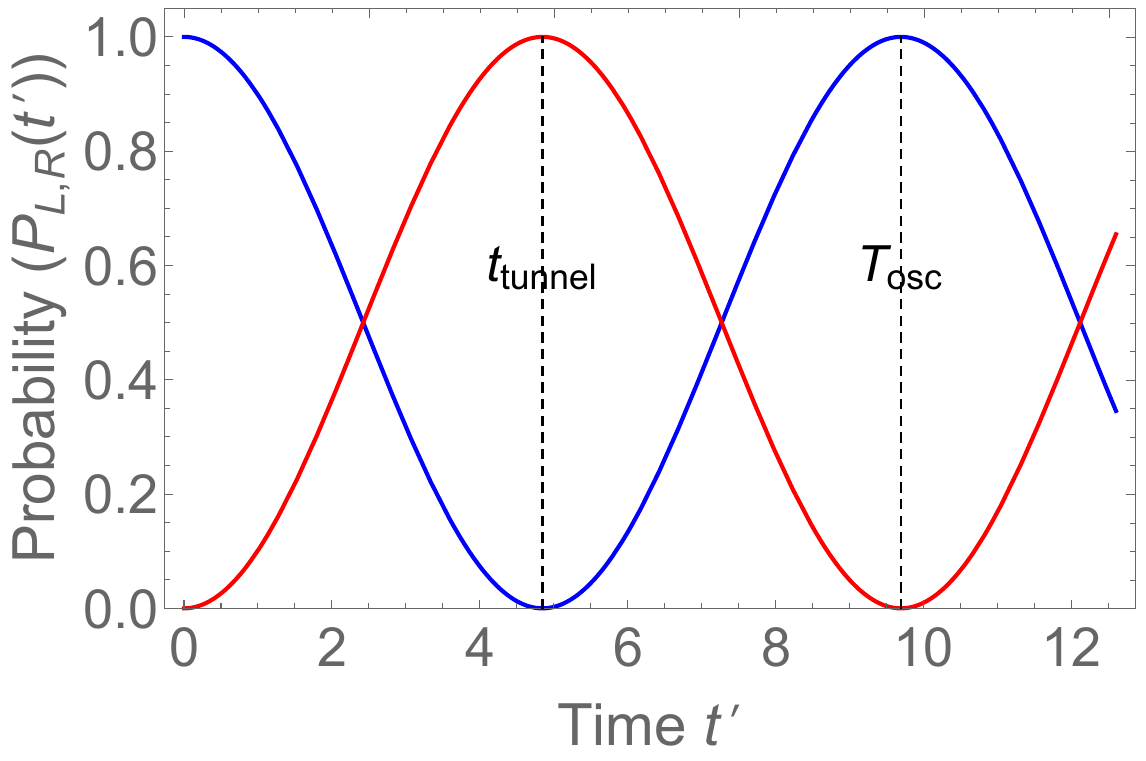}
  \includegraphics[width=.52\linewidth, height=1.25 in]{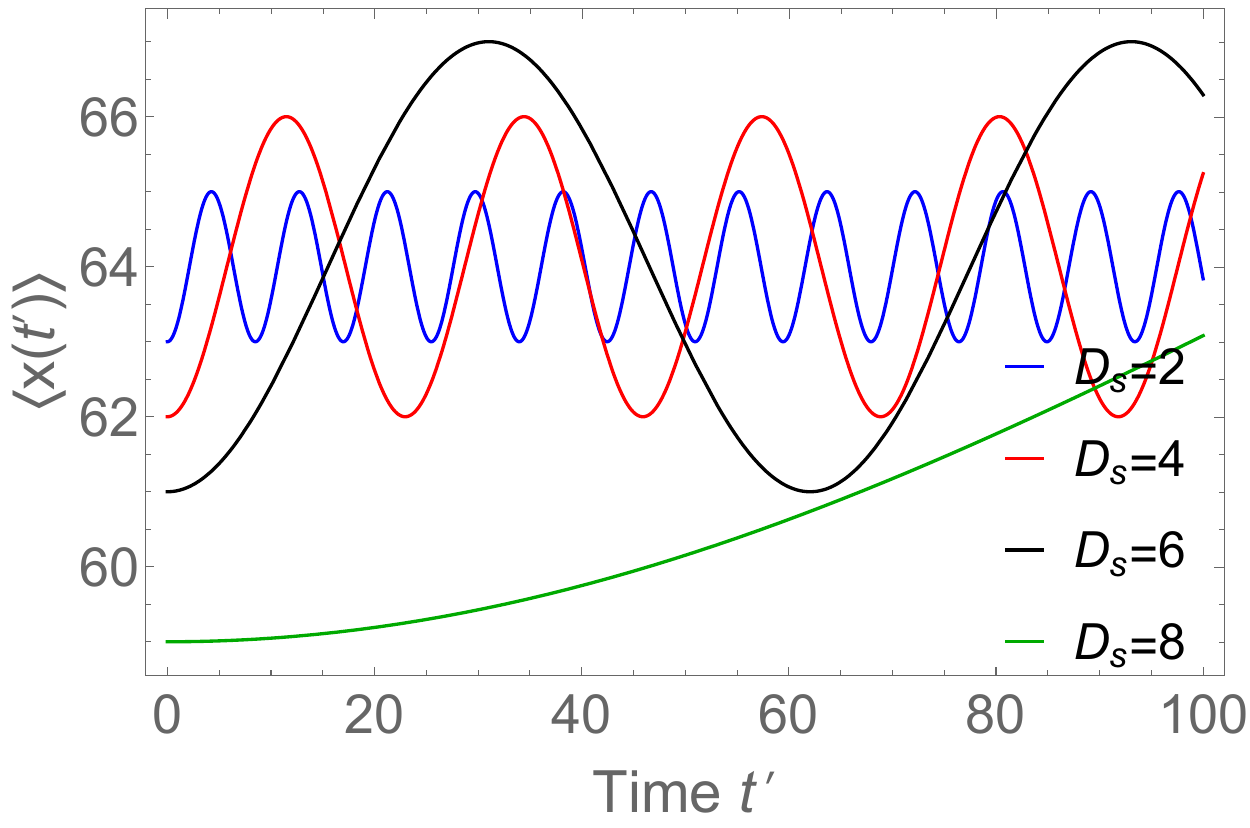}}
     \put(-50,78){(a)}
     \put(110,78){(b)}
    \end{picture}
\caption{(a) {\bf Probability} of JRMs as a function of time and (b) {\bf time evolution of the position expectation value $\langle x(t^{\prime})\rangle$}. In (a), the blue curve ($P_{L}$)  is the probability of finding the JRM at the left DW, while the red curve ($P_{R}$) indicates the same at the right DW. For (a), we fix $\Delta_{0}=12.5$, and $\xi/a=10$.} 
\label{figzesc}
\end{figure}
\begin{figure*}[t]
\centering
    \includegraphics[width=.43\linewidth, height=2.2 in]{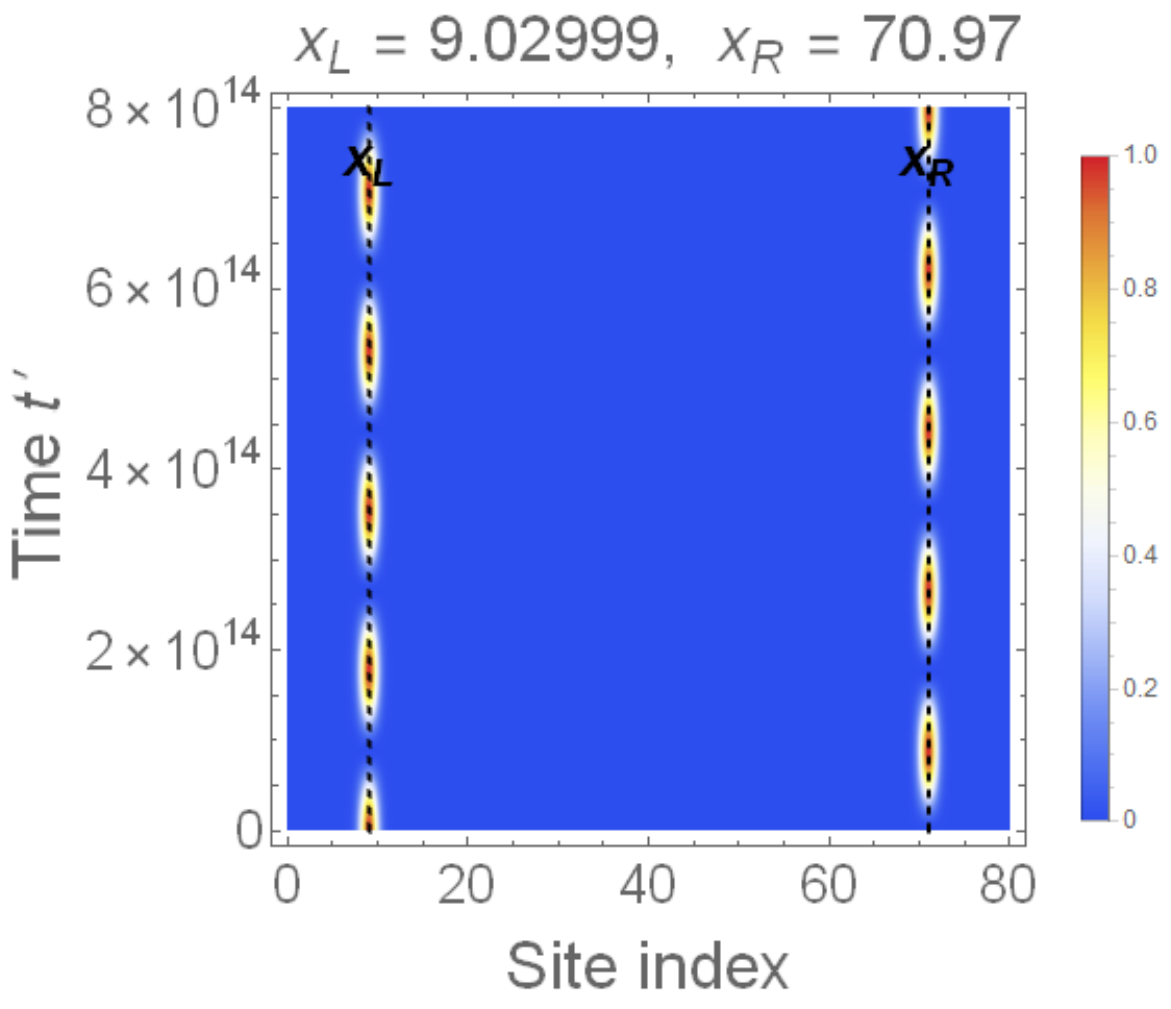}
  \includegraphics[width=.4\linewidth, height=2.2 in]{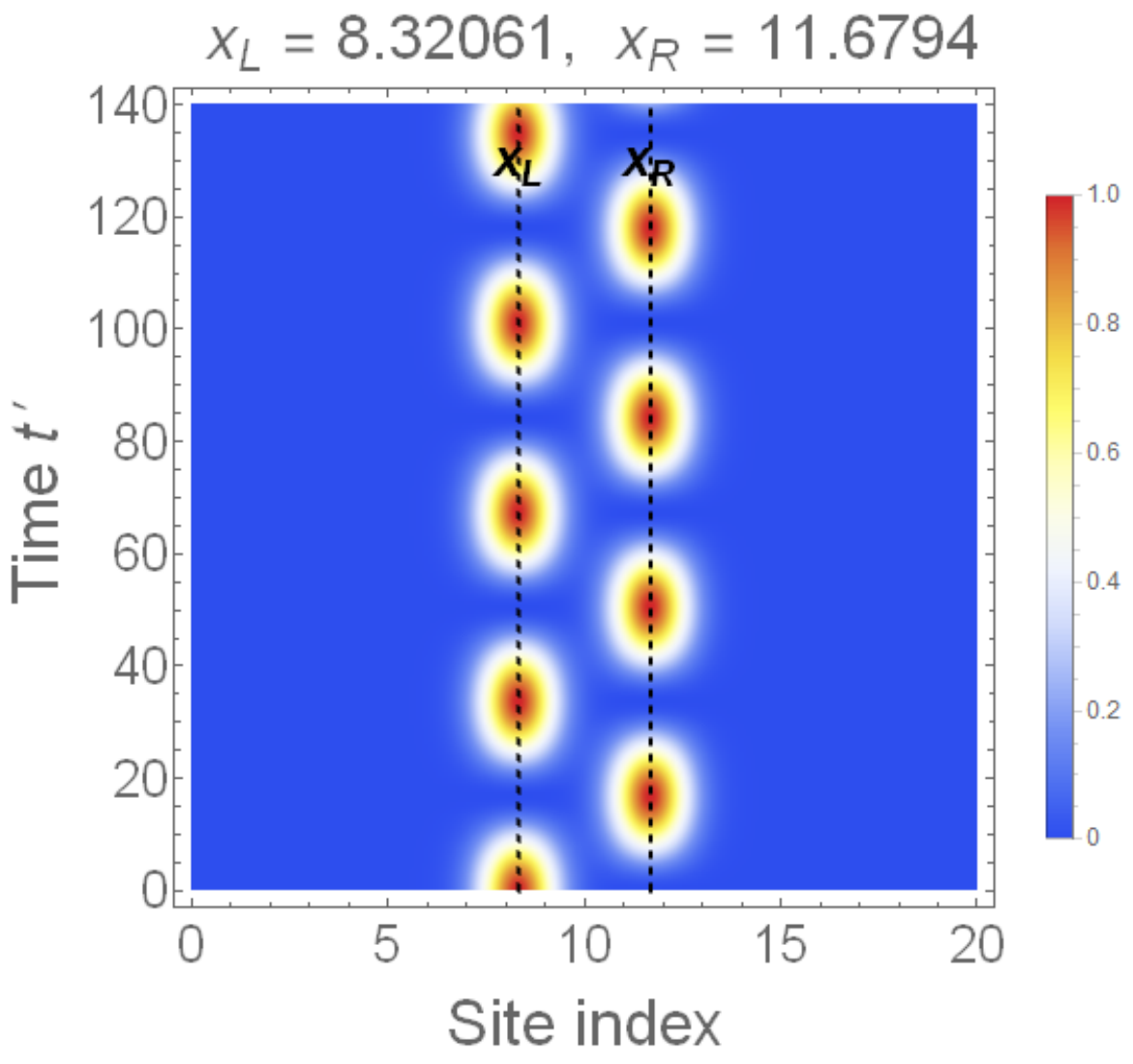}
     \put(-320,66){(a)}
     \put(-150,66){(b)}\\
    \includegraphics[width=.4\linewidth, height=2.2 in]{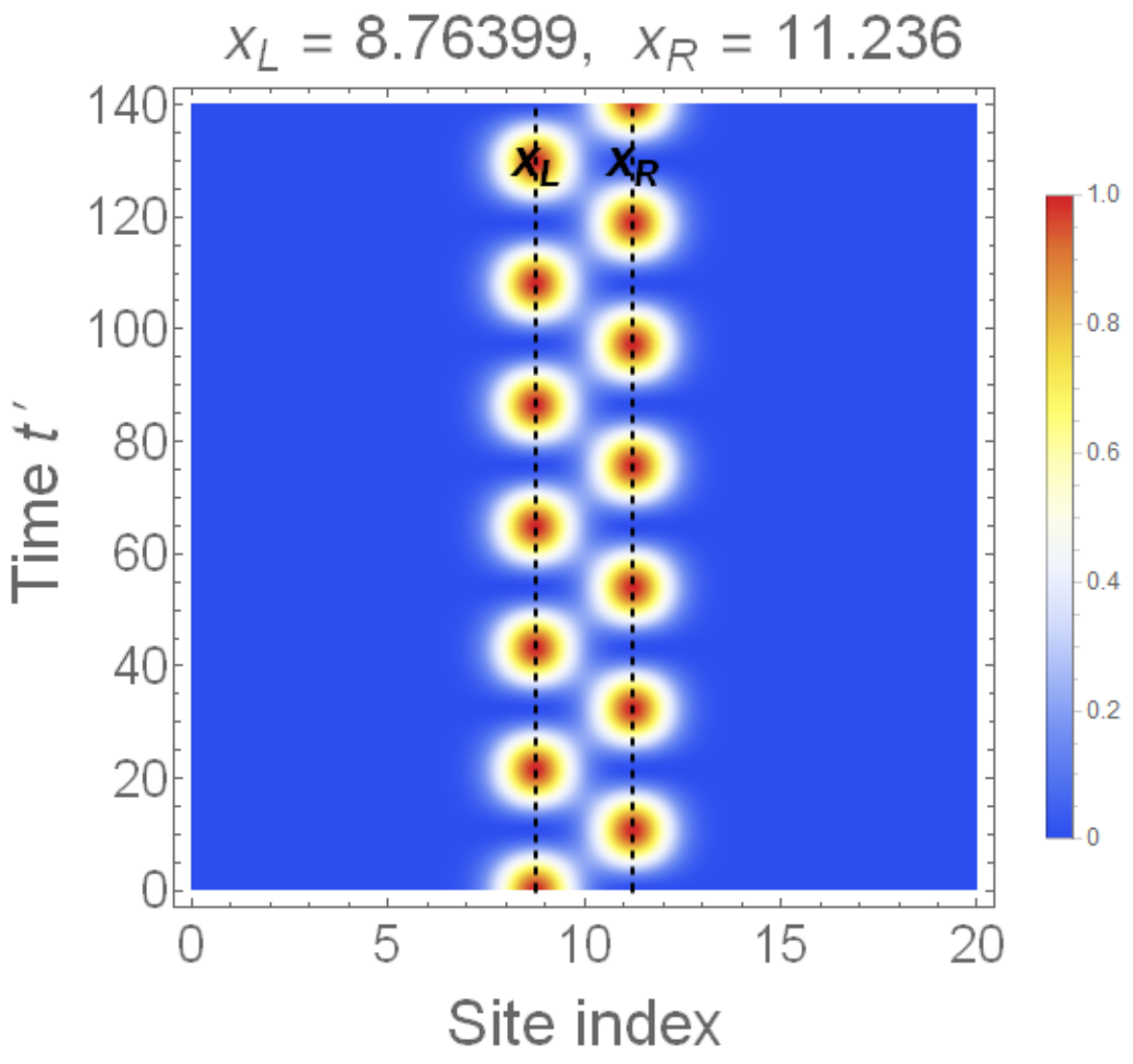}
    \includegraphics[width=.4\linewidth, height=2.2 in]{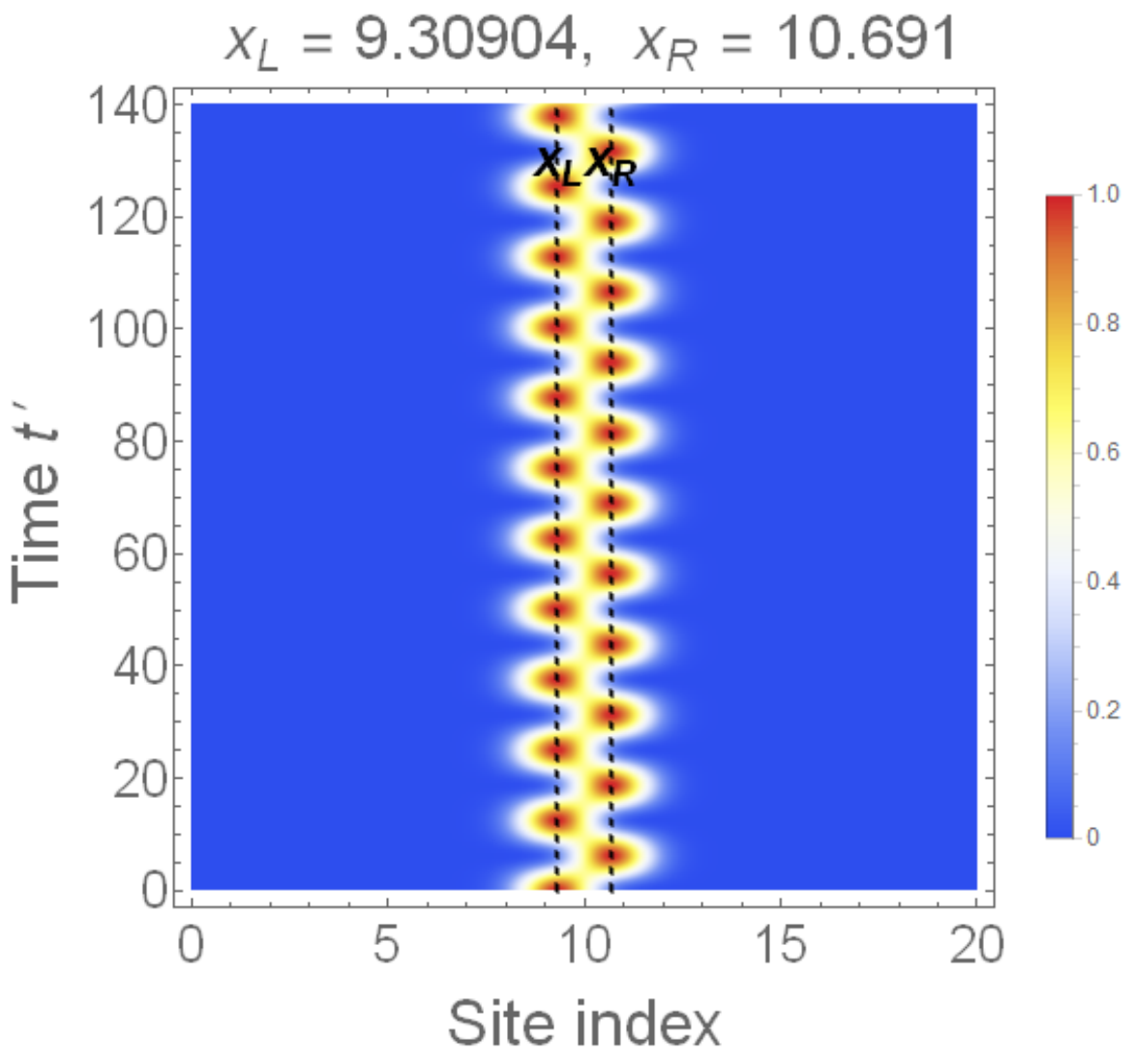}
   \put(-350,66){(c)}
    \put(-150,68){(d)}
\caption{ {\bf Probability density} $\rho(x,t^{\prime})$ illustrating coherent tunneling between two JRMs localized at the DWs $x_{L}$ and $x_{R}$ for varying $\Delta_{0}$ and fixed $\xi/a=10$. The vertical dashed lines mark the DW positions, $x_{L}$ and $x_{R}$. Other parameters are: (a) $L=80$, $\Delta_{0}=1.42$, (b) $L=20$, $\Delta_{0}=8.5$, (c) $L=20$, $\Delta_{0}=11.5$, and (d) $L=20$, $\Delta_{0}=20.5$.} 
\label{tunnel}
\end{figure*}
where $T$ denotes the tunneling matrix element between the two JRMs. In the regime considered here, we use the overlap-induced form $
T\simeq T_0 e^{-R^2/4}$, whereas, in the asymptotic large-separation regime, the tunneling amplitude is more generally governed by the exponential tails of the JRMs and takes the form $
T\simeq T_0 e^{-R (l_G/l_D)}$ (as discussed earlier; also see Appendix~\ref{asymptotic} for detailed calculation).

Thus, at $t^{\prime}=0$, the JRMs are entirely localized at the left DW. Then, they tunnel to the right DW with tunneling time $t^{\prime}=t_{tunnel}=\frac{\pi}{2T}$ and periodically back to the left DW with a full oscillation period given by $t^{\prime}=T_{osc}=\frac{\pi}{T}=\frac{2\pi}{\Delta E}$.  This periodic behavior provides a direct dynamical manifestation of topological quantum
tunneling in the modified SSH chain, exactly like a Rabi oscillation in a two-level quantum-mechanical double-well system\cite{shankar,qm,dynamic}. To demonstrate the nature of dynamical tunneling oscillations between the JRMs, we plot Eq. (\ref{prob}) for $\Delta_{0}=12.5$ and $\xi/a=10$ in Fig. \ref{figzesc}(a). For the chosen set of parameter values, the system exhibits $t_{tunnel}\approx 4.845$ and $T_{osc}\approx 9.691$, which are correctly marked by dashed vertical lines. We also notice that $P_{L}(t^{\prime})$ and $P_{R}(t^{\prime})$ are out of phase, and the total probability density is conserved: $P_{L}(t^{\prime})+P_{R}(t^{\prime})=1$. On the other hand, one can also look at the expectation value, $\langle x(t^{\prime})\rangle=x_{0}-\frac{D_{s}}{2}\cos(2Tt^{\prime})$, the characteristic curve plotted in Fig. \ref{figzesc}(b),  which measures the average position of the probability density of JRMs at time $t^{\prime}$. We can notice faster oscillations for decreasing $D_s$. The oscillatory nature of $\langle x(t^{\prime})\rangle$ between $x_{L}$ and $x_{R}$ is a direct signature of coherent tunneling and oscillations becoming almost frozen for large $\operatorname{D_{s}}$, as expected\cite{comment}. 

If we start with an initial state localized at the right DW, \(|\Psi(x,0)\rangle=|\Psi_R(x)\rangle\), the corresponding probabilities of finding the state at the left and right DWs are
\begin{equation}
\left.
\begin{aligned}
P_{L}(t^{\prime})&=|\langle\Psi_{L}(x)|\Psi_R(x,t^{\prime})\rangle|^2=\sin^2(Tt^{\prime})\\
P_{R}(t^{\prime})&=|\langle\Psi_{R}(x)|\Psi_R(x,t^{\prime})\rangle|^2=\cos^2(Tt^{\prime})
\end{aligned}
\right\}
\label{prob1}
\end{equation}
Thus, the initially right-localized JR state undergoes coherent oscillations between the two DWs, with \(P_L(t^{\prime})+P_R(t^{\prime})=1\).

In the above analysis, we have used $E_{\pm}=\pm T$ with the localized JR states $|\Psi_L\rangle$ and $\Psi_R\rangle$ forming an orthonormal basis for the isolated JR doublet. Under these conditions, Eqs.~\eqref{time}--\eqref{prob1} describe coherent population transfer between the two DWs, with the oscillation frequency set by the tunnel coupling $T$. In particular, the complete population transfer predicted by Eqs.~\eqref{prob} and (\ref{prob1}) corresponds to the ideal two-level limit; deviations from $P_L(t')+P_R(t')=1$ or from the sinusoidal oscillations signal the admixture of higher-energy bulk states beyond the isolated JR doublet.

To shed light on the coherent tunneling oscillations more clearly, we can now investigate the probability density of the time-evolved wave function of the JRMs, $\rho(x,t^{\prime})=|\Psi_L(x,t^{\prime})|^2=\rho_{\operatorname{AC}}(x,t^{\prime})+\rho_{\operatorname{BD}}(x,t^{\prime})$ (with $\rho_{\operatorname{AC}}(x,t^{\prime})=\cos^2(Tt^{\prime})|\Psi_{L}(x)|^2$ and $\rho_{\operatorname{BD}}(x,t^{\prime})=\sin^2(Tt^{\prime})|\Psi_{R}(x)|^2$). Here, $\rho_{\mathrm{AC}}$ and $\rho_{\mathrm{BD}}$ represent the probability densities associated with the left and right JRMs, respectively, which are predominantly localized on the (A,~C) and (B,~D) sublattices. The complementary $\cos^2(Tt^{\prime})$ and $\sin^2(Tt^{\prime})$ factors describe the coherent transfer of probability between the two DWs. The density function $\rho(x,t')$ against site index is plotted for different values of $\Delta_{0}$ at fixed $\xi$ in Fig. \ref{tunnel}. The plot suggests that the probability density remains exactly confined at the two DWs, confirming that the JRMs are topological interface modes rather than extended bulk states. In all plots, the periodic alteration of dense spots is a characteristic hallmark of coherent tunneling oscillations between the two JRMs. Furthermore, the figure clearly illustrates the gradual evolution from independent JRMs to strongly hybridized JRMs as the DW separation is progressively reduced. 

We notice in Fig. \ref{tunnel}(a) that the tunneling period, $T_{osc}$, associated with the density transfer between the two JRMs becomes extremely high for exceedingly large DW separations, indicating negligible density transfer between the modes. This reflects that the exponentially suppressed spatial overlap between the localized modes corresponds to the regime $R\gg1$, where $S_{\rm env}\approx0$. Consequently, the density distribution exhibits nearly static vertical stripes localized at $x_L$ and $x_R$, indicating effectively decoupled JRMs. The tunneling period decreases when DW separation decreases, reflecting faster probability density transfer between two JRMs, as can be seen from Figs. \ref{tunnel}(b)--(d). Nevertheless, this behavior again indicates that the decreasing DW separation markedly enhances the overlap between the JRMs, thereby strengthening the tunneling and giving rise to pronounced coherent tunneling oscillations. 

In particular, Fig. \ref{tunnel}(c) shows a substantial increase in the oscillation frequency, accompanied by faster transfer of probability between the two DWs. However, in Fig. \ref{tunnel}(d), the density distributions associated with the two DWs become strongly overlapped, and the oscillations become very rapid, making the two JRMs less distinguishable as isolated localized modes. This behavior is consistent with the strong-hybridization regime discussed above. A similar tunneling behavior can be found in a quantum mechanical double-well system\cite{dynamic}. 

\begin{figure}[htbp]
\centering
 \vskip 1.2 in
   \begin{picture}(100,100)
     \put(-70,0){
\includegraphics[width=.98\linewidth,height=2.6in]{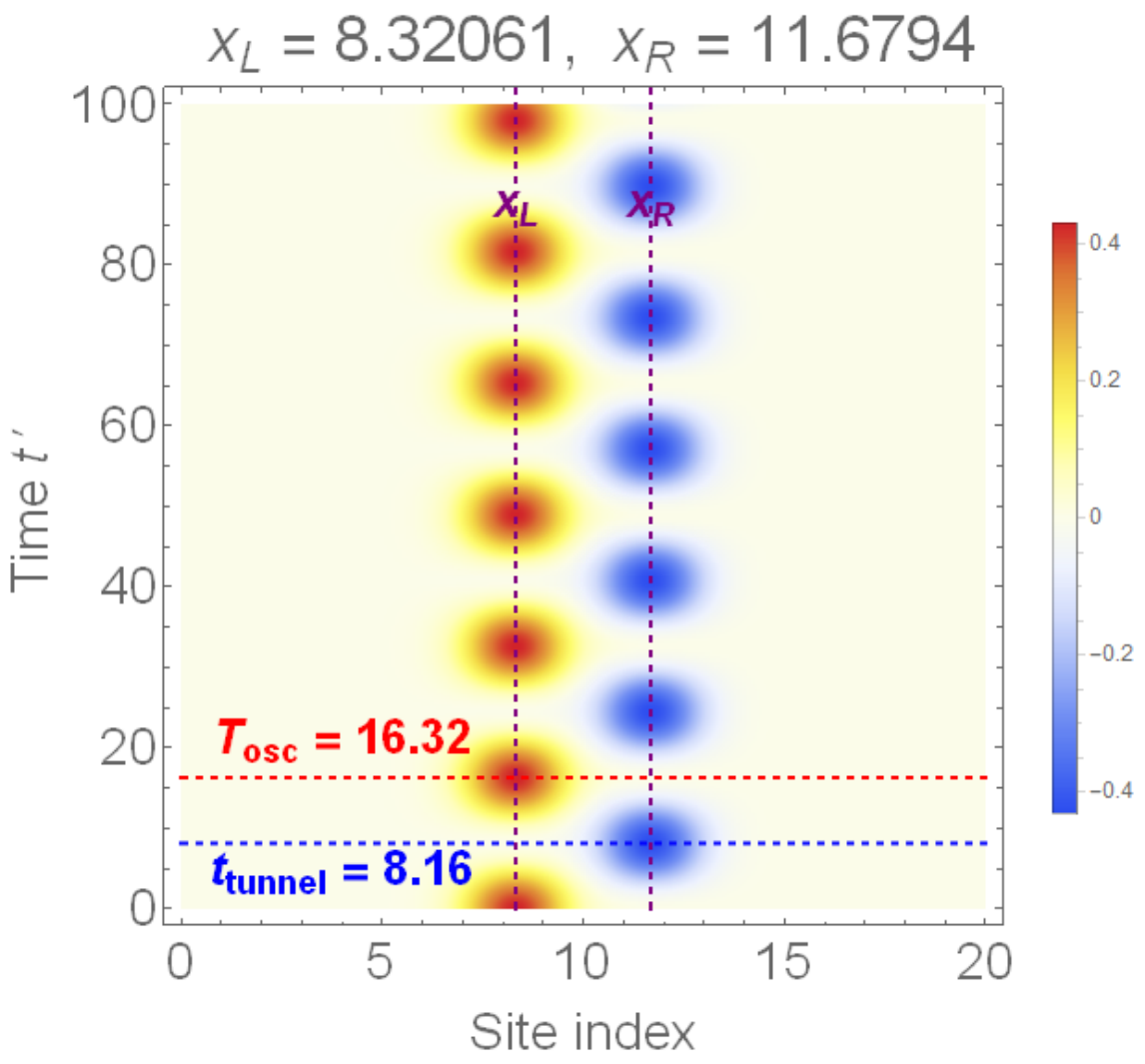}}
  \put(5,85){(a)}
  \end{picture}\\
  \vskip 1.2 in
   \begin{picture}(100,100)
    \put(-70,0){
\includegraphics[width=.98\linewidth,height=2.6in]{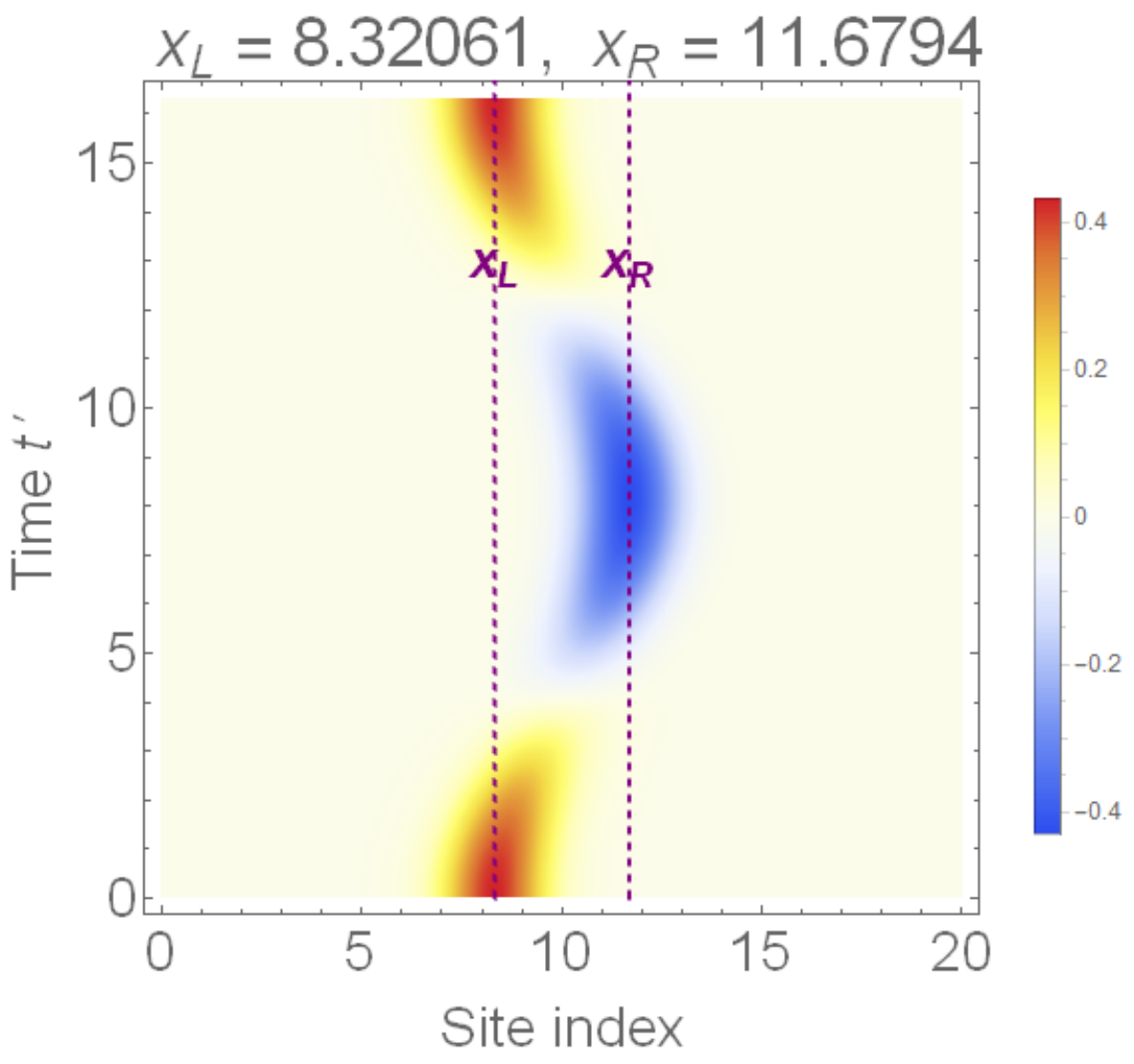}}
     \put(5,85){(b)}
   \end{picture}
\caption{(a) {\bf Sublattice polarization based oscillations between JRMs} and (b) {\bf spatio-temporal evolution of polarization density $m(x,t^{\prime})$}. The vertical dashed lines mark the DW positions $x_{L}$ and $x_{R}$. In both plots, red/yellow ($m>0$) in the color bar represents that JRMs at the left DW have weight only on (A,~C) sublattices, whereas blue ($m<0$) dictates that JRMs at the right DW predominantly occupy the (B,~D) sublattices and equal occupation of the two sectors is marked by the white ($m\simeq 0$) color. Here, we set $\Delta_{0}=8.5$, and $\xi/a=10$.} 
\label{sublattice}
\end{figure}
As is intuitively obvious, the tunneling period shows an increasing tendency (exponential nature) with increasing $R$ as well as with increasing $\xi$ and is plotted in Fig. \ref{timep} (see Appendix~\ref{tunneling}). As the tunneling period is governed by the spatial overlap of the exponentially localized JR wavefunctions, it therefore encodes information about the localization length and spatial overlap of the JR wavefunctions. Interestingly, we can observe the same coherent tunneling oscillations upon varying $\xi$ while keeping $\Delta_{0}$ fixed, as shown in Fig. \ref{tunnela} of Appendix~\ref{tunneling}.

In order to realize the change in sublattice character during tunneling, we need to study the sublattice polarization density of time-evolved JR states, $m(x,t)=\rho_{\operatorname{AC}}(x,t^{\prime})-\rho_{\operatorname{BD}}(x,t^{\prime})$, which plays the role of magnetization or pseudospin density\cite{ryu,su1,s1,s2,s3}. Thus, $m(x,t)$ in Fig. \ref{sublattice}(a) not only shows tunneling between JRMs but also depicts their sublattice polarization near each DW. At $t^{\prime}=0$, $m(x,0)>0$ ($\rho_{\operatorname{AC}}(x,t^{\prime})>\rho_{\operatorname{BD}}(x,t^{\prime})$) indicates that the JRMs at the left DW are entirely (A,~C) sublattice polarized. At $t_{tunnel}\approx 8.16$, $m(x,t_{tunnel})<0$ ($\rho_{\operatorname{BD}}(x,t^{\prime})>\rho_{\operatorname{AC}}(x,t^{\prime})$) reflects that JRMs at the right DW are (B,~D) sublattice polarized. After a full oscillation period $T_{osc}\approx 16.32$, the JRMs return to the initial configuration. The red and blue dashed horizontal lines in Fig. \ref{sublattice}(a) correctly mark these characteristic times. Furthermore, for the oscillatory center $x_{c}(t^{\prime})$, we get the polrization density as $m(x,t^{\prime})=\sqrt{\alpha/\pi}\cos(2Tt^{\prime})e^{-\alpha(x-x_{c}(t^{\prime}))^2}$ such that $x_{c}(t^{\prime})$ oscillates between the two DWs, mimicking the motion of the expectation value $\langle x(t^{\prime})\rangle$ (discussed before). Figs. \ref{sublattice}(a) and (b) clearly illustrate the periodic exchange of both localization and sublattice polarization of hybridized JRMs between the two DWs: this forces us to consider the pair of oscillatory JRMs as a ``{\it Jackiw-Rebbi molecule}'', analogous to tunneling in coupled quantum dots and hybridization of Majorana zero modes in topological superconductors\cite{alicea,majorana,majorana3}. Further, such behavior provides a topological realization of molecular bonding and antibonding orbitals\cite{griffiths,grosso} through the hybridization of JRMs bound to the DWs.

It is noteworthy to note here that, since the JRMs belong to different sublattice sectors, the oscillations are not only spatial but also involve the periodic transfer of sublattice polarization between two DWs. This makes the dynamics richer than that of a simple Schr{\"o}dinger double-well system\cite{shankar,qm,dynamic}.


\section{Summary and conclusions}\label{summary}
To summarize, in this article, we have investigated the formation, hybridization, and coherent tunneling dynamics of zero-energy JRMs in a modified SSH chain. Employing an analytical continuum description around the quadratic point $k_0=0$ does not produce topologically protected JRMs, as the effective Dirac mass $M(x)=-\Delta^2(x)$ does not undergo a topological sign inversion under the kink profile, and the result is consistent with the previous study\cite{mandal}. Interestingly, employing a continuum description around the Dirac-type points $k_0=\pm\pi/4a$, we showed that an effective Dirac Hamiltonian governs the low-energy excitations with a position-dependent mass term, $M(x)=2t^{2}-\Delta^2(x)$, where $\Delta(x)$ assumes a kink-type domain-wall profile like that shown in Eq. (\ref{kink}) of Appendix \ref{app:analytical}. Unlike the conventional SSH model\cite{kane}, here the effective Dirac mass $M(x)$ (corresponding to the low-energy Dirac theory around $k_0=\pm\pi/4a$) shows double sign reversal around the chain center $x_{0}$ (specifically at $x_{L,R}=x_0\pm\frac{\xi}{a}\,\operatorname{arctanh}\!\left(\frac{\sqrt{2}\,t}{\Delta_0}\right)$) when the DW amplitude exceeds the critical value ($\Delta_{0}>\sqrt{2}t$). Consequently, two spatially separated topological DWs emerge, each supporting a localized JR bound state.

The appearance of individual zero-energy JRMs and near-zero-energy hybridized JRMs at the DWs is entirely determined by the DW separation $D_{s}=\frac{2\xi}{a}\operatorname{arctanh}(\frac{\sqrt{2}t}{\Delta_{0}})$. Using both analytical (effective two-state description) and numerical (diagonalization of real-space Hamiltonian) calculations, we demonstrate that the two exact zero-energy JRMs are infinitely far apart when the DW separation is infinite. However, for finite separation, the spatial overlap of their exponentially decaying tails produces a finite quantum tunneling that lifts the exact zero-energy degeneracy and generates symmetric and antisymmetric combinations of JRMs separated by finite energy splitting $\Delta E$, which are identical to the bonding-antibonding states in a double-well quantum system\cite{q1,q2,griffiths,shankar}. Moreover, the tunneling strength in the spatial overlap regime typically scales as $T\sim e^{-R^2/4}$, while in the asymptotic tail regime it typically scales as $T \sim e^{-R (l_G/l_D)}$.

To elucidate the coherent tunneling dynamics of hybridized JRMs, we analyzed the real-time evolution of JRMs initially localized at the left DW. The dynamics is governed by tunnel splitting $\Delta E$, and the corresponding occupation probabilities oscillate as $P_{L}(t^{\prime})=\cos^2(Tt^{\prime})$  and $P_{R}(t^{\prime})=\sin^2(Tt^{\prime})$, demonstrating that the JRMs, initially localized at the left DW, undergo coherent quantum tunneling to the right DW with tunneling time $t_{tunnel}$ and periodically return to the left DW with full oscillation period $T_{osc}$. However, looking at the average position $\langle x(t^{\prime})\rangle$ of the probability density of JRMs at time $t^{\prime}$, we notice the oscillatory nature of it between $x_{L}$ and $x_{R}$, which is a hallmark of coherent tunneling oscillations. This behavior constitutes a topological realization of Rabi oscillations in a quantum mechanical double-well system\cite{shankar,qm,dynamic} and provides a direct manifestation of quantum coherence between spatially separated defect states (JRMs). The resulting dynamics is formally equivalent to those of a quantum particle in a double-well potential\cite{shankar,qm,dynamic}. Furthermore, choosing an initial state localized at the right DW ensures that the oscillatory dynamics originate from the right DW. Moreover, the probability density function of time-evolving JRMs $\rho(x,t^{\prime})=\rho_{\operatorname{AC}}(x,t')+\rho_{\operatorname{BD}}(x,t')$ nicely portrays the evolution
from independent JRMs (with high $T_{osc}$) to strongly hybridized JRMs (with low $T_{osc}$) with the decrease of DW separation.

To visualize sublattice polarization during the tunneling process, we introduce the sublattice polarization density, $m(x,t^{\prime})=\rho_{\operatorname{AC}}(x,t')-\rho_{\operatorname{BD}}(x,t')$, which acts as a local pseudospin polarization density in the effective (AC,~BD) sublattice basis\cite{ryu,su1,s1,s2,s3}. The spatiotemporal evolution of $m(x,t^{\prime})$ directly visualizes the coherent exchange of both localization and sublattice character of hybridized JRMs between the two DWs, providing a real-space signature of coherent tunneling and hybridization. Thus, the periodic exchange of localization, sublattice polarization, and probability density establishes the hybridized JRMs pair as a topological analogue of a diatomic molecular system, which may be viewed as a ``{\it Jackiw–Rebbi molecule}".

Our results establish a unified connection between topological defect states, quantum tunneling, and molecular-state formation in 1D topological systems. The modified SSH chain, therefore, provides a controllable platform for experimentally investigating the coherent dynamics of topological bound states, since the localization length, DW separation, and tunnel coupling can be tuned through the parameters of the domain-wall profile. 
Furthermore, our predicted tunnel splitting and coherent oscillations of JRMs should be experimentally accessible in several synthetic topological platforms, particularly in femtosecond-laser-written photonic waveguide arrays\cite{wave,wave1}, cold-atom realization of SSH chains\cite{gadway,a1,leder}, graphene-based engineered Dirac systems\cite{a2}, topological photonic or acoustic lattices\cite{li,esmann,amo,a3} and quantum gas microscopy\cite{quantum}. In all such systems, the spatially varying hopping amplitudes can be engineered with high precision.  Specifically, in photonic systems, the propagation distance acts as an effective time coordinate, allowing direct observation of coherent oscillations of light intensity between the two interface modes, while quantum gas microscopy enables site-resolved imaging of analogous tunneling dynamics in cold-atom lattices. In particular, the two JRMs could be identified through spatially resolved measurements of the local intensity, density, or spectral response around the two DWs, as experimentally established in SSH-type photonic and acoustic systems, where localized DW or edge states have been directly observed through their spatial profiles\cite{li,amo,tang,tang}. Notably, the hybridization of the two JRMs could be verified experimentally by measuring the near zero-energy splitting through controlled variation of the topological DW separation $D_s$\cite{devices1}. Moreover, SSH photonic lattices enable controlled repositioning and interaction of DWs through tunable inter-site couplings\cite{gabriel}, providing a feasible route to systematically vary $D_s$ and test the predicted Gaussian-overlap and asymptotic exponential regimes of the tunnel coupling. The coherent tunneling dynamics can be experimentally probed by selectively exciting a JRM at one DW and monitoring the time-resolved population transfer between the two DWs, with $P_L(t')=\cos^2(Tt')$ and $P_R(t')=\sin^2(Tt')$ providing a direct measure of the tunnel coupling. Sublattice-resolved detection can further reveal the associated polarization based oscillations, offering an independent signature of JRM hybridization, as demonstrated by selective excitation and dynamical control of topological states in SSH waveguide lattices \cite{tang}.

These experimentally accessible signatures not only provide direct tests of the predicted JRM hybridization and coherent tunneling dynamics, but also highlight broader physical implications offered by the spatially engineered mass profile. Beyond the tunneling phenomena discussed in this work, the effective mass profile naturally creates a finite topological cavity bounded by two mass-inversion points, suggesting possible connections to resonant cavity modes, quasi-bound states, and analogue horizon physics\cite{cavity,cavity1,cavity2,bh1,bh2}. Moreover, understanding the hybridization of spatially separated JRMs is therefore crucial for designing the topological quantum devices\cite{devices,devices1}, where controlled coupling between JRMs serves as a key resource.

The present work opens several promising directions for future research, including Floquet engineering of topological states and defect modes \cite{a5,kitagawa,lindner,rudner}, disorder-induced modifications of topological bound states and coherent dynamics \cite{hughes,meier}, interaction effects in topological zero-mode systems \cite{rachel}, non-Hermitian extensions of topological interface physics \cite{yao,shiozaki,gong}, and many-body realizations in synthetic quantum platforms \cite{atala,gadway,lohse,tomita}. More broadly, the demonstrated interplay between topology, localization, and quantum tunneling highlights the potential of engineered domain-wall structures as versatile building blocks for topological quantum-state manipulation and transport\cite{kiteav,majorana2}. Moreover, the result of the paper is interesting and plausible in connection to experiments on quantum dot SSH models\cite{pan,pham}.  Lastly, we should mention here that in the future, we have a plan to study the out-of-equilibrium behavior of such a modified SSH model, subjected to a quantum quench that leads to an effective metal-insulator transition for the conventional SSH model\cite{a4}.
\section*{Acknowledgements}
The author thanks S. Kar and S. Mandal (IOP, Bhubaneswar) for the fruitful discussions. The author acknowledges financial support from DST-SERB (ANRF), Government of India under grant no. CRG/2022/002781.
\section*{DATA AVAILABILITY}
The data that support the findings of this study are available from the corresponding author upon reasonable request.

\appendix
\section{}
\label{app:A}
\subsection{Analytical Jackiw-Rebbi Method: Continuum limit of Momentum-space Hamiltonian}
\label{app:analytical}
In the seminal work by JR\cite{rebbi}, one-dimensional Dirac fermions are coupled with a solitonic field. The solitonic field was introduced in the low-energy effective Dirac Hamiltonian as a position-dependent mass term, the specific structure of which determines the topology of the system. A change of sign in the mass term about the DW yields localized ZES at its position. In this respect, it is noted that intermediate-order topological interfacial states can be found in an $n$-dimensional SSH model as elaborated in Ref.\cite{feng}. The topological defect of the solitonic field is known as a kink. Polyacetylene is the condensed-matter realization of such a situation at any DW junction between two possible dimerizations ($\Delta/t<0$ and $\Delta/t>0$)\cite{su1,su2}. The general recipe for the JR method is as follows: 

Expand the Bloch Hamiltonian Eq. (\ref{11}) around the gap-closing point such that the low-energy Hamiltonian follows a 1D Dirac equation where the mass term becomes position-dependent (spatially inhomogeneous), then change its sign by introducing a topological defect. The typical choice of mass profile, centered at site $x=x_{0}$, is defined as
\begin{equation}\label{kink}
M(x)=M_{0}\tanh\Big[\frac{x-x_{0}}{\xi/a}\Big],
\tag{A1}\end{equation}
in which $M_{0}$ and $\xi$ denote the amplitude and the width of the mass profile, respectively. Now, $M(x)$ should satisfy $\lim \limits_{x-x_{0}\to-\infty}M(x)=-M_{0}$ and  $\lim \limits_{x-x_{0}\to\infty}M(x)=M_{0}$. For $M_{0}>0$, two constant fields $\pm M_{0}$ represent the two different phases of our system, and only the topological texture holds a single localized state with energy $E = 0$\cite{dw,su1,su2}. To describe the zero-energy JR mode, we allow the kink-shaped mass profile $M(x)$ to interpolate between $M_{0}$ and $-M_{0}$ and create a DW at $x=x_{0}$\cite{jackiw}. Using this mechanism, one can analytically investigate the JR zero modes in the presence of DWs.

\subsection{Determinant-based and Projection Method}\label{hamila}
{\bf Determinant-based Method:} The Bloch Hamiltonian Eq. (\ref{11}) is in the basis $\Psi_{k}=\begin{pmatrix}
\operatorname{A}& \operatorname{B}&\operatorname{C}&\operatorname{D}\end{pmatrix}^T$. The Bloch Hamiltonian can be changed to block off-diagonal form, as it preserves the chiral symmetry\cite{mandal}. We first apply the unitary transformation with the unitary matrix 
\begin{equation}\label{8}
U=
\begin{pmatrix}
1 & 0 & 0 & 0 \\\
0 & 0 & 1 & 0\\\
0 & 1 & 0 & 0\\\
0 & 0 & 0 & 1
\end{pmatrix}
\tag{A2}\end{equation}
makes the Hamiltonian $\mathcal{H}(k)$ in Eq. (\ref{11}) into the block off-diagonal form 
\begin{equation}\label{7aa}
\mathcal{\widetilde H}(k)\rightarrow U\mathcal{H}(k)U^{-1}=\begin{pmatrix}
0 & V (k)\\
V^{\dagger}(k) & 0 
\end{pmatrix},
\tag{A3}
\end{equation}
where the upper off-diagonal block is the $2\times2$ square matrix and is read as
\begin{equation}\label{7a}
V(k)=
\begin{pmatrix}
 (t+\Delta) & te^{-i4ak} \\
 t & (t-\Delta)
\end{pmatrix}
\tag{A4}\end{equation}
with ${\bf Det}[V(k)]=(t^2-\Delta^2)-t^2e^{-i4ak}$ is a complex quantity which is an effective two-dimensional topological vector. The applied unitary matrix also reorders the basis $\Psi_k$ as $\phi=U\Psi_k=\begin{pmatrix}
\phi_{\rm {AC}}& \phi_{\rm {BD}}\end{pmatrix}^T$ with $\phi_{\rm AC}=\begin{pmatrix}
\rm {A}& \rm {C}\end{pmatrix}^T$ and $\phi_{\rm BD}=\begin{pmatrix}
\rm {B}& \rm {D}\end{pmatrix}^T$. Therefore, the block off-diagonal Hamiltonian is confined in this new basis. One can write ${\bf Det}[V(k)]=h_{x}(k)-ih_{y}(k)$, where $h_{x}(k)=\operatorname{Re}[{\bf Det}[V(k)]]=(t^2-\Delta^2)-t^2 \cos 4ak$ and $h_{y}(k)=\operatorname{Im}[{\bf Det}[V(k)]]=-t^2 \sin 4ak$. Thus, the topological information can then be characterized in terms of this complex quantity. Accordingly, the block off-diagonal form can then be written as an effective two-band Hamiltonian associated with the topological vector as:
\begin{equation}\label{blockoff}
\mathcal{H_{\mathrm{eff}}}(k)=h_{x}(k)\sigma_{x}+h_{y}(k)\sigma_{y},
\tag{A5}\end{equation}
in which $\sigma_x$ and $\sigma_y$ refer to the Pauli matrices. Here, $\mathcal{H}_{\mathrm{eff}}(k)$ provides an effective two-level representation of the topological structure encoded in ${\bf Det}[V(k)]$.

Now, expanding around $k_0=\pm\pi/4a$, we get $h_x(k)$ and $h_y(k)$ as
\begin{equation}\label{dirac}
  h_{x}(k)=2t^2-\Delta^2-\frac{8a^2t^2q^2}{\hbar^2}+...,~\  \ h_y(k)=\frac{4at^2q}{\hbar}+..,
\tag{A6}\end{equation}
with $(k-k_0)=q/\hbar$ and $|q|/\hbar\ll1/a$, in which $q$ and $ \hbar$ denote the momentum operator and the reduced Planck constant, respectively. Therefore, performing a low-energy description of Eq. (\ref{blockoff}) near $k_0=\pm\pi/4a$, we arrive at an effective Dirac Hamiltonian as following
\begin{equation}\label{blockoff1}
\mathcal{H}_{\mathrm{eff}}^D(q)=(2t^2-\Delta^2)\sigma_x+\frac{4at^2}{\hbar}q\sigma_y+\mathcal{O}(q^2)
\tag{A7}\end{equation}
Upon substituting $q=-i\hbar\partial_{x}$, we obtain the effective continuum Hamiltonian as given below 
\begin{equation}\label{blockoff2}
\mathcal{H}_{\mathrm{eff}}^D(x)=M(x)\sigma_x-i\hbar v\sigma_y\partial_x+\mathcal{O}(\partial_x^2)
\tag{A8}\end{equation}
with the effective Dirac mass $M(x)=2t^2-\Delta^2(x)$ and $v=4at^2/\hbar$ is the effective Dirac velocity. Crucially, the $2\times2$ continuum Dirac Hamiltonian Eq.~\eqref{blockoff2} and the resulting quadratic form of effective mass are not assumed phenomenologically; they are obtained from the low-energy expansion of the effective topological Hamiltonian (Eq. (\ref{blockoff})) around each of the gap-closing points $k_0=\pm\pi/4a$.

$\bullet$ To determine the JRMs in the microscopic four-component basis, we now embed the Dirac envelopes into the appropriate null-space vectors. Now, the zero-energy equation of the block off-diagonal Hamiltonian (Eq. \eqref{7aa}) in the reordered basis, $\mathcal{\widetilde H}(k)\phi=0$, gives $V(k)\phi_{\rm BD}=0$ and $V(k)^{\dagger}\phi_{\rm AC}=0$. At a non-trivial gap-closing point, ${\bf Det}[V(k)]=0$ and ${\bf Det}[V^\dagger(k)]=0$, so both $V(k)$ and $V^\dagger(k)$ possess nontrivial null spaces, respectively. The corresponding two-component null vectors, $\phi_{\rm BD}$ and $\phi_{\rm AC}$, determine the four-component ZESs. 

t either critical momenta $k_0=\pm\pi/4a$, $e^{-i4ak_0}=-1$. Thus
\begin{equation}\label{7a}
V(k)=
\begin{pmatrix}
 (t+\Delta(x)) & -t \\
 t & (t-\Delta(x))
\end{pmatrix}
\tag{A9}\end{equation}
At the left DW $\Delta(x_L)=-\sqrt{2}t$, the solution of $V(k)^{\dagger}\phi_{\rm AC}=0$ results in $\phi_{\rm AC}=N~\begin{pmatrix}
\rm {1}& \sqrt{2}-1\end{pmatrix}^T$. Similarly, solving $V(k)\phi_{\rm BD}=0$ at the right DW $\Delta(x_R)=\sqrt{2}t$ produces $\phi_{\rm BD}=N'~\begin{pmatrix}
\rm {1}& 1+\sqrt{2}\end{pmatrix}^T$. After embedding the two-component null vectors $\phi_{\rm AC}$ and $\phi_{\rm BD}$ into the appropriate sector, the normalized four-component zero-energy eigenvectors, in the basis $\Psi_{k}=\begin{pmatrix}
\operatorname{A}& \operatorname{B}&\operatorname{C}&\operatorname{D}\end{pmatrix}^T$, are obtained as $u_{\operatorname{AC}}^{(-)}=N~\begin{pmatrix}
1 & 0&(\sqrt{2}-1)&0\end{pmatrix}^T$ and $u_{\operatorname{BD}}^{(+)}=N^{\prime}~\begin{pmatrix}
0 & 1&0&(1+\sqrt{2})\end{pmatrix}^T$ with $N=[1+(\sqrt{2}-1)^2]^{-1/2}$ and $N'=[1+(1+\sqrt{2})^2]^{-1/2}$ denoting the normalization constant. It is noticed that $u_{\rm AC}^{(-)}$ and $u_{\rm BD}^{(+)}$ have a nonzero amplitude on the $(\rm A,C)$ and $(\rm {B,D})$ sublattices, respectively. They are normalized and orthogonal, as they follow $\langle u_{\rm AC}^{(-)}|u_{\rm AC}^{(-)}\rangle=\langle u_{\rm BD}^{(+)}|u_{\rm BD}^{(+)}\rangle=1$ and $\langle u_{\rm AC}^{(-)}|u_{\rm BD}^{(+)}\rangle=\langle u_{\rm AC}^{(+)}|u_{\rm BD}^{(-)}\rangle=0$. Importantly, the same null-space procedure also produces $u_{\operatorname{AC}}^{(+)}=N'~\begin{pmatrix}
1 & 0&-(1+\sqrt{2})&0\end{pmatrix}^T$ and $u_{\operatorname{BD}}^{(-)}=N''~\begin{pmatrix}
0 & 1&0&-(\sqrt{2}-1)\end{pmatrix}^T$ with normalization constant $N'=[1+(1+\sqrt{2})^2]^{-1/2}$ and $N''=[1+(\sqrt{2}-1)^2]^{-1/2}$, respectively, at the other critical value of $\Delta(x)$.

The two-component normalized envelope spinors derived in Eq. \eqref{envelope} refer to the effective Dirac basis, not directly to the physical A,B,C,D sites. To obtain the four-component JR states, we map this effective component onto the appropriate non-trivial gap-closing basis vectors. Consequently, we get
\begin{equation}
\boxed{
\begin{aligned}
\Psi_L(x)&=f_L(x)u_{AC}^{(-)},\\
\Psi_R(x)&=f_R(x)u_{BD}^{(+)}.
\end{aligned}}
\tag{A10}
\label{JRs}
\end{equation}

{\bf Projection Procedure:}
The behavior of the JRMS along with their microscopic sublattice-based polarization can also be investigated by the projection method. To elucidate the sublattice structure of these modes, we express the Bloch Hamiltonian in Eq.~(\ref{11}) in the four-sublattice basis $\Psi_k=(\rm A~B~C~D)^T$ and construct the low-energy basis at the Dirac-type gap-closing points $k_0=\pm\pi/4a$. For the critical point $k_0=\pm\pi/4a$ and the critical modulation \(\Delta_c=s\sqrt{2}t\) with $s=\pm1$, the zero-energy basis states are obtained by solving the eigenvalue equation
$\mathcal H(k_0,\Delta_c)u_j^{(s)}=0$.
The resulting two linearly independent and orthonormal eigenvectors, for $s=+1$, can be written as
$u_{\operatorname{AC}}^{(+)}
=C'\begin{pmatrix}
1&0&-(1+\sqrt{2})&0
\end{pmatrix}^{T},$ $u_{\operatorname{BD}}^{(+)}
=C'\begin{pmatrix}
0&1&0&1+\sqrt{2}
\end{pmatrix}^{T}$,
where \(C'=[1+(1+\sqrt{2})^2]^{-1/2}\) is the normalization constant. While for $s=-1$, the corresponding orthonormal zero-energy eigenvectors are $u_{\operatorname{AC}}^{(-)}
=C''\begin{pmatrix}
1&0&(\sqrt{2}-1)&0
\end{pmatrix}^{T},$ $u_{\operatorname{BD}}^{(-)}
=C''\begin{pmatrix}
0&1&0&-(\sqrt{2}-1)
\end{pmatrix}^{T}$,
where \(C''=[1+(\sqrt{2}-1)^2]^{-1/2}\) is the normalization constant. The states $u_{\operatorname{AC}}^{(s)}$ and $u_{\operatorname{AC}}^{(s)}$ have support exclusively on the (A,~C) and (B,~D)
sublattices, respectively. Importantly, they satisfy \(\bigl(u_i^{(s)}\bigr)^\dagger u_j^{(s)}=\delta_{ij}\) and therefore span the two-dimensional zero-energy eigenspace at the critical points.

We introduce the $4\times2$ projection matrix $U_s=(u_{\operatorname{AC}}^{(s)}~u_{\operatorname{BD}}^{(s)})$, whose columns are the two orthonormal basis states. It satisfies $U_s^\dagger U_s=I_2$, where $I_2$ is $2\times 2$ identity matrix, and the projector onto the low-energy subspace is
\begin{equation}
\mathcal{P}=U_s U_s^\dagger
=|u_{\operatorname{AC}}^{(s)}\rangle \langle u_{\operatorname{AC}}^{(s)}|
+|u_{\operatorname{BD}}^{(s)}\rangle \langle u_{\operatorname{BD}}^{(s)}|.
\tag{A11}
\end{equation}
Within the low-energy approximation, the slowly varying part of the lattice wavefunction can consequently be expanded as
\begin{equation}
\Psi(x)=U_s\phi(x)
=\phi_1(x)u_{\operatorname{AC}}^{(s)}
+\phi_2(x)u_{\operatorname{BD}}^{(s)},
\tag{A12}
\end{equation}
where \(\phi(x)=(\phi_1(x)~\phi_2(x))^T\) is the two-component envelope spinors. The envelope amplitudes are obtained by projecting the wave function onto the corresponding basis states, \(\phi_j(x)=u_j^{(s)\dagger}\Psi(x)\), provided that \(\Psi(x)\) is restricted to the low-energy subspace. Here, $\{u_{\rm AC}^{(s)}~u_{\rm BD}^{(s)}\}$ form a two-dimensional zero-energy subspace, which is embedded in the microscopic four-sublattice space. 

To derive the effective continuum Hamiltonian, we project the Bloch Hamiltonian in Eq.~(\ref{11}) onto the two-dimensional zero-energy subspace at the Dirac-type gap-closing points \(k_0=\pm\pi/4a\). We consider both critical modulations, $\Delta_c=s\sqrt{2}\,t$. Defining \((k-k_0)=q/\hbar\), with $|q|/\hbar\ll1/a$, and writing $\Delta=\Delta_c+\delta\Delta$, we expand the Bloch Hamiltonian (Eq.~(\ref{11})) as $\mathcal H(k,\Delta)\simeq\mathcal H_c+\delta\mathcal H$, retaining terms only to first order in $q$ and $\delta\Delta$. Since $e^{-i4ak_0}=-1$, we have $e^{-i4ak}\simeq-1+i\frac{4aq}{\hbar}$ and $e^{i4ak}\simeq-1-i\frac{4aq}{\hbar}$. Consequently, the first-order perturbation is
\begin{equation}
\delta\mathcal H=
\begin{pmatrix}
0&\delta\Delta&0&i\frac{4atq}{\hbar}\\
\delta\Delta&0&0&0\\
0&0&0&-\delta\Delta\\
-i\frac{4atq}{\hbar}&0&-\delta\Delta&0
\end{pmatrix}.
\tag{A13}
\end{equation}
Projecting the Bloch Hamiltonian in Eq. \eqref{11} onto the two-dimensional zero-energy subspace spanned by $\{u_{\rm AC}^{(s)}~u_{\rm BD}^{(s)}\}$, we obtain the projected Hamiltonian $\mathcal H_{\mathrm{proj}}^{(s)}=U_s^\dagger\mathcal H(k,\Delta)U_s$.  At the critical point $(k_c,\Delta_c)$, $U_s^\dagger\mathcal H_cU_s=0$. Therefore, the diagonal matrix elements vanish, and the off-diagonal element becomes nonzero. Hence, the projection gives $
\mathcal{H}_{\mathrm{proj}}^{(s)}=U_s^\dagger\delta\mathcal H U_s
=\begin{pmatrix}
0&h_s\\
h_s^*&0
\end{pmatrix}$, where
$h_s=u_{\operatorname{AC}}^{(s)\dagger}
\delta\mathcal H\,u_{\operatorname{BD}}^{(s)}
=\delta\Delta+is\frac{\sqrt2\,atq}{\hbar}$. Consequently, near either critical momentum, the effective projected Hamiltonian takes the form of
\begin{equation}\label{two-a}
\mathcal{H}_{\mathrm{proj}}^{(s)}(q,\Delta)
=\delta\Delta\,\sigma_x-s\frac{\sqrt{2}atq}{\hbar}\sigma_y+\mathcal{O}(q^2,q\delta\Delta,
(\delta\Delta)^2).
\tag{A14}
\end{equation}
Thus, the sign of the kinetic term depends on the sign of the critical modulation, whereas the mass term in this basis is $\delta\Delta$. For convenience, we now adopt an equivalent representation of the projected Hamiltonian. We perform the constant unitary transformation $U_s\mapsto\widetilde U_s=U_sW_s$, where $W_{+1}=\sigma_z$ and $W_{-1}=I_2$. The resulting Hamiltonian is
\begin{equation}\label{two-d}
\widetilde{\mathcal H}_{\rm proj}^{(s)}(q,\delta\Delta)=-s\delta\Delta\,\sigma_x+\frac{\sqrt2\,atq}{\hbar}\sigma_y.
\tag{A15}
\end{equation}
Now, near each non-trivial critical point $\Delta_c=s\sqrt{2}t$, expanding $(2t^2-\Delta^2(x))$ gives $(2t^2-\Delta^2(x))=-s2\sqrt{2}t\delta\Delta(x)-(\delta\Delta(x))^2$. Therefore, the position-dependent effective Dirac mass, up to the leading order, is $M_D(x)=-s\delta\Delta(x)\simeq\frac{(2t^2-\Delta^2(x))}{2\sqrt{2}t}$. We then obtain the continuum Dirac Hamiltonian by substituting \(q\rightarrow-i\hbar\partial_x\) as
\begin{equation}
\mathcal H_{\rm proj}^{D}(x)
=M_D(x)\sigma_x-i \hbar v_D\sigma_y\partial_x+\mathcal{O}(\partial_x^2,(\delta\Delta)^2),
\tag{A16}
\end{equation}
where \(v_D=\frac{\sqrt2\,at}{\hbar}\) is the effective Dirac velocity. The resulting continuum Hamiltonian provides a low-energy description near either a Dirac-type gap-closing point or a non-trivial critical modulation. The procedure for getting the resulting two-component JR envelope has been discussed in Sec. \ref{ana}.

$\bullet$ The corresponding four-component JRMs can also be obtained. At the left DW, where $\Delta(x_L)=-\sqrt{2}t$, we use the basis $u_{\operatorname{AC}}^{(-)}$, while at the right DW, where $\Delta(x_R)=+\sqrt{2}t$, we use the basis $u_{\operatorname{AC}}^{(+)}$. The four-sublattice microscopic wavefunctions are reconstructed from the two-component envelope spinors through $\Psi_j(x)=\widetilde U_s\phi_j(x)$, where $\widetilde U_s$ denotes the appropriate basis transformation at each DW. Specifically, the continuum representation uses $\widetilde U_-=U_-$ at the left DW and $\widetilde U_+=U_+\sigma_z$ at the right DW. Combining these transformations with the corresponding chirality eigenvectors, $\chi_L=(1~0)^T$ and $\chi_R=(0~1)^T$, and absorbing the overall phase factor, yields the DW states in the microscopic four-sublattice basis as written in Eq. \eqref{JRs}.

$\bullet$ Importantly, the continuum Dirac Hamiltonian, \(\mathcal{H}_{\mathrm{proj}}^D(x)\), obtained via the projection method is characterized by the effective mass $M_D(x)=\frac{2t^2-\Delta^2(x)}{2\sqrt{2}t}$ and the Dirac velocity $v_D=\frac{\sqrt{2}at}{\hbar}$. Hence, the continuum Hamiltonians obtained from the two methods differ only by a constant normalization factor, as they are related via $\mathcal{H}_{\mathrm{proj}}^D(x)=\frac{1}{2\sqrt{2}t}\mathcal{H}_{\mathrm{eff}}^D(x)$. Therefore, the overall low-energy Dirac physics, including the same mass-sign reversal, DW structure, zero-energy JRMs, localization properties, and importantly tunneling dynamics, remained unchanged; it only rescales the associated energy and velocity scales consistently.

$\bullet$ It is instructive to discuss the physical origin and uniqueness of the resulting quadratic mass profile, which consistently emerges from both our topological determinant formulation and the subspace projection method.  The bulk energy gap is uniquely determined by the eigenvalues of the Bloch Hamiltonian (Eq. \eqref{11}), which remain invariant under any unitary basis transformations. Accordingly, the zero-energy phase boundaries at $|\Delta/t| = \sqrt{2}$ are independent of the chosen representation. In the chiral off-diagonal basis, the determinant of the off-diagonal block takes the form ${\bf Det} [V(k)]=(t^2-\Delta^2)-t^2e^{-i4ak}$. Evaluating its real part around the gap-closing momentum $k_0=\pm\pi/4a$ yields the quadratic combination $(2t^2-\Delta^2)$, which determines the vanishing of the bulk gap. The same quadratic dependence also follows from the subspace projection method. At the two gap-closing points $k_0=\pm\pi/4a$, projection of the Bloch Hamiltonian (Eq. \eqref{11}) onto the corresponding low-energy subspaces yields effective mass terms proportional to $(2t^2-\Delta^2)$. Although the two projections may differ in their basis-dependent signs or Pauli-matrix representations, both reproduce the same gap-closing condition, $ \Delta_c=\pm\sqrt {2} t$. 

Nevertheless, the explicit matrix representation of the continuum Hamiltonian, including the assignment of the mass term to a particular Pauli matrix, will alter the explicit assignment of the Pauli matrices (e.g., transforming a mass term from $\sigma_x$ to $\sigma_z$), the functional quadratic form of the physical mass profile $M(x)$ and its spatial crossover points are uniquely fixed by the lattice topology itself. Thus, the quadratic dependence on hopping modulation is intrinsic to the bulk gap-closing structure of the underlying lattice model rather than being an artifact of the chosen basis.


\subsection{Effective Two-Level Hamiltonian and the Tunneling Matrix element}\label{projection}
{\bf Effective Two-Level Hamiltonian:} Here, we derive the effective two-level Hamiltonian by projecting the
continuum Dirac Hamiltonian onto the subspace spanned by the two
localized JRMs. The continuum Hamiltonian
given in Eq. \eqref{low} possesses chiral
symmetry, generated by the effective chiral operator $\Gamma_{\rm eff}=\sigma_z$, such that
$\left\{\mathcal{H}_{\rm eff}^{D},\Gamma_{\rm eff}\right\}=0,
\quad \Gamma_{\rm eff}^2=I_2$.

However, the two-component normalized and orthogonal JRMs, namely $|\phi_L\rangle$ and $|\phi_R\rangle$, are eigenstates of the chiral operator $\Gamma_{\rm eff}$ with opposite eigenvalues:
\begin{equation}
\begin{aligned}
\Gamma_{\rm eff}|\phi_L\rangle&=+|\phi_L\rangle,\\
\Gamma_{\rm eff}|\phi_R\rangle&=-|\phi_R\rangle,
\end{aligned}
\qquad
\langle\phi_i|\phi_j\rangle=\delta_{ij}.
\label{JR_chiral_basis}
\tag{A17}
\end{equation}
Additionally, the orthogonality of the two states follows from their opposite chiral eigenvalues. We now introduce the projector onto the two-dimensional subspace spanned by these states, $\{|\phi_L\rangle,|\phi_R\rangle\}$, as
\begin{equation}
\mathcal{P}'=|\phi_L\rangle\langle\phi_L|
 +|\phi_R\rangle\langle\phi_R|.
\label{JR_projector-a}
\tag{A18}
\end{equation}
Now, projecting the continuum Hamiltonian $\mathcal{H}_{\rm eff}^{D}$ onto this subspace produces the projected two-level Hamiltonian
$\mathcal{H}_{\rm eff}^{P}
=\mathcal{P}'\mathcal{H}_{\rm eff}^{D}\mathcal{P}'$. In the orthonormal basis $\{|\phi_L\rangle,|\phi_R\rangle\}$, its matrix elements are
\begin{equation}
\left(\mathcal{H}_{\rm eff}^{P}\right)_{ij}
=\langle\phi_i|\mathcal{H}_{\rm eff}^{D}|\phi_j\rangle,
\qquad i,j\in\{L,R\}.
\label{projected_matrix_elements}
\tag{A19}
\end{equation}
Thus,
\begin{equation}
\mathcal{H}_{\mathrm{eff}}^{P}=
\begin{pmatrix}
\langle\phi_L|\mathcal{H}_{\mathrm{eff}}^{D}|\phi_L\rangle&\langle\phi_L|\mathcal{H}_{\mathrm{eff}}^{D}|\phi_R\rangle\\[6pt]
\langle\phi_R|\mathcal{H}_{\mathrm{eff}}^{D}|\phi_L\rangle&\langle\phi_R|\mathcal{H}_{\mathrm{eff}}^{D}|\phi_R\rangle
\end{pmatrix}
\tag{A20}
\end{equation}
The diagonal matrix elements vanish as a consequence of chiral symmetry. For either $i=L$ or $R$, using
$\Gamma_{\rm eff}|\phi_i\rangle=\gamma_i|\phi_i\rangle$, where
$\gamma_i=\pm1$, and $\{\mathcal{H}_{\rm eff}^{D},\Gamma_{\rm eff}\}=0$, we obtain the following:
\begin{equation}
\langle\phi_i|\mathcal{H}_{\rm eff}^{D}|\phi_i\rangle
=\langle\phi_i|
\Gamma_{\rm eff}\mathcal{H}_{\rm eff}^{D}\Gamma_{\rm eff}
|\phi_i\rangle \nonumber\\
=-\langle\phi_i|
\mathcal{H}_{\rm eff}^{D}|\phi_i\rangle.
\tag{A21}
\end{equation}
Consequently,
\begin{equation}
\langle\phi_L|\mathcal{H}_{\rm eff}^{D}|\phi_L\rangle
=\langle\phi_R|\mathcal{H}_{\rm eff}^{D}|\phi_R\rangle=0.
\label{zero_diagonal}
\tag{A22}
\end{equation}
In contrast, chiral symmetry permits a nonzero off-diagonal matrix element, which describes tunneling between the two localized JRMs. Now, defining\cite{d1,devices1}
\begin{equation}
T=\langle\phi_L|\mathcal{H}_{\rm eff}^{D}|\phi_R\rangle,
\label{tunneling_matrix_element}
\tag{A23}
\end{equation}
Hermiticity gives
$\langle\phi_R|\mathcal{H}_{\rm eff}^{D}|\phi_L\rangle=T^*$.
The effective two-state Hamiltonian therefore takes the form:
\begin{equation}
\mathcal{H}_{\rm eff}^{P}=
\begin{pmatrix}
0&T\\
T^*&0
\end{pmatrix}.
\label{projected_two_level}
\tag{A24}
\end{equation}
The off-diagonal matrix element $T$ can be evaluated explicitly by the overlap integral $T=\langle\phi_L|\mathcal{H}_{\mathrm{eff}}^D|\phi_R\rangle
=\int dx\phi_L^\dagger(x)\mathcal{H}_{\mathrm{eff}}^D(x)\phi_R(x)$,
where $\phi_{L,R}(x)=f_{L,R}(x)\chi_{L,R}$ denote the normalized localized JRMs, with $f_{L,R}(x)$ their spatial envelopes and $\chi_{L,R}$ their internal spinors. Although the envelope functions generally have a finite spatial overlap, $S_{\mathrm{env}}=\int dxf_L^*(x)f_R(x)$, the full-state overlap vanishes due to the orthogonality of their internal spinors, $\chi_L^\dagger\chi_R=0$. Hence, the projected Hamiltonian is represented in an orthonormal basis satisfying
$\langle\phi_i|\phi_j\rangle=\delta_{ij},
\quad \text{with} \quad i,j\in\{L,R\}$ and $\delta_{ij}$ is the Kronecker delta function, and no overlap-matrix correction is required within this two-state subspace.


{\bf Tunneling Matrix Element:} The tunneling strength between the JRMs can be obtained by looking at the spatial overlap region where the JRMs exhibit Gaussian core behavior. The tunneling strength can be estimated either by a detailed overlap integral or by using a dimensional estimation based on the characteristic kinetic-energy scale of the continuum Hamiltonian. First, we proceed by evaluating the line integral. The tunneling matrix element becomes\cite{d1,devices1}
\begin{equation}
T=\langle\phi_L|\mathcal{H}_{\rm eff}^D|\phi_R\rangle=\int_{-\infty}^{\infty}\phi_L^\dagger(x)\mathcal{H}_{\rm eff}^D\phi_R(x)dx
\tag{A25}
\end{equation}
Now, using $\phi_L^\dagger=f_L^*(x)\chi_L^\dagger$, $\sigma_x\chi_R=\chi_L$ and $\sigma_y\chi_R=-i\chi_L$, we obtain
\[
\begin{aligned}
\phi_L^\dagger\mathcal H_{\rm eff}^D\phi_R(x)
&=f_L^*(x)\chi_L^\dagger\left[M(x)\sigma_x
-i\hbar v_D\sigma_y\partial_x\right]
f_R(x)\chi_R\\
&=f_L^*(x)\left[M(x)f_R(x)
-\hbar v_D f_R'(x)\right]\chi_L^\dagger\chi_L.
\end{aligned}
\]
Now, plugging $\chi_L^\dagger\chi_L=1$, and $f_R'(x)=-\alpha(x-x_R)f_R(x)$, we get
\begin{equation}\label{Ta}
T=\int_{-\infty}^{\infty} dx\,f_L^*(x)f_R(x)
\left[M(x)+\hbar v_D\alpha(x-x_R)\right].
\tag{A26}
\end{equation}
Substituting the Gaussian envelope product $f_L^*(x)f_R(x)=\sqrt{\frac{\alpha}{\pi}}e^{-\alpha x_*^2}e^{-\alpha(x-x_0)^2}$ in Eq. \eqref{Ta} gives
\begin{multline}\label{Tb}
T=\sqrt{\frac{\alpha}{\pi}}e^{-\alpha x_*^2}\Bigg[\int_{-\infty}^{\infty} M(x)e^{-\alpha(x-x_0)^2}
dx\\{+\hbar v_D\alpha\int_{-\infty}^{\infty}xe^{-\alpha(x-x_0)^2}
dx}\\-\hbar v_D\alpha x_R\int_{-\infty}^{\infty}e^{-\alpha(x-x_0)^2}
dx\Bigg].
\tag{A27}
\end{multline}
Now we evaluate each term one by one. The first term is the integration of the mass term. As the overlap is predominantly discernible around the midpoint ($x \approx x_0$) due to the steepness of the factor $e^{-\alpha (x-x_0)^{2}}$, one can approximate the mass profile by its constant central plateau value $M(x)=M(x_0)\sim2t^2$. This leaves: $\int_{-\infty}^{\infty} M(x_0)e^{-\alpha(x-x_0)^2}
dx=M(x_0)\sqrt{\frac{\pi}{\alpha}}$. For the second integration term, we consider $y=x-x_0$, so $x=y+x_0$ and $dx=dy$, and the integrand $xe^{-\alpha(x-x_0)^2}=(y+x_0)e^{-\alpha y^2}$. Here, $ye^{-\alpha y^2}$ is an odd function evaluated over perfectly symmetric limits $[-\infty, \infty]$. Therefore, the second integral $\int_{-\infty}^{\infty}xe^{-\alpha(x-x_0)^2}dx=\int_{-\infty}^{\infty}ye^{-\alpha y^2}dy+x_0\int_{-\infty}^{\infty}e^{-\alpha y^2}dy=x_0\sqrt{\frac{\pi}{\alpha}}$. While the third term is the standard Gaussian integral, which produces $\int_{-\infty}^{\infty}e^{-\alpha(x-x_0)^2}
dx=\sqrt{\frac{\pi}{\alpha}}$. Combining all these results, Eq. \eqref{Tb} reduces to
\begin{equation}\label{a22}
\begin{aligned}
T&=\Big(M(x_0)-\hbar v_D\alpha (x_R-x_0)\Big)e^{-\alpha x_*^2}\\&=(M(x_0)-\hbar v_D\alpha x_*)e^{-\alpha x_*^2}
\end{aligned}
\tag{A28}
\end{equation}
Using $l_G=\alpha^{-1/2}$, we can rewrite Eq. \eqref{a22} as
\begin{equation}
T=g\hbar\frac{v_D}{l_G}e^{-\alpha x_*^2}
\tag{A29}
\end{equation}
with $g=\Big(\frac{M(x_0)l_G}{\hbar v_D}-\frac{x_*}{l_G}\Big)$ is a dimensionless quantity. Thus, the tunneling strength can be scaled as $T\sim \hbar \frac{v_D}{l_G} e^{-\alpha x_*^2}$ for $g$ of order unity, provided that it does not introduce additional significant dependence on the DW separation. Now, substituting $v_D=4at^2/\hbar$, and $l_G=\sqrt{\frac{\sqrt2\,t\,\xi\Delta_0}
{\left(\Delta_0^2-2t^2\right)}}$, we arrive at
\begin{equation}\label{Tmatrix-a}
T\sim\sqrt{\frac{8\sqrt{2}a^2\Delta_{0}t^3}{\xi}\Big(1-\frac{2t^2}{\Delta_{0}^2}\Big)}\exp\Big[-\frac{\Delta_{0}}{\sqrt{2}t\xi}\Big(1-\frac{2t^2}{\Delta_{0}^2}\Big)x_{\star}^2\Big],
\tag{A30}
\end{equation}

Secondly, rather than evaluating a line integral, the total tunnel coupling $T$ is determined via a semi-classical scaling approximation\cite{scaling-a}. The kinetic part of the continuum Hamiltonian $\mathcal{H}_{\rm eff}^D$ is given by
$\mathcal{H}_{\rm k}=-i\hbar v_D\sigma_y\partial_x$. A wavefunction localized over a length $l_G$ varies predominantly on that scale, so its partial derivative ($\partial_x$) is of the order $1/l_G$. Consequently, the characteristic kinetic-energy scale associated with the localized mode is estimated as $E_{\rm k}\sim \hbar\frac{v_D}{l_G}$. Therefore, the pre-exponential coefficient is governed exclusively by the intrinsic quantum confinement energy scale of an isolated zero-energy JRMs. However, the spatial overlap of the two Gaussian envelopes gives $S_{\rm env} = e^{-\alpha x_\star^2}$. Combining the localized energy scale with the envelope overlap factor produces the scaling relation $T \sim \hbar \frac{v_D}{l_G} e^{-\alpha x_\star^2}$. Notice that the exponential dependence arises from the overlapping of Gaussian envelopes of the JRMs.

\subsection{Asymptotic form of $T$ and $\Delta E$}
\label{asymptotic}
For widely separated DWs, the two JRMs are localized near $x_L$ and $x_R$, respectively, while their exponentially small overlap is determined by the tails extending into the region between the two DWs. In particular, when $D_s$ is sufficiently large, there is a broad central region around the midpoint $x_0$ that lies inside the interval $x_L<x<x_R$ and is far from both DWs. In this region, the consideration of linearized mass approximation near the individual DWs is no longer applicable. Instead, since $\Delta(x_0)=0,$ and $|\Delta(x)|\ll\sqrt{2}t$ over a sufficiently broad region around $x_0$, the effective mass approaches the constant value $M_{\mathrm{asym}}=2t^2$. Thus, in this approximately constant-mass region, the Dirac equation admits exponentially decaying solutions. The Gaussian JR envelopes near the DWs, therefore, develop asymptotic exponential tails as they extend away from their respective DWs toward the central region. Following Eq. (\ref{low}), the asymptotic JR solutions are therefore obtained as
\begin{equation}\label{envelope3}
 \Psi_{j}(x)\sim e^{-\kappa|x-x_{j}|}
\tag{A31} \end{equation}
where $\kappa=\frac{|M_{\mathrm{asym}}|}{v}$, $v=\frac{4at^2}{\hbar}$ $j=L/R$ and asymptotic localization length becoming $l_D=\kappa^{-1}$.

Consequently, the leading exponential dependence of the spatial overlap between two asymptotic tails is $S_{\rm env}=\int dx f_{L}^*(x)f_{R}(x)\sim e^{-\kappa R l_{G}}$. Therefore, the tunneling matrix element and tunnel splitting behave as
\begin{equation}\label{asymp}
T\sim T_0(\Delta_0)e^{- R (l_G/l_D)},
\quad
\Delta E=2T
\simeq 2T_0(\Delta_0)e^{- R (l_G/l_D)}.
\tag{A32}\end{equation}
with $R$ is the dimensionless ratio mentioned in the main text. This exponential form (Eq. \eqref{asymp}) is distinct from the Gaussian-overlap expression of Eq.~\eqref{Tmatrix} and describes the large-$D_s$ regime in which the spatial overlap is governed by the far tails of the JRMs.

We can also observe the explicit dependence of $T$ and $\Delta E$ on $\xi$ both in the spatial overlap and asymptotic tail regime.\\ 
$\bullet$~~{\bf Spatial overlap regime:} The tunneling strength, given in Eq. (\ref{Tmatrix}), has the form: $T\sim e^{-D_{s}^2/(2l_G)^2}$ and $\Delta E\sim e^{-D_{s}^2/(2l_G)^2}$. Plugging $D_s$ and $l_G$ into these equations, we now get
\begin{equation}\label{ccc}
T\sim e^{-c_G\xi},
\quad \text{with}\quad
c_G=
\frac{\Delta_0^2-2t^2}{\sqrt{2}\,t\Delta_0}
\operatorname{arctanh}^2
\left(\frac{\sqrt{2}\,t}{\Delta_0}\right).
\tag{A33}
\end{equation}\\
$\bullet$~~{\bf Asymptotic-tail regime:} In this regime, the form of $T$ and $\Delta E$ is given in Eq. (\ref{asymp}). Substituting $D_s$ and $l_D$ in Eq. (\ref{asymp}), we now arrive at
\begin{equation}\label{ddd}
T\sim e^{-c_D\xi},
\quad \text{with}\quad
c_D=\operatorname{arctanh}
\left(\frac{\sqrt{2}\,t}{\Delta_0}\right).
\tag{A34}
\end{equation}

\section{}
\label{app:B}

\subsection{Real-space Analysis: Numerical Lattice Study}
\label{numerical}
To simulate the JR mode numerically, we need to consider the spatially dependent hopping modulation such that they flip their relative magnitude across the center of the DW. Following\cite{mandal,scollon}, we begin by introducing a spatially varying modulation parameter $\Delta(x)$ for the numerical diagonalization of the real-space Hamiltonian. Thus, the modulation in hopping strength now takes the following form
\begin{equation}\label{domain}
\Delta(x)=\Delta_{0}\tanh\Big[\frac{i-i_{0}}{\xi/a}\Big]
\tag{B1}\end{equation}
in which $\Delta_{0}$, and $\xi$ represent, respectively, the amplitude and the width of the DW. Here, the index $i$ refers to the location of the $i$-th site of the chain, and the parameter $i_{0}$ fixes the position of the DW. In this study, we fix the position of the DW at $i_{0}=L/2$ in a way that the left side ($i<i_{0}$) is in one topological phase and the right side ($i>i_{0}$) lies in another topological phase. A smooth (sharp) DW can be achieved for large (small) $\xi$. With this, the diagonalization of the Hamiltonian given in Eq. (\ref{1}) under OBC results in two zero modes. 

We should emphasize here that, unlike the physical edge of the chain, this type of DW creates an effective or artificial boundary inside the chain, separating and connecting two distinct topological phases of the system. It is noted that physical boundaries or edges of the SSH chain can be considered as DWs with the vacuum or surrounding and the DWs, similar to the physical edge of the SSH chain, host zero-energy localized states\cite{mandal}.

\subsection{Local Winding Number Calculation and JR Mechanism}
\label{wind}
The system without DW contains uniform modulation $\Delta$ and therefore preserves  translation invariance. Thus, the topological property of the uniform chain can be described by the global bulk Bloch winding number, as discussed in  Section \ref{sec2} of the main text. In contrast, the JR mechanism considers a spatially varying modulation parameter $\Delta(x)$, which explicitly breaks the translational invariance\cite{dw}. Therefore, a single global bulk winding number cannot be assigned to the entire inhomogeneous chain in the same sense as for a uniform system. Instead, the JR mechanism is characterized by the local topology and by the associated effective mass inversion at the DWs. Thus, one may define a local (or a position-dependent) winding number by treating $\Delta(x)$ as a slowly varying parameter (or as a locally uniform parameter) and evaluating the winding number of the corresponding locally homogeneous Hamiltonian. We have therefore avoided interpreting the full inhomogeneous system as globally trivial or nontrivial and instead characterize it through the spatially resolved local topological invariant.

As $\Delta(x)$ varies slowly, we can use a local/frozen block off-diagonal Hamiltonian given in Eq. \eqref{7aa} at each position $x$: $\mathcal{\widetilde H}_x(k,\Delta)=\mathcal{\widetilde H}_x(k,\Delta(x))$. This is a hypothetical homogeneous Hamiltonian obtained by freezing the slowly varying parameter at its value at $x$. We can then define a local winding number as\cite{mandal,bdi}
\begin{equation}
{\mathcal W}_{\mathrm{loc}}(x)=\frac{1}{2\pi i}\int_{BZ}dk \partial_{k}\ln[{\bf Det}[V(k,\Delta(x))]]
 \tag{B2}\end{equation}
with ${\bf Det}[V(k,\Delta(x))]=t^2-\Delta^2(x)-t^2e^{i4ak}$. This is what allows us to discuss the local topology of the spatially varying system. Notice that the local Hamiltonian is gapless at the three points $\Delta(x)=-\sqrt{2}t$, $\Delta(x)=\sqrt{2}t$ and $\Delta(x)=0$. Then, for a gapped local Hamiltonian, the winding number can be defined as
\begin{equation}\label{winding}
{\mathcal W}_{\mathrm{loc}}(x) = \begin{cases}
     1, & 0<|\Delta(x)|<\sqrt{2}t,\\[2pt]
    0, & |\Delta(x)|>\sqrt{2}t,\quad \Delta(x)=0,\\[2pt]
    \text{undefined},& |\Delta(x)|=\sqrt{2}t
  \tag{B3}\end{cases}
\end{equation}

\begin{figure}
     \vskip -.1 in
   \begin{picture}(100,100)
     \put(-90,0){
  \includegraphics[width=.54\linewidth, height=1.15 in]{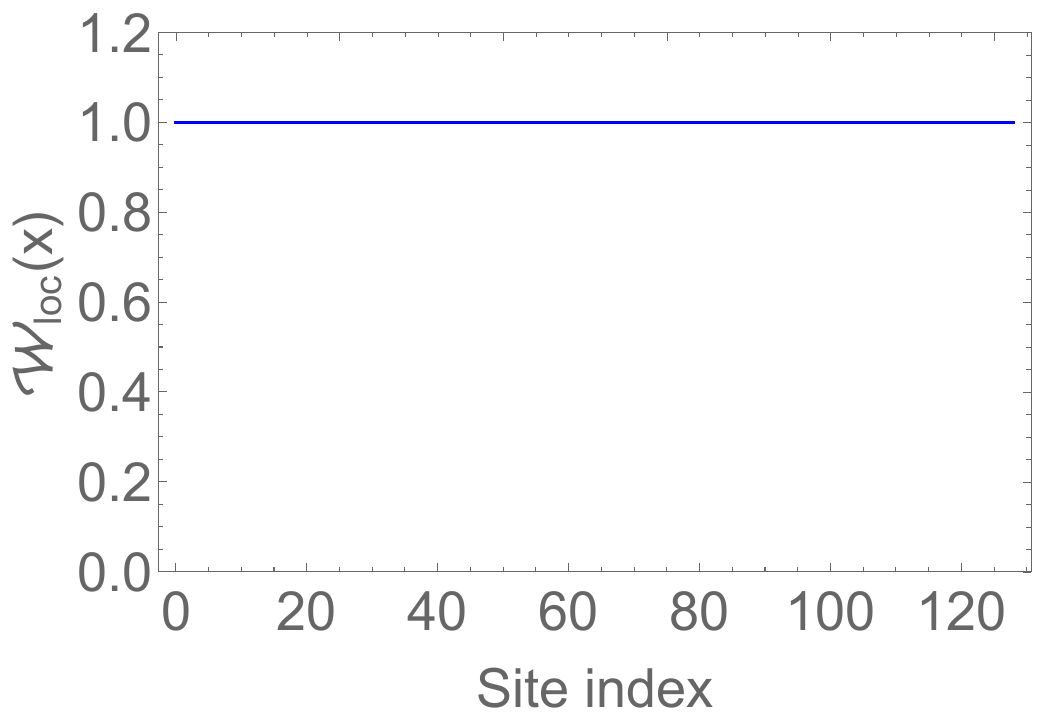}
  \includegraphics[width=.52\linewidth, height=1.15 in]{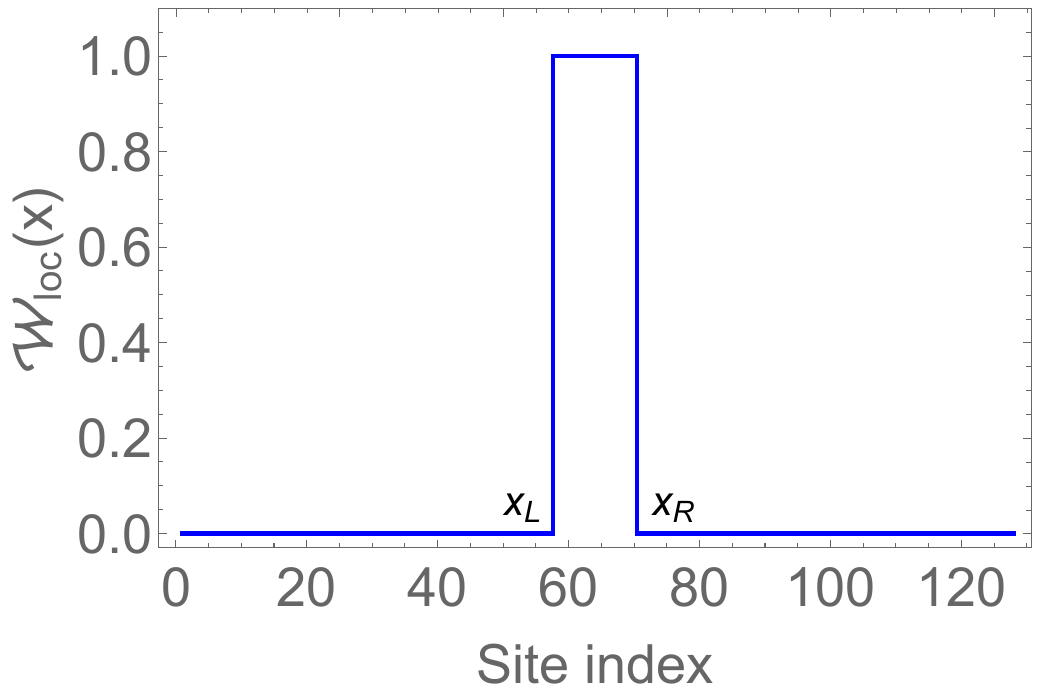}}
     \put(-40,58){(a)}
     \put(80,60){(b)}
    \end{picture}
\caption{{\bf Local winding number} for (a) $\Delta_{0}<\sqrt{2}t$ and (b) $\Delta_{0}>\sqrt{2}t$. Here, we set $L=128$, $\xi/a=10$.}
\label{winda}
\end{figure}
Let us now connect this local winding number with the JR mechanism around $k_0=\pm\pi/4a$ where the effective Dirac mass is $M(x)=2t^2-\Delta^2(x)$, with $\Delta(x)$ can be taken from Eq. \eqref{kink}. 

{\bf Case I:} for $\Delta_{0}<\sqrt{2}t$, $0<|\Delta(x)|<\Delta_{0}<\sqrt{2}t$, and no mass inversion is noticeable (see Fig. \ref{dw2}(b)). So ${\mathcal W}(x)=1$ $\forall x$ (see Fig. \ref{winda}(a)) and non-existence of JRMs. 

{\bf Case II:} For $\Delta_{0}>\sqrt{2}t$, the condition $M(x)=0$ gives two mass inversion points at the DWs $x_{L}=x_{0}- x_{\star}$ and $x_{R}=x_{0}+x_{\star}$, corresponding to $\Delta(x)=-\sqrt{2}t$ and $\Delta(x)=\sqrt{2}t$, respectively (notice Fig. \ref{dw2}(d)). Importantly, $\Delta(x)=0$ at the point $x=x_0$ but with $M(x)=2t^2$. Hence, for the spatially varying profile, $\Delta(x_0)=0$ is simply an ordinary positive mass point for the continuum theory around $k_0=\pm\pi/4a$, which therefore lies within the position mass region (or within $x_L<x<x_R$, particularly at $x=x_0$) and does not constitute a DW. The DWs arise, instead, from the two spatial zeros of $M(x)$ located at $\Delta(x_L)=-\sqrt2t$ and $\Delta(x_L)=-\sqrt2t$. Within the two DWs $x_{L}<x<x_{R}$; the condition $|\Delta(x)|<\sqrt{2}t$ gives ${\mathcal W}_{\mathrm{loc}}(x)=1$. While outside the region, the condition $|\Delta(x)|>\sqrt{2}t$ provides ${\mathcal W}_{\mathrm{loc}}(x)=0$ (see Fig. \ref{winda}(b)). Thus, we arrive at
\begin{equation}\label{winding}
\mathcal{W}_{\mathrm{loc}}(x)=
\begin{cases}
0, & x<x_L,\quad x>x_R,\\[2pt]
1, & x_L<x<x_R,\\[2pt]
\text{undefined}, & |\Delta(x)|=\sqrt{2}t,
\end{cases}
\qquad
\tag{B4}
\end{equation}
where the latter points correspond to the local bulk gap closings. Equivalently, within the continuum Dirac description, the local topological character is determined by the sign of the effective Dirac mass $M(x)$: the region with $M(x)<0$ corresponds to $\mathcal{W}_{\mathrm{loc}}(x)=0$, while the region with $M(x)>0$ corresponds to $\mathcal{W}_{\mathrm{loc}}(x)=1$, within the convention adopted here. Thus, for \(\Delta_0>\sqrt{2}t\), the DW profile produces a spatial sequence: $\mathcal{W}_{\mathrm{loc}}(x):0\rightarrow1\rightarrow0$. The negative effective mass regions outside \(x_L\) and \(x_R\) are locally trivial, whereas the positive effective mass region within \(x_L<x<x_R\) is locally non-trivial. Consequently, the two DWs correspond to the changes in local winding number as \(0\rightarrow1\) and \(1\rightarrow0\), and each DW supports a JRM. Therefore, the two JRMs originate from the two local topological DWs, rather than from a single global bulk winding number of the entire spatially inhomogeneous chain. The finite region between the two DWs can therefore be regarded as a (locally) finite non-trivial topological segment embedded in a locally (trivial) topological background.

Although the regions outside the two DWs have the same trivial winding number, $\mathcal{W}_{\rm loc}(x<x_L)=\mathcal{W}_{\rm loc}(x>x_R)=0$, this does not eliminate the existence of DW states. Each DW separately connects two regions with different local topological character and is therefore associated with a unit change in the local winding number $\Delta \mathcal{W}_{\rm loc}(x)=\pm1$\cite{teo}. Therefore, $$\Delta {\mathcal W}_{\rm loc}(x)\Big|_{x=x_L}=+1,\quad \Delta {\mathcal W}_{\rm loc}(x)\Big|_{x=x_R}=-1.$$ The signs depend not only on the direction in which the DW is traversed, but also depend on the chiral eigenvalues at the respective DWs\cite{bdi}, and the magnitude is one. Similarly to the bulk-boundary correspondence, here the bulk-defect correspondence provides $N(x_{L})=|\Delta {\mathcal W}(x_{L})|=1$ and $N(x_{R})=|\Delta {\mathcal W}(x_{R})|=1$\cite{teo}. Thus, each DW supports only a JRM. Hence, the DW states exemplify the bulk-defect correspondence, which states that the change in the local bulk winding number across the DWs guarantees localized JRMs\cite{bdi,teo}.

\begin{figure}[h]
     \vskip -.1 in
   \begin{picture}(100,100)
     \put(-90,0){
  \includegraphics[width=.54\linewidth, height=1.245 in]{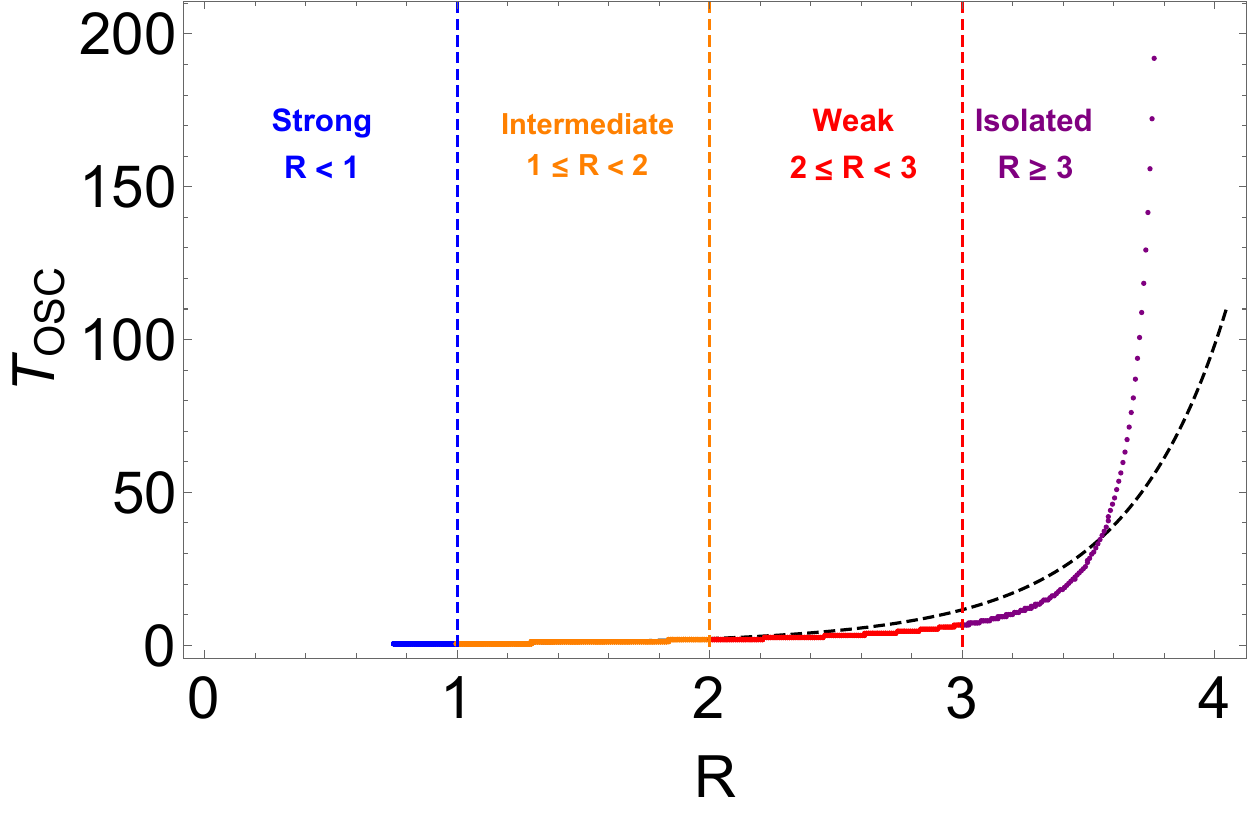}
  \includegraphics[width=.52\linewidth, height=1.255 in]{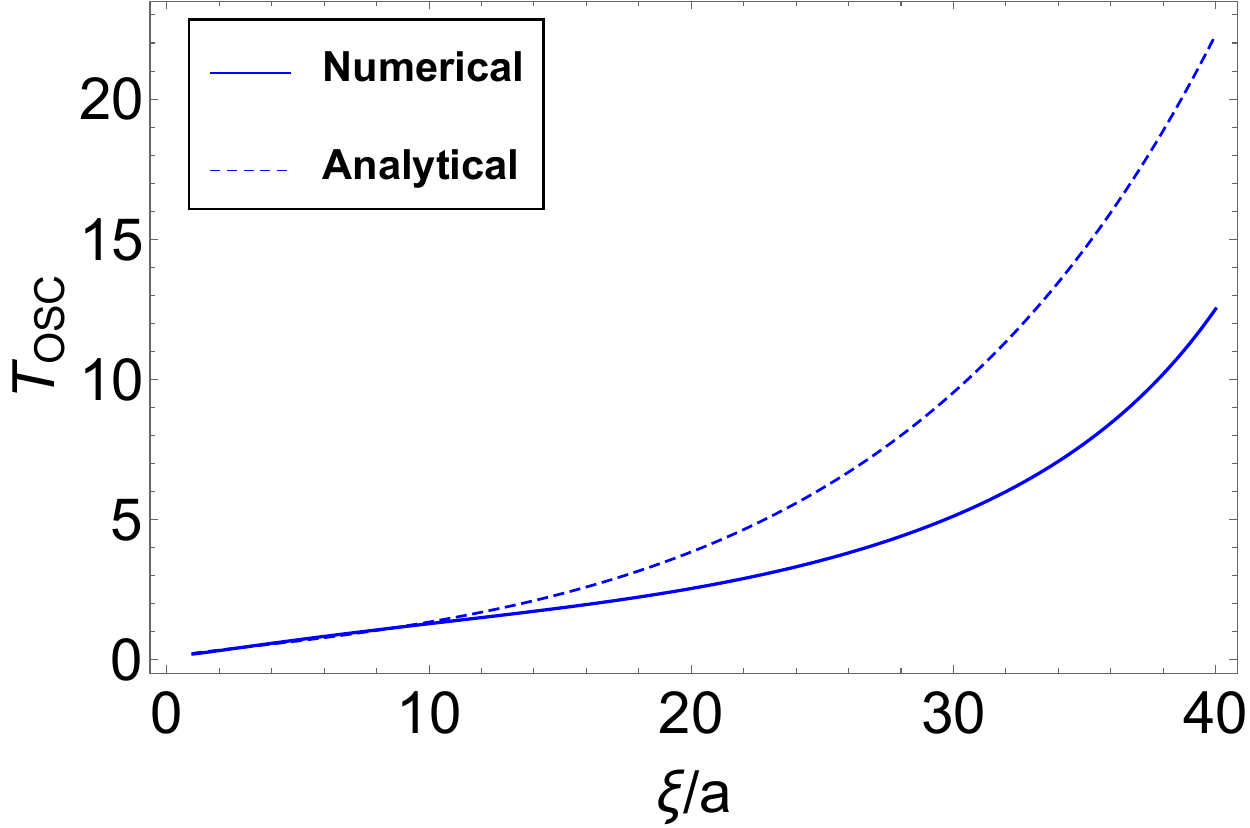}}
     \put(-30,50){(a)}
     \put(120,65){(b)}
    \end{picture}
\caption{{\bf Tunneling period} vs $R$ with $\xi/a=10$ (a) and vs DW width $\xi/a$ for $\Delta_{0}=20$ (b).} 
\label{timep}
\end{figure}
\begin{figure*}[t]
\centering
    \includegraphics[width=.43\linewidth, height=2.2 in]{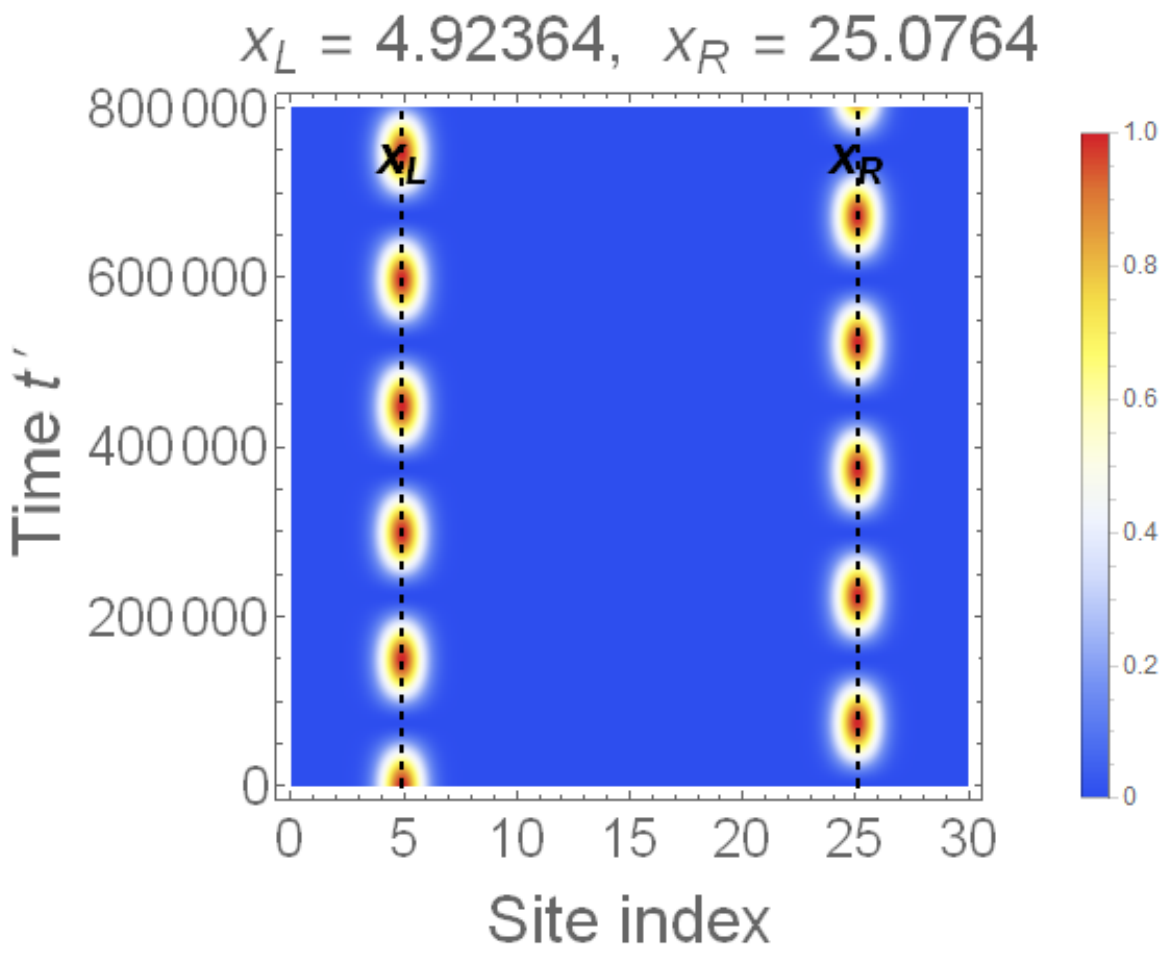}
  \includegraphics[width=.4\linewidth, height=2.2 in]{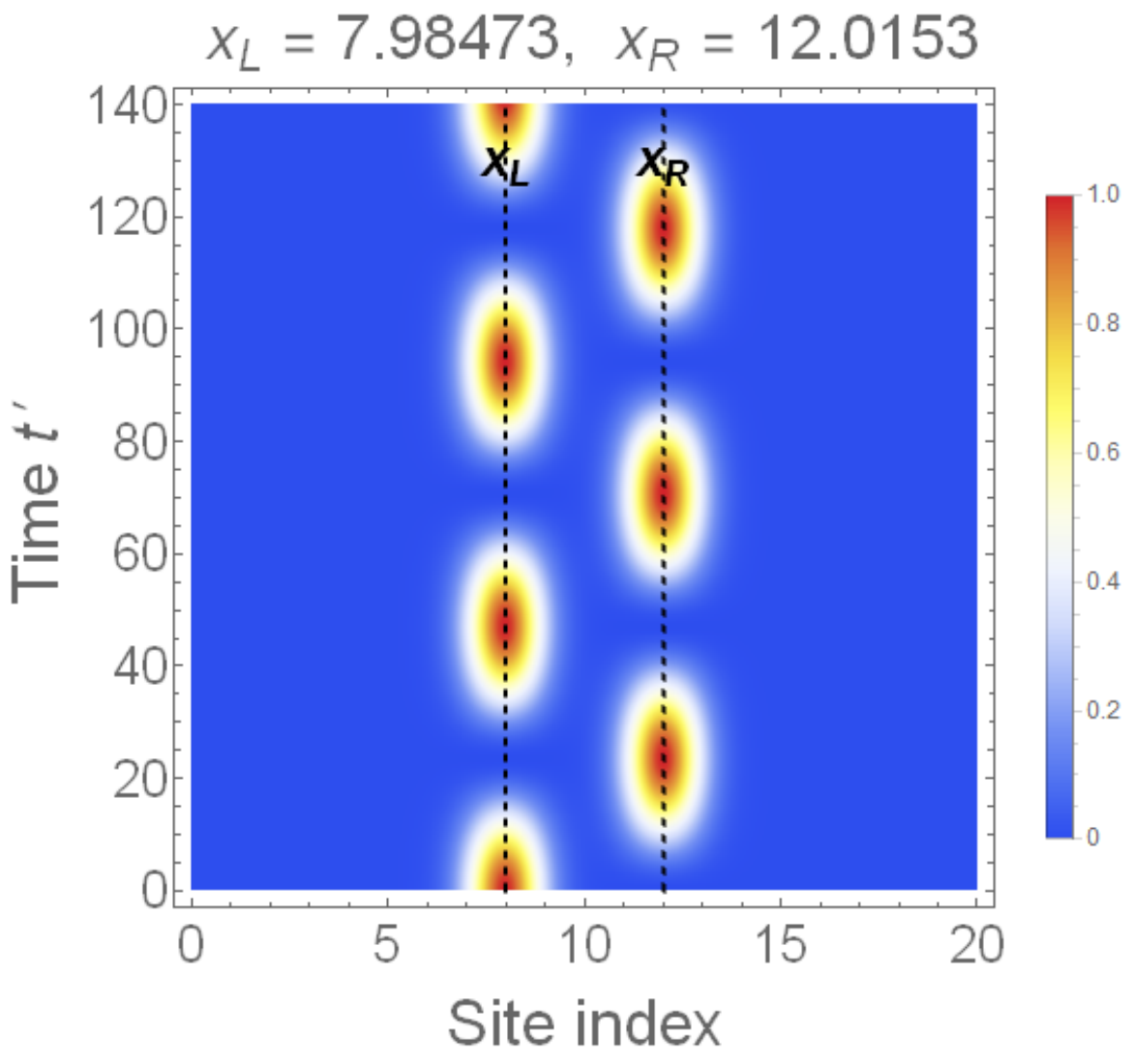}
     \put(-320,66){(a)}
     \put(-150,66){(b)}\\
    \includegraphics[width=.4\linewidth, height=2.2 in]{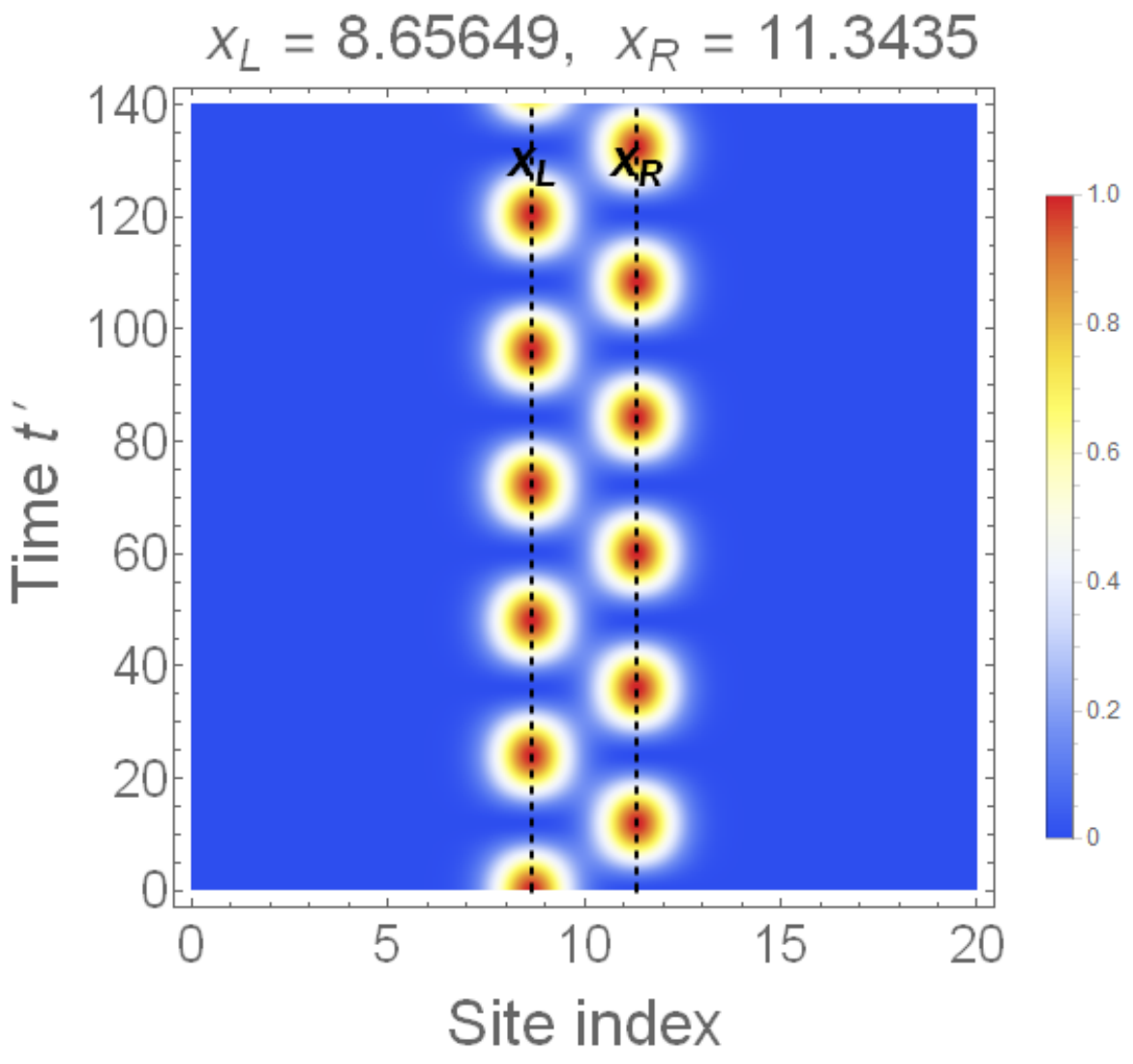}
    \includegraphics[width=.4\linewidth, height=2.2 in]{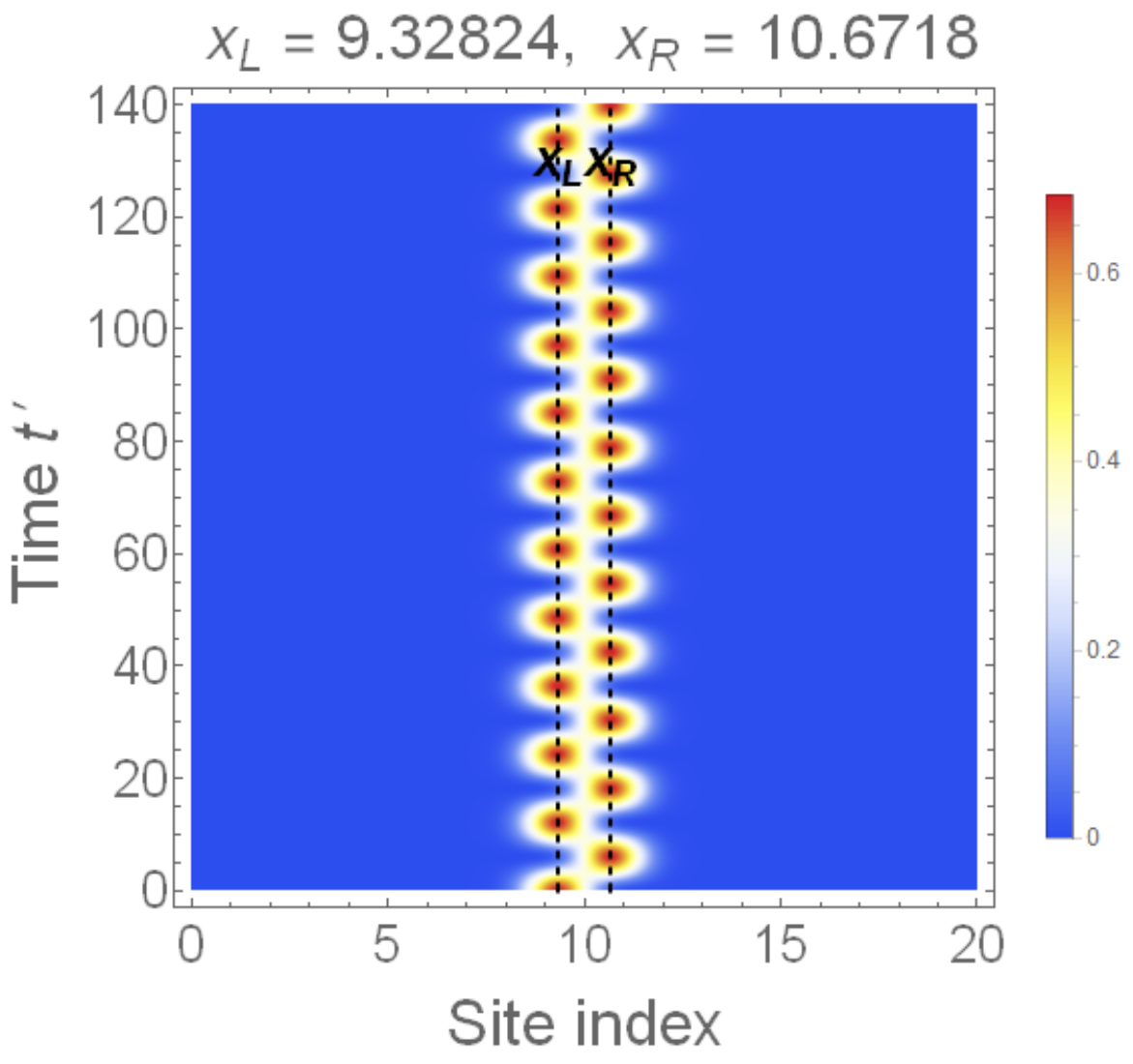}
   \put(-350,66){(c)}
    \put(-150,68){(d)}
\caption{ {\bf Probability density} $\rho(x,t)$ illustrating coherent tunneling between two JRMs localized at the DWs $x_{L}$ and $x_{R}$ for varying $\xi/a$ and fixed $\Delta_{0}=8.5$. The vertical dashed lines indicate the DW positions $x_{L}$ and $x_{R}$. Other parameters are: (a) $L=30$, $\xi/a=60$, (b) $L=20$, $\xi/a=12$, (c) $L=20$, $\xi/a=8$, and (d) $L=20$, $\xi/a=4$.}
\label{tunnela}
\end{figure*}
\subsection{Graph for Oscillation Period and Probability Density}
\label{tunneling}
The oscillation period is related with the tunneling matrix via $T_{osc}=\frac{\pi}{T}=\frac{2\pi}{\Delta E}$. This tunneling matrix has been elaborately derived in Eq. \eqref{Tmatrix} for the spatial overlap regime and in Appendix \ref{asymptotic} for the asymptotic-tail regime. Therefore, based on these equations, the nature of the tunneling period with $R$ and $\xi/a$ is plotted in Fig. \ref{timep}, which shows the exponential growth of $T_{osc}$ with $R$ and $\xi/a$.

In Fig. \ref{tunnela}, we depict the probability density function for different $\xi$, keeping $\Delta_{0}$ fixed. The plot illustrates that density transfer becomes progressively slower as $\xi/a$ increases (see Fig. \ref{tunnela}(a)). One can notice the strong hybridization for a very sharp domain wall (Fig. \ref{tunnela}(d)). The linear correspondence between $D_{s}$ and $\xi$ is expected as $D\propto\xi$.


\begin{thebibliography}{00}
\bibitem{comment} The oscillation of $\langle x(t^{\prime})\rangle$ does not mean that a JR mode physically moves from one DW to the other. Instead, it means that the occupation probability tunnels back and forth between two fixed JR modes.
\end{thebibliography}
\end{document}